\documentclass[aap,nameyear,seceqn,MSNbibl,dvips]{arximspdf}
\usepackage{graphicx}
\doi{10.1214/09-AAP643}
\volume{20}
\issue{4}
\pubyear{2010}
\firstpage{1253}
\lastpage{1302}

\makeatletter

\newtheorem{lemma}{Lemma}[section]
\newtheorem{proposition}[lemma]{Proposition}
\newproclaim{remark}[lemma]{Remark}
\newproclaim{definition}[lemma]{Definition}
\newtheorem{corollary}[lemma]{Corollary}
\newtheorem{theorem}[lemma]{Theorem}
\newproclaim{example}[lemma]{Example}

\makeatother

\begin{document}
\begin{frontmatter}

\title{Optimal investment policy and dividend payment strategy in an
insurance company}
\runtitle{Optimal investment policy and dividend payment strategy}

\begin{aug}
\author[A]{\fnms{Pablo} \snm{Azcue}\corref{}\ead[label=e1]{pazcue@utdt.edu}} and
\author[A]{\fnms{Nora} \snm{Muler}\ead[label=e2]{nmuler@utdt.edu}}
\runauthor{P. Azcue and N. Muler}
\affiliation{Universidad Torcuato Di Tella and Universidad Torcuato Di Tella}
\address[A]{Departamento de Matem\'{a}ticas\\
Universidad Torcuato Di Tella\\
Mi\~{n}ones 2159\\
C1428ATG Buenos Aires\\
Argentina\\
\printead{e1}\\
\phantom{E-mail: }\printead*{e2}} 
\end{aug}

\received{\smonth{12} \syear{2007}}
\revised{\smonth{9} \syear{2009}}

%
\begin{abstract}
We consider in this paper the optimal dividend problem for an insurance
company whose uncontrolled reserve process evolves as a classical
Cram\'{e}r--Lundberg process. The firm has the option of investing part
of the
surplus in a Black--Scholes financial market. The objective is to find a
strategy consisting of both investment and dividend payment policies which
maximizes the cumulative expected discounted dividend pay-outs until
the time
of bankruptcy. We show that the optimal value function is the smallest
viscosity solution of the associated second-order integro-differential
Hamilton--Jacobi--Bellman equation. We study the regularity of the
optimal value
function. We show that the optimal dividend payment strategy has a band
structure. We find a method to construct a candidate solution and
obtain a
verification result to check optimality. Finally, we give an example
where the
optimal dividend strategy is not barrier and the optimal value function
is not
twice continuously differentiable.
\end{abstract}

%
\begin{keyword}[class=AMS]
\kwd[Primary ]{91B30}
\kwd[; secondary ]{91B28}
\kwd{91B70}
\kwd{49L25}.
\end{keyword}
\begin{keyword}
\kwd{Cram\'{e}r--Lundberg process}
\kwd{insurance}
\kwd{dividend payment strategy}
\kwd{optimal investment policy}
\kwd{Hamilton--Jacobi--Bellman equation}
\kwd{viscosity solution}
\kwd{risk control}
\kwd{dynamic programming principle}
\kwd{band strategy}
\kwd{barrier strategy}.
\end{keyword}

\pdfkeywords{91B30, 91B28, 91B70, 49L25, Cramer--Lundberg process,
insurance, dividend payment strategy,
optimal investment policy, Hamilton--Jacobi--Bellman equation,
viscosity solution, risk control,
dynamic programming principle,
band strategy,
barrier strategy}

\end{frontmatter}

\section{Introduction}\label{sec1}

A classical problem in actuarial mathematics is to maximize the
cumulative expected discounted dividend pay-outs. In the
Cram\'{e}r--Lundberg setting, this optimization problem was introduced
by De Finetti (\citeyear{Finetti57}); Gerber (\citeyear{G69}) proved the
existence of an optimal dividend payment strategy and showed that it
has a band structure. The cumulative expected discounted dividend
pay-outs is a way to value a company as it can be seen, for instance, in
the classical paper by Miller and Modigliani (\citeyear{MM61}) for the
deterministic case and more recently in Sethi, Derzko and Lehoczky
(\citeyear{SDL84a}, \citeyear{SDL84b}) and Sethi (\citeyear{S96})
for the stochastic case.

In this paper we consider this optimization problem in the classical
Cram\'{e}r--Lundberg setting, but we allow the management the
possibility of
controlling the stream of dividend pay-outs and of investing part of the
surplus in a Black and Scholes financial market. We impose a borrowing
constraint: short-selling of stocks or to borrow money to buy stocks is not
allowed. Technically, the unconstrained optimization problem is simpler.

Azcue and Muler (\citeyear{AM05}) consider the problem of maximizing
the cumulative
expected discounted dividend pay-outs of an insurance company when the
management has the possibility of controlling the risk exposure by
reinsurance. In this case, the optimal value function was characterized
as the
smallest viscosity solution of the first-order integro-differential
Hamilton--Jacobi--Bellman equation, and the optimal dividend payment
strategy was found.

In this paper, the optimization problem is more complex than the one we
treated before. One difference is that the associated Hamilton--Jacobi--Bellman
equation is a nonlinear degenerate second-order integro-differential equation
subject to a differential constraint. The possibility that the
ellipticity of
the second-order operator involved in this equation can degenerate at any
point together with the fact that there is an integral term, makes it more
difficult to prove the existence and regularity of solutions. However,
when we
obtain the solution of this operator in Section~\ref{sec6}, we see that the ellipticity
only degenerates at zero and so the degeneracy is not as serious as it could
be (the solution turns out to be twice continuous differentiable). Another
difference is that, since in this case the controlled surplus involves a
Brownian motion, there is not an optimal strategy. Nevertheless, we
prove that
the optimal value function can be written explicitly as a limit of value
functions of strategies. So, we introduce the notion of limit dividend
strategies and prove that the optimal limit strategy has a \textit{band
structure}.

In a diffusion setting, which means that the surplus is modeled as a Brownian
motion, different cases were studied; we can mention, for instance,
Asmussen and
Taksar (\citeyear{AT97}) for the problem of dividend optimization and
H\o jgaard and
Taksar (\citeyear{HT04}) for the case of dividend, reinsurance and portfolio
optimization. The main difference between the two settings is that the HJB
equation in the diffusion case is a differential equation and not an
integro-differential one. Other differences are that in the diffusion setting
the optimal strategies are always barrier strategies, that there is a natural
boundary condition at zero for the associated HJB equation and that this
equation has always classical concave solutions; these properties might not
occur in the Cram\'{e}r--Lundberg setting.

Avram, Palmowski and Pistorius (\citeyear{APP07}) study the problem of
maximizing the
discounted dividend pay-outs when the uncontrolled surplus of the company
follows a general spectrally negative L\'{e}vy process in absence of
investment. The HJB equation associated with this optimization problem
is also
a second-order integro-differential equation but its ellipticity does
not degenerate.

In both, H\o jgaard and Taksar (\citeyear{HT04}) and Avram, Palmowski
and Pistorius
(\citeyear{APP07}), the corresponding HJB equations are second-order
equations whose
ellipticity does not degenerate at zero, so to characterize the optimal value
function among the solutions of the HJB equation they use the natural boundary
condition at zero. In this paper, we do not have a natural condition at zero
but we do not need this boundary condition because the ellipticity of
the HJB
degenerates at this point. The lack of a boundary condition at zero
makes more
difficult to obtain a numerical scheme.

The main results of this paper are the following:

In the first part of the paper, we obtain the optimal value function as the
smallest viscosity solutions of the associated HJB equation, and we
prove a
verification theorem that allows us, since the optimal value function
has not
a natural boundary condition at zero, to recognize the optimal value function
among the many viscosity solutions of the associated HJB equation.

From Section \ref{sec6} on, we assume that the claim-size distribution has a bounded
density; this allows us to show that the optimal value function is twice
continuously differentiable except possibly for some points. We find the
optimal value function for small surpluses, and we prove that the optimal
strategy is \textit{stationary}, that is, the decision of what
proportion of
the surplus is invested in the risky asset, and how much to pay out as
dividends at any time depends only on the current surplus. We also
prove that
the optimal dividend payment policy has a band structure. In
particular, the
optimal dividend payment policy for large surpluses is to pay out immediately
the surplus exceeding certain level as dividends. We also obtain the best
barrier strategy and show both an example where the optimal dividend payment
policy is barrier as well as an example where it is not. The second example
shows that, even for claim-size distributions with bounded density, the
optimal value function could be neither concave nor twice continuously
differentiable.

This paper is organized as follows. In Section \ref{sec2}, we state the optimization
problem and prove some properties about the regularity and growth of the
optimal value function. In Section \ref{sec3}, we state the dynamic programming
principle and show that the optimal value function is a viscosity
solution of
the HJB equation associated with the optimization problem. In Section
\ref{sec4}, we
prove the uniqueness of viscosity solutions of the HJB equation with a
boundary condition at zero. In Section \ref{sec5}, we prove that the optimal value
function is the smallest supersolution of the HJB equation and give a
verification theorem that states that a supersolution which can be
obtained as
a limit of value functions of admissible strategies is the optimal value
function. In Section \ref{sec6}, we construct via a fixed-point operator a classical
solution of the second-order integro-differential equation involved in
the HJB
equation. In Section \ref{sec7}, we use the solution obtained in Section \ref{sec6} to obtain
the value function of the optimal barrier strategy. In Section \ref{sec8}, we
find the
optimal value function for small surpluses, show that the optimal
strategy is
stationary and prove that the optimal dividend payment policy has a band
structure. In Section \ref{sec9}, we show some numerical examples. We have
placed some
technical lemmas in the \hyperref[app]{Appendix} to improve the
readability of the main text.

\section{The stochastic control problem}\label{sec2}

We assume that the surplus of an insurance company in the absence of control
of dividends payment and investment follows the classical Cram\'{e}r--Lundberg
process; that is, the surplus $X_{t}$ of the company is described by
%
%
\begin{equation} \label{XtOriginal}%
X_{t}=x+p t-{ \sum_{i=1}^{N_{t}}}
U_{i},
\end{equation}
where $x$ is the initial surplus, $p$ is the premium rate, $N_{t}$ is a
Poisson process with claim arrival intensity $\beta>0$ and the claim sizes
$U_{i}$ are i.i.d. random variables with distribution $F$. We assume
that the
distribution $F$ has finite expectation $\mu$ and satisfies $F(0)=0$.

We consider that the financial market is described as a classical
Black--Scholes model where we have a risk-free asset with price process $B_{t}
$ and a risky asset with price process $S_{t}$ satisfying%
\[
\cases{
dB_{t}=r_{0}B_{t}\,dt,\cr
dS_{t}=rS_{t}\,dt+\sigma S_{t}\,dW_{t},}
\]
where $W_{t}$ is a standard Brownian motion independent to the process $X_{t}$.
We consider for simplicity $r_{0}=0$.

We define $\Omega$ as the set of paths with left and right limits and
$(\Omega,\mathcal{F},P)$ as the complete probability space with filtration
$ ( \mathcal{F}_{t} ) _{t\geq0}$ generated by the processes $X_{t}$
and $W_{t}$. A \textit{control strategy} is a process $\pi=(\gamma_{t},L_{t})$
where $\gamma_{t}\in[0,1]$ is the proportion of the surplus
invested in
stocks at time $t$, and $L_{t}$ is the cumulative dividends the company has
paid out until time $t$. The control strategy $(\gamma_{t},L_{t})$ is
\textit{admissible} if the process $\gamma_{t}$ is predictable and the process
$L_{t}$ is predictable, nondecreasing and c\`{a}gl\`{a}d (left continuous
with right limits).

We are considering the case where $\gamma_{t}\in[0,1]$ because we
are allowing neither short-selling of stocks nor borrowing money from other
sources to buy stocks.

Denote by $\Pi_{x}$ the set of all the admissible control strategies with
initial surplus~$x$. For any $\pi\in\Pi_{x}$, the controlled risk process
$X_{t}^{\pi}$ can be written as
%
%
\begin{equation}\label{Xt}%
X_{t}^{\pi}=x+pt+r\int_{0}^{t} X_{s}^{\pi}\gamma_{s}\,ds+\sigma
\int_{0}^{t} X_{s}^{\pi}\gamma_{s}\,dW_{s}-
{ \sum_{i=1}^{N_{t}}}
U_{i}-L_{t}.
\end{equation}

All the jumps of the process $X_{t}^{\pi}$ are downward, $X_{t^{-}}^{\pi
}-X_{t}^{\pi}>0$ if there is a claim at time $t$ and $X_{t}^{\pi
}-X_{t^{+}%
}^{\pi}>0$ only at the discontinuities of $L_{t}$. We also ask $\Delta
L_{t}:=L_{t^{+}}-L_{t}\leq X_{t}^{\pi}$ for any $t\geq0$; this means
that the
company cannot pay immediately an amount of dividends exceeding the surplus.

Given an admissible strategy $\pi\in\Pi_{x}$, let $\tau^{\pi}=\inf
\{t\geq0\dvtx X_{t}^{\pi}<0\}$ be the \textit{ruin time} of the
company, note
that it can only occur at the arrival of a claim. We define the value function
of $\pi$ by%
%
%
\begin{equation}\label{Vpi}%
V_{\pi}(x)=E_{x}\biggl(\int_{0}^{\tau^{\pi}}e^{-cs}\,dL_{s}\biggr),
\end{equation}
where $c$ is the discount factor. The integral is interpreted pathwise
in a
Lebesgue--Stieltjes sense.

We consider the following optimization problem:
%
%
\begin{equation}\label{V}%
V(x)=\sup\{V_{\pi}(x)\mbox{ with }\pi\in\Pi_{x}\}\qquad\mbox{for }x\geq0.
\end{equation}
For technical reasons, we define $V(x)=0$ for $x<0$. We restrict
ourselves to
the case $c>r>0$; we will see in Remark \ref{Vinfinito} that in the case
$c<r$, the optimal value function is infinite.

To show that the optimal value function $V$ is well defined and to describe
some of its basic properties, we first state some results of the related
controlled risk process without claims and without paying dividends.
\begin{lemma}
\label{LemmaPYt} Given $x\geq0$ and any
admissible investment strategy $\gamma_{t}\in[0,1]$ consider
the process,
\[
Y_{t}=x+mt+r\int_{0}^{t} Y_{s}\gamma_{s}\,ds+\sigma\int
_{0}%
^{t} Y_{s}\gamma_{s}\,dW_{s}.
\]

\begin{enumerate}
\item[(a)] If $m\geq0$, then $E_{x} ( Y_{t}%
e^{-ct} ) \leq e^{-(c-r)t} ( x+m(1-e^{-rt})/r ) $.

\item[(b)] If $x>0$ and $\widetilde{\tau}=\inf\{
t\dvtx Y_{t}%
<0\}$, then $\lim_{h\rightarrow0}P(\widetilde{\tau}%
<h)=0$.

\item[(c)] If $\gamma_{t}\equiv1$, then $E_{x} (
Y_{t}e^{-ct} ) =e^{-(c-r)t} ( x+m(1-e^{-rt})/r ) $
for any $m\in\mathbf{R}$.
\end{enumerate}
\end{lemma}
\begin{pf}
We can write $Y_{t}=xU_{t}+U_{t}\int_{0}^{t} mU_{s}^{-1}\,ds$ where
%
%
\begin{equation} \label{Ut}%
U_{t}=e^{\int_{0}^{t}( r\gamma_{s}-{\sigma^{2}}/{2}\gamma
_{s}^{2})\,ds+\int_{0}^{t} \sigma\gamma_{s}\,dW_{s}}.
\end{equation}
The process $e^{-%
{ \int_{0}^{t}}
r\gamma_{s}\,ds}U_{t}$ is a martingale [see, for instance, Karatzas and Shreve
(\citeyear{KS91})]. Then the results follow using elementary
computations for linear
diffusion processes.
\end{pf}

In the next two propositions, we prove that $V$ has linear growth, and
we give
bounds on the increments of $V$ using the value functions of some simple
admissible strategies.
\begin{proposition}
\label{1}
The optimal value function $V$ is well
defined and
satisfies
\[
x+p/ ( \beta+c ) \leq V(x)\leq rx/(c-r)+p/(c-r)
\qquad\mbox{for }x\geq0.
\]
\end{proposition}
\begin{pf}
Consider an initial surplus $x\geq0$. Given any $\pi=(\gamma
_{s},L_{s})\in
\Pi_{x}$, consider the controlled process $X_{t}^{\pi}$ for $t\geq0$, and
define $X_{t}^{\pi}=0$ for $t<0$. Then
\begin{eqnarray*}
\widetilde{L}_{s} &=& L_{s}-\sigma\int_{0}^{s} X_{u}^{\pi}\gamma
_{u}\,dW_{u} \\
&\leq& x+ps+r\int_{0}^{s} X_{u}^{\pi}\gamma
_{u}\,du-%
{ \sum_{i=1}^{N_{s}}}
U_{i}\\
&\leq& x+ps+r\int_{0}^{s} X_{u}^{\pi}\gamma_{u}\,du.
\end{eqnarray*}
Consider the process $Y_{t}$ defined as in Lemma \ref{LemmaPYt} with $m=p$
and the investment strategy $\gamma_{s}$ corresponding to $\pi$. Since
$X_{t}^{\pi}\leq Y_{t}$, we obtain from Lemma \ref{LemmaPYt}(a) that
$E_{x} ( X_{t}^{\pi}e^{-ct} ) \leq e^{-(c-r)t} (
x+p(1-e^{-rt})/r ) $. Since $r<c$ and $e^{-cs}$ is a positive and
decreasing function, we have that
\begin{eqnarray*}
V_{\pi}(x)&=&E_{x}\biggl(\int_{0}^{\tau}e^{-cs}\,dL_{s}\biggr) = E_{x}%
\biggl(\int_{0}^{\tau}e^{-cs}\,d\widetilde{L}_{s}\biggr)\\
& \leq& E_{x}\biggl(\int_{0}^{\infty}e^{-cs}\,d\biggl(x+ps+r\int
_{0}%
^{s} X_{u}^{\pi}\gamma_{u}\,du\biggr)\biggr)\\
& \leq& \int_{0}^{\infty}e^{-cs}p\,ds+r\int_{0}^{\infty
}%
E_{x}(e^{-cs}X_{s}^{\pi})\,ds\\
& \leq& rx/(c-r)+p/(c-r).
\end{eqnarray*}
So $V(x)={\sup_{\pi\in\Pi_{x}}}V_{\pi}(x)$ is well defined and
satisfies the second inequality.

Let us prove now the first inequality. Given an initial surplus $x\geq0$,
consider the admissible strategy $\pi_{0}$ which pays immediately the whole
surplus $x$ and then pays the incoming premium $p$ as dividends until the
first claim which in this strategy means ruin. Define $\tau_{1}$ as the time
arrival of the first claim; we have
\[
V_{\pi_{0}}(x)=x+pE_{x}\biggl(\int_{0}^{\tau
_{1}}e^{-ct} \,dt\biggr)=x+p/(\beta
+c),
\]
but by definition $V(x)\geq V_{\pi_{0}}(x)$, so we get the
result.
\end{pf}
\begin{proposition}
\label{2}
If $y>x\geq0$, the function $V$ satisfies:

\textup{(a)} $V(y)-V(x)\geq y-x$;

\textup{(b)} $V(y)-V(x)\leq( e^{(c+\beta)(y-x)/p}-1 )
V(x)$.
\end{proposition}
\begin{pf}
(a) Given $\varepsilon>0$, consider an admissible strategy $\pi\in\Pi
_{x} $
with $V_{\pi}(x)\geq V(x)-\varepsilon$. We define a new strategy
$\overline{\pi}\in\Pi_{y}$ in the following way, pay immediately $y-x$ as
dividends and then follow the strategy $\pi\in\Pi_{x}$; this new
strategy is
admissible. We have that
\[
V(y)\geq V_{\overline{\pi}}(y)=V_{\pi}(x)+(y-x)\geq V(x)-\varepsilon+(y-x)
\]
and the result follows.

(b) Given $\varepsilon>0$, take an admissible strategy $\pi\in\Pi_{y}$ such
that $V_{\pi}(y)\geq V(y)-\varepsilon$. Let us define the strategy
$\overline{\pi}\in\Pi_{x}$ that starting at $x$, pay no dividends and invest
all the surplus in bonds if $X_{t}^{\overline{\pi}}<y$ and follow strategy
$\pi$ when the current surplus reaches $y$. This strategy is
admissible. If
there is no claim up to time $t_{0}=(y-x)/p$, the surplus $X_{t_{0}%
}^{\overline{\pi}}=y$. The probability of reaching $y$ before the first claim
is $e^{-\beta t_{0}}$, so we obtain
\[
V(x)\geq V_{\overline{\pi}}(x)\geq V_{\pi}(y)e^{- ( c+\beta)
t_{0}}\geq\bigl( V(y)-\varepsilon\bigr) e^{- ( c+\beta)
(y-x)/p}
\]
and we get the result.
\end{pf}

As a direct consequence of the previous proposition we have that
$V$ is increasing and locally Lipschitz in $[0,+\infty)$, this
implies that $V$ is absolutely continuous, that $V^{\prime}(x)$ exists a.e.
and that $1\leq V^{\prime}(x)\leq V(x)(c+\beta)/p$ at the points where the
derivative exists. We will prove later in this paper that $V$ is continuously
differentiable with bounded derivative and that the linear growth condition
given by Proposition \ref{1} can be improved to $V(x)\leq x+p/c$ for
$x\geq0$.
\begin{remark}
\label{Vinfinito}
The value function $V$ is infinite in the case
that $c<r$. To see this, let us consider the worst possible case, that is
$p\leq\beta\mu$. We can assume that $x>x_{0}:= ( \beta\mu-p+1 )
/r>0$ because, if the initial surplus $x$ is smaller than $x_{0}$ there
is a
positive probability that the surplus surpass the level $x_{0}$ [take, for
instance, the strategy which pays no dividends and keeps all the
surplus in
bonds up to time $T= ( x_{0}-x ) /p+1$]. Given $t_{0}>0$, consider
the following admissible strategy $\pi_{t_{0}}\in\Pi_{x}$: divide the company
in two departments, one of them deals only with the investment and the payment
of dividends and the other with the insurance business. The investment
department starts with capital $x$, invest all the surplus on risky
assets and
diverts to the insurance department a constant flow $p_{0}=\beta\mu
-p+1$ up to
time $t_{0}\wedge\widetilde{\tau}_{1}$ when the whole surplus is paid as
dividends. Here $\widetilde{\tau}_{1}$ is the first time the surplus of the
investment department reaches zero. Let $X_{t}^{(1)}$ be the surplus process
of the investment department, we have that $X_{t\wedge t_{0}\wedge
\widetilde{\tau}_{1}}^{(1)}\geq Y_{t\wedge t_{0}}$ where $ Y_{t}$ is the
process described in Lemma \ref{LemmaPYt}(c) with $m=-p_{0}$. The insurance
department starts with no surplus, pays no dividends and receives a constant
flow $p_{0}+p$ that is larger than $\beta\mu$ up to time $t_{0}\wedge
\widetilde{\tau}_{1}\wedge\widetilde{\tau}_{2}$, where $\widetilde{\tau}_{2}$
is the ruin time of the insurance department (assuming that the insurance
department keeps always receiving the constant flow $p_{0}+p$). The stopping
time $\widetilde{\tau}_{2}$ is independent of both $\widetilde{\tau
}_{1}$ and
the process $Y_{t}$. Call $\tau=t_{0}\wedge\widetilde{\tau}_{1}\wedge
\widetilde{\tau}_{2}$, the value function of this admissible strategy
satisfies%
\begin{eqnarray*}
V_{\pi_{t_{0}}}(x) & \geq& E_{x}\bigl(X_{\tau}^{(1)}e^{-c\tau}\chi_{\{
\widetilde
{\tau}_{1}\geq t_{0},\widetilde{\tau}_{2}\geq t_{0}\}}\bigr)\geq
E_{x}\bigl(Y_{t_{0}%
}e^{-ct_{0}}\chi_{\{\widetilde{\tau}_{1}\geq t_{0},\widetilde{\tau
}_{2}\geq
t_{0}\}}\bigr)\\
& = & E_{x}\bigl(Y_{t_{0}}e^{-ct_{0}}\chi_{\{\widetilde{\tau}_{1}\geq t_{0}%
\}}\bigr)P(\{\widetilde{\tau}_{2}\geq t_{0}\})\geq E_{x}(Y_{t_{0}}e^{-ct_{0}
})P(\{\widetilde{\tau}_{2}=\infty\}),
\end{eqnarray*}
because $Y_{t_{0}}<0$ for $t_{0}>\widetilde{\tau}_{1}$. We can compute the
survival probability of the insurance department [see, for instance, Teugels
(\citeyear{T03})] as $P(\{\widetilde{\tau}_{2}=\infty\})=1-\beta\mu/ (
p_{0}+p ) >0$. So, from Lemma \ref{LemmaPYt}(c), we conclude that
$V(x)\geq\lim_{t_{0}\rightarrow\infty}V_{\pi
_{t_{0}}}(x)=\infty$.
\end{remark}

\section{The Hamilton--Jacobi--Bellman equation}\label{sec3}

In this section we associate a Hamilton--Jacobi--Bellman equation to the
optimization problem (\ref{V}) and we prove that the optimal value function
$V$ is a viscosity solution of this equation.

The notion of viscosity solution was introduced by Crandall and Lions
(\citeyear{CL83})
for first order Hamilton--Jacobi equations and by Lions (\citeyear
{L83}) for second-order
partial differential equations. Nowadays, it is a standard tool for studying
HJB equations [see, for instance, Fleming and Soner (\citeyear{FS93})
and Bardi and
Capuzzo-Dolcetta (\citeyear{BCd97})].

We first state the \textit{dynamic programming principle}; the proof is
similar to the one in Azcue and Muler (\citeyear{AM05}).
\begin{proposition}
\label{PropositionDPP}
For any $x\geq0$ and any stopping
time $\tau$, we can write
\[
V(x)=\sup_{\pi=(\gamma_{t},L_{t})\in\Pi_{x}}E_{x} \biggl( \int
_{0}^{\tau\wedge\tau^{\pi}}e^{-cs}\,dL_{s}+e^{-c ( \tau\wedge
\tau^{\pi} ) }V(X_{\tau\wedge\tau^{\pi}}^{\pi}) \biggr).
\]
\end{proposition}

The HJB equation associated to the optimization problem (\ref{V}) is the
following fully nonlinear second-order degenerate integro-differential
equation with derivative constraint:
%
%
\begin{equation} \label{InicDif}%
\max\{1-u^{\prime}(x),\mathcal{L}^{\ast}(u)(x)\}=0,
\end{equation}
where%
%
%
\begin{equation} \label{Lestrella}%
\mathcal{L}^{\ast}(u)(x)=\sup_{\gamma\in[0,1]}\mathcal{L}%
_{\gamma}(u)(x)
\end{equation}
and
%
%
\begin{eqnarray} \label{Lu}%
\mathcal{L}_{\gamma}(u)(x) &=& \sigma^{2}\gamma^{2}x^{2}u^{\prime\prime
}(x)/2+ ( p+r\gamma x ) u^{\prime}(x)\nonumber\\[-8pt]\\[-8pt]
&&{} - (c+\beta)u(x)+\beta
\int_{0}^{x}u(x-\alpha)\,dF(\alpha).\nonumber
\end{eqnarray}

This equation is obtained assuming that the optimal value function $V$ is
twice continuously differentiable. We will show in Section \ref{sec9} that this
is not
always the case, so we consider \textit{viscosity solutions} of this equation.
\begin{definition}
\label{NuevaDefinicionSubySuper}
A continuous function
$\underline{u}\dvtx[0,\infty)\rightarrow\mathbf{R}$ is a viscosity
subsolution of
(\ref{InicDif}) at $x\in(0,\infty)$ if any twice continuously
differentiable function $\psi$ defined in $(0,\infty)$
with $\psi(x)=\underline{u}(x)$ such that
$\underline{u}-\psi$ reaches the maximum at $x$ satisfies
$\max\{1-\psi^{\prime}(x),\mathcal{L}^{\ast}(\psi)(x)\}\geq0$,
and a continuous function $\overline{u}\dvtx[0,\infty)\rightarrow
\mathbf{R}$ is a viscosity supersolution of (\ref{InicDif}) at
$x\in(0,\infty)$ if any twice continuously differentiable function
$\varphi$ defined in $(0,\infty)$ with $\varphi
(x)=\overline{u}(x)$ such that $\overline
{u}-\varphi
$ reaches the minimum at $x$ satisfies $\max
\{1-\varphi^{\prime}(x),\mathcal{L}^{\ast}(\varphi)(x)\}\leq0$.

Finally, a continuous function $u\dvtx[0,\infty
)\rightarrow\mathbf{R}$ is a viscosity solution of (\ref
{InicDif}) if
it is both a viscosity subsolution and a viscosity supersolution at any
$x\in(0,\infty)$.
\end{definition}

In addition to Definition \ref{NuevaDefinicionSubySuper}, there are two other
equivalent formulations of viscosity solutions. The proof of the equivalence
of these definitions is standard [see, for instance, Benth, Karlsen and Reikvam
(\citeyear{BKR02})]. We use the three definitions indistinctly.
\begin{definition}
\label{DefinicionSubySuper}
Given a twice continuously differentiable
function $f$ and a continuous function $u$, let us define
the operator,
%
%
\begin{eqnarray} \label{LVF}%
\mathcal{L}_{\gamma}(u,f)(x)&=&\sigma^{2}\gamma^{2}x^{2}f^{\prime\prime
}(x)/2+ ( p+r\gamma x ) f^{\prime}(x)\nonumber\\[-8pt]\\[-8pt]
&&{}-(c+\beta)u(x)+\beta
\int_{0}^{x}u(x-\alpha)\,dF(\alpha).\nonumber
\end{eqnarray}

A continuous function $\underline{u}\dvtx[0,\infty)\rightarrow
\mathbf{R}%
$ is a viscosity subsolution of (\ref{InicDif}) at
$x\in(0,\infty)$ if any twice continuously differentiable function
$\psi$ defined in $(0,\infty)$ such that
$\underline{u}-\psi$ reaches the maximum at $x$ satisfies
$\max\{1-\psi^{\prime}(x),\sup_{\gamma\in[0,1]}\mathcal
{L}%
_{\gamma}(\underline{u}$, $\psi)(x)\}\geq0$, and a twice continuous
function $\overline{u}\dvtx[0,\infty)\rightarrow\mathbf{R}%
$ is a viscosity supersolution of (\ref{InicDif}) at $x\in
(0,\infty)$ if any twice continuously differentiable function
$\varphi$ defined in $(0,\infty)$ such that
$\overline{u}-\varphi$ reaches the minimum at $x$ satisfies
$\max\{1-\varphi^{\prime}(x),\sup_{\gamma\in[0,1]}%
\mathcal{L}_{\gamma}(\overline{u},\varphi)(x)\}\leq0$.
\end{definition}
\begin{definition}
\label{NovisimaDefinicionSubySuper}
Given any continuous function
$u\dvtx[0,\infty)\rightarrow\mathbf{R}$ and any \mbox{$x>0$},
the set of second superdifferentials of $u$ at $x$
is defined as
\[
D^{+}u(x)= \biggl\{ (d,q)\mbox{ such that }\limsup_{h\rightarrow0}%
\frac{u(x+h)-u(x)-hd-h^{2}q/2}{h^{2}}\leq0 \biggr\}
\]
and the set of second subdifferentials of $u$ at
$x$ is defined as
\[
D^{-}u(x)= \biggl\{ (d,q)\mbox{ such that }\liminf_{h\rightarrow0}%
\frac{u(x+h)-u(x)-hd-h^{2}q/2}{h^{2}}\geq0 \biggr\}.
\]
Let us call
%
%
\begin{eqnarray}\label{Ludq}%
\mathcal{L}_{\gamma}(u,d,q)(x)&=&\sigma^{2}\gamma^{2}x^{2}q/2+ ( p+r\gamma
x ) d-(c+\beta)u(x)\nonumber\\[-8pt]\\[-8pt]
&&{}+\beta\int_{0}^{x}u(x-\alpha)\,dF(\alpha).\nonumber
\end{eqnarray}

A continuous function $\underline{u}\dvtx[0,\infty)\rightarrow
\mathbf{R}%
$ is a viscosity subsolution of (\ref{InicDif}) at $x\in
(0,\infty
)$ if $\max\{1-d,\sup_{\gamma\in[0,1]}\mathcal
{L}%
_{\gamma}(\underline{u},d,q)(x)\}\geq0$ for all $(d,q)\in
D^{+}\underline{u}(x)$ and $\overline{u}\dvtx[0,\infty
)\rightarrow\mathbf{R}$ is a viscosity supersolution of
(\ref{InicDif}) at $x\in(0,\infty)$ if $\max\{1-d,\sup
_{\gamma\in[0,1]}\mathcal{L}_{\gamma}(\overline{u}%
,d,q)(x)\}\leq0$ for all $(d,q)\in D^{-}\overline{u}(x)$.
\end{definition}

The next proposition states the semiconcavity of the viscosity
solutions of
the HJB equation.
\begin{proposition}
\label{supersemiconcave}
Any absolutely continuous and nondecreasing
supersolution of $\mathcal{L}^{\ast}(u)=0$  in $(0,\infty)$
is semiconcave in any interval $[x_{0},x_{1}]\subset(0,\infty)$.
\end{proposition}
\begin{pf}
It is enough to prove that there exists a constant $K$ and a sequence of
semiconcave functions $v_{n}$ in $[0,x_{1}]$ such that $v_{n}^{\prime
\prime
}\leq K$ a.e. and $v_{n}\rightarrow\overline{u}$ uniformly in $[0,x_{1}]$.

Since $\overline{u}$ is an absolutely continuous function, there exists
$k_{0}\geq1$ such that $ \vert\overline{u}(x)-\overline{u}(y) \vert
\leq k_{0} \vert x-y \vert$ for all $x,y\in[0,x_{1}]$. Let us
define, for any $x\in[0,x_{1}]$,
%
%
\begin{equation} \label{Definicionu1vn}%
v_{n}(x)=\inf_{y\in[0,x_{1}]} \{ \overline{u}(y)+n^{2}%
(x-y)^{2}/2 \} .
\end{equation}
It can be proved, as in Lemma 5.1 of Fleming and Soner (\citeyear
{FS93}), that
$v_{n}$ is
semiconcave and the inequality $0\leq\overline{u}(x)-v_{n}(x)\leq2k_{0}%
^{2}/n^{2}$ holds for all $x\in[0,x_{1}]$, so $v_{n}\rightarrow
\overline{u}$ uniformly. We have that if $x+h\leq x_{1}$, then $v_{n}%
(x+h)-v_{n}(x)\leq k_{0}h$ for $h\leq x_{1}-x$. In effect, take $y_{0}
\in[0,x_{1}]$ such that $v_{n}(x)=\overline{u}(y_{0})+n^{2}%
(x-y_{0})^{2}/2$, we have
\begin{eqnarray*}
v_{n}(x+h)-v_{n}(x) & \leq& \bigl( \overline{u}(y_{0}+h)+n^{2}(x-y_{0}%
)^{2}/2 \bigr) - \bigl( \overline{u}(y_{0})+n^{2}(x-y_{0})^{2}/2 \bigr) \\
& = & \overline{u}(y_{0}+h)-\overline{u}(y_{0})\\
& \leq& k_{0}h.
\end{eqnarray*}

Since $v_{n}$ is semiconcave, the set%
\[
A= \{ x\in[0,x_{1}]\dvtx v_{n}^{\prime}(x)\mbox{ and }%
v_{n}^{\prime\prime}(x)\mbox{ exist for all }n\in\mathbf{N}\mbox{ and
}F(x)=F(x^{-}) \}
\]
has full measure.

We want to prove that
%
%
\begin{equation} \label{AcotacionDerivadaSegundadevn}%
v_{n}^{\prime\prime}(x)\leq8(c+\beta)\overline{u}(x_{1})/(\sigma
^{2}x_{0}%
^{2})\qquad\mbox{in }[x_{0},x_{1}]\cap A.
\end{equation}
Take $\overline{x}\in[ x_{0},x_{1}]\cap A$, and consider $\overline{y}_{n}%
\in[0,x_{1}]$ such that
%
%
\begin{equation}\label{vn}%
v_{n}(\overline{x})=\overline{u}(\overline{y}_{n})+n^{2}(\overline{x}-\overline{y}_{n})^{2}/2.
\end{equation}
It can be proved that
%
%
\begin{equation} \label{Cercanoyax}%
x_{0}/2\leq\overline{y}_{n}\leq\overline{x}\quad\mbox{and}\quad\overline{x}-\overline{y}_{n}\leq
2k_{0}/n^{2}.
\end{equation}
By (\ref{Definicionu1vn}), we have
\[
v_{n}(\overline{x}+h)\leq\overline{u}(\overline{y}_{n}+h)+n^{2}(\overline{x}-\overline
{y}_{n}%
)^{2}/2,
\]
so we obtain from (\ref{vn}) that
\begin{eqnarray*}
&&
\liminf_{h\rightarrow0}\frac{\overline{u}(\overline
{y}_{n}+h)-\overline
{u}(\overline{y}_{n})-hv_{n}^{\prime}(\overline{x})-h^{2}v_{n}^{\prime\prime}(\overline
{x})/2}{h^{2}}\\
&&\qquad\geq\liminf_{h\rightarrow0}\frac{v_{n}(\overline{x}%
+h)-v_{n}(\overline{x})-hv_{n}^{\prime}(\overline{x})-h^{2}v_{n}^{\prime\prime
}(\overline
{x})/2}{h^{2}}=0.
\end{eqnarray*}
Then we have that $ ( v_{n}^{\prime}(\overline{x}),v_{n}^{\prime\prime}%
(\overline{x}) ) \in D^{-}\overline{u}(\overline{y}_{n})$.

Since $\overline{u}$ is a viscosity supersolution of (\ref{InicDif}) at
$\overline{y}_{n}$, we have from Definition \ref{NovisimaDefinicionSubySuper}
that
%
%
\begin{equation}\label{desig-u1}%
\mathcal{L}_{1}(\overline{u},v_{n}^{\prime}(\overline{x}),v_{n}^{\prime\prime
}%
(\overline{x}))(\overline{y}_{n})\leq0.
\end{equation}

If $v_{n}^{\prime\prime}(\overline{x})\leq0$,  inequality
(\ref{AcotacionDerivadaSegundadevn}) holds, and if $v_{n}^{\prime
\prime
}(\overline{x})>0$, from (\ref{Cercanoyax}) and (\ref{desig-u1}) we get that
\[
\sigma^{2}x_{0}^{2}v_{n}^{\prime\prime}(\overline{x})/8\leq\sigma^{2}\overline{y}%
^{2}v_{n}^{\prime\prime}(\overline{x})/2\leq(c+\beta)\overline{u}(\overline{y}%
)\leq(c+\beta)\overline{u}(x_{1})
\]
and so we have (\ref{AcotacionDerivadaSegundadevn}).
\end{pf}

The next proposition states that the optimal value function of our control
problem is a viscosity solution of equation (\ref{InicDif}).
We will
show in the next section that this result is not enough to characterize
univocally the optimal value function.
\begin{proposition}
\label{Vviscositysolution}
The optimal value function
$V$ is a viscosity solution of (\ref{InicDif}) in
$(0,\infty)$.
\end{proposition}
\begin{pf}
We prove first that $V$ is a viscosity supersolution. Let us call $\tau_{1}$
and $U_{1}$ the time and the size of the first claim. For fixed
$l_{0}\geq0$
and $\gamma_{0}\in[0,1]$, consider the admissible strategy $\pi
_{0}=(\gamma_{0},tl_{0})\in\Pi_{x}$.

Assume first that $l_{0}>p$. Given any $h>0$, consider the process $Y_{t}$
defined in Lemma \ref{LemmaPYt} with $m=p-l_{0}$ and $\gamma_{t}=\gamma
_{0} $.
Let us consider $\widetilde{\tau}=\inf\{t\dvtx Y_{t}<0\}$. Using Proposition
\ref{PropositionDPP} with $\tau=\tau_{1}\wedge h$, we obtain that
%
%
\begin{equation} \label{lomayp0}%
V(x)\geq E_{x} \biggl( \int_{0}^{\tau\wedge\tau^{\pi_{0}}}e^{-cs}%
l_{0}\,ds+e^{-c ( \tau\wedge\tau^{\pi_{0}} ) }V(X_{\tau\wedge
\tau^{\pi_{0}}}^{\pi_{0}}) \biggr) .
\end{equation}
Note that $\tau\wedge\tau^{\pi_{0}}=\tau\wedge\widetilde{\tau}$, so we have
%
%
\begin{eqnarray}\label{lomayp1}\hspace*{32pt}
E_{x} \biggl( \int_{0}^{\tau\wedge\widetilde{\tau}}e^{-cs}%
l_{0}\,ds \biggr) & \geq& E_{x} \biggl( \chi_{ \{ \tau\leq\widetilde{\tau
} \} }\int_{0}^{\tau}e^{-cs}l_{0}\,ds \biggr) \nonumber\\
& = & E_{x} \biggl( \int_{0}^{\tau}e^{-cs}l_{0}\,ds \biggr)
-E_{x} \biggl( \chi_{ \{ \widetilde{\tau}<\tau\} }\int
_{0}^{\tau}e^{-cs}l_{0}\,ds \biggr) \\
& \geq& E_{x} \biggl( \int_{0}^{\tau}e^{-cs}l_{0}\,ds \biggr)
-hl_{0}P(\widetilde{\tau}<h)\nonumber
\end{eqnarray}
and
%
%
\begin{eqnarray}\label{lomayp2}\qquad
&&E_{x} \bigl( e^{-c ( \tau\wedge\widetilde{\tau} ) }V(X_{\tau
\wedge\widetilde{\tau}}^{\pi_{0}}) \bigr) \nonumber\\%
&&\qquad= E_{x} \bigl( \chi_{ \{ \tau\leq\widetilde{\tau} \} }e^{-c\tau
}V(X_{\tau}^{\pi_{0}}) \bigr) +E_{x} \bigl( \chi_{ \{ \widetilde{\tau
}<\tau\} }e^{-c\widetilde{\tau}}V(0) \bigr) \nonumber\\
&&\qquad= E_{x} \bigl( \chi_{ \{ \tau_{1}\neq\tau\leq\widetilde{\tau} \}
}e^{-ch}V(Y_{h}) \bigr) +E_{x} \bigl( \chi_{ \{ \tau_{1}=\tau
\leq\widetilde{\tau} \} }e^{-c\tau_{1}}V(Y_{\tau_{1}}-U_{1}) \bigr) \nonumber\\
&&\qquad\quad{} +E_{x} \bigl( \chi_{ \{ \widetilde{\tau}<\tau\} }e^{-c\widetilde
{\tau}}V(0) \bigr) \nonumber\\[-8pt]\\[-8pt]
&&\qquad\geq E_{x} \bigl( \chi_{ \{ \tau_{1}\neq\tau\} }e^{-ch}%
V(Y_{h}) \bigr) -E_{x} \bigl( \chi_{ \{ \tau_{1}\neq\tau>\widetilde
{\tau} \} }e^{-ch}V(Y_{h}) \bigr) \nonumber\\
&&\qquad\quad{} +E_{x} \bigl( \chi_{ \{ \tau_{1}=\tau\} }e^{-c\tau_{1}}%
V(Y_{\tau_{1}}-U_{1}) \bigr)\nonumber\\
&&\qquad\quad{} -E_{x} \bigl( \chi_{ \{ \tau_{1}%
=\tau>\widetilde{\tau} \} }e^{-c\tau_{1}}V(Y_{\tau_{1}}-U_{1}) \bigr)
\nonumber\\
&&\qquad\geq E_{x}\bigl( \chi_{ \{ \tau_{1}\neq\tau\} }e^{-ch}%
V(Y_{h}) \bigr) +E_{x} \bigl( \chi_{ \{ \tau_{1}=\tau\}
}e^{-c\tau_{1}}V(Y_{\tau_{1}}-U_{1}) \bigr),\nonumber
\end{eqnarray}
because $V(y)=0$ for $y<0$ and $Y_{\tau}-U_{1}<Y_{\tau}<0$ for $\tau
>\widetilde{\tau}$. Then, from (\ref{lomayp0}), (\ref{lomayp1}) and
(\ref{lomayp2}), we get that%
%
%
\begin{eqnarray}\label{lomayp}\hspace*{35pt}
V(x) & \geq& l_{0}\bigl(1-e^{-h ( c+\beta) }\bigr)/(c+\beta)-hl_{0}%
P(\widetilde{\tau}<h)+e^{-(c+\beta)h}E_{x} ( V(Y_{h}) )
\nonumber\\[-8pt]\\[-8pt]
& &{} +\beta\int_{0}^{h} \biggl( \int_{0}^{\infty}E_{x}%
\bigl(V_{x}(Y_{s}-\alpha)\bigr)\,dF(\alpha) \biggr)
e^{-(c+\beta)s}\,ds.\nonumber
\end{eqnarray}

Assume now that $l_{0}\leq p$, we obtain with a simpler argument that
%
%
\begin{eqnarray}\label{lomenp}\qquad
V(x) & \geq& l_{0}\bigl(1-e^{-h ( c+\beta) }\bigr)/(c+\beta)+e^{-(c+\beta
)h}E_{x}(V(Y_{h}))\nonumber\\[-8pt]\\[-8pt]
& &{} +\beta\int_{0}^{h} \biggl( \int_{0}^{\infty}%
E_{x}\bigl(V(Y_{s}-\alpha)\bigr)\,dF(\alpha) \biggr)
e^{-(c+\beta)s}\,ds.\nonumber
\end{eqnarray}

Dividing by $h$, we get from (\ref{lomayp}) and (\ref{lomenp}) that
\begin{eqnarray*}
0 & \geq& l_{0} \bigl( 1-e^{-h ( c+\beta) } \bigr) /\bigl( (
c+\beta) h\bigr)+e^{-h ( \beta+c ) } \bigl( E_{x}(V(Y_{h}%
))-V(x) \bigr) /h\\
& &{} +\bigl(e^{-h ( c+\beta) }-1\bigr)V(x)/h\\
& &{} + (\beta/h)\int_{0}%
^{h} \biggl( \int_{0}^{\infty}E_{x} \bigl( V(Y_{s}-\alpha
)-V(x) \bigr)\, dF(\alpha) \biggr) e^{-(c+\beta)s}\,ds\\
& &{} + V(x)(\beta/h)\int_{0}^{h}e^{-(c+\beta
)s}\,ds-l_{0}P(\widetilde
{\tau}<h)
\end{eqnarray*}
and so
%
%
\begin{eqnarray}\label{estainequality}\qquad
0 & \geq& \bigl( 1-e^{-h ( c+\beta) } \bigr) l_{0}/\bigl( (
c+\beta) h\bigr)+e^{-h ( \beta+c ) } \bigl( E_{x}\bigl(V(Y_{h}%
)-V(x)\bigr) \bigr) /h\nonumber\\
& &{} +c \bigl( e^{-h ( c+\beta) }-1 \bigr) V(x)/\bigl( (
c+\beta) h\bigr)\nonumber\\[-8pt]\\[-8pt]
& &{} +(\beta/h)\int_{0}^{h} \biggl( \int_{0}^{\infty}%
E_{x} \bigl( V(Y_{s}-\alpha)-V(x) \bigr)\, dF(\alpha) \biggr) e^{-(c+\beta
)s}\,ds\nonumber\\
&&{} -l_{0}P(\widetilde{\tau}<h).\nonumber
\end{eqnarray}

Let $\varphi\dvtx(0,\infty)\rightarrow\mathbf{R}$ be a twice continuously
differentiable function such that $V-\varphi$ reaches the minimum in
$(0,\infty)$ at $x$ with $\varphi(x)=V(x)$. Since $x>0$, we can assume without
loss of generality that $\varphi$ is defined in $\mathbf{R}$ and that
$\varphi(y)\leq0$ for $y<0$. From (\ref{estainequality}) we get
%
%
\begin{eqnarray}\label{DesigDivH}\quad
0 & \geq& \bigl( 1-e^{-h ( c+\beta) } \bigr) l_{0}/\bigl( (
c+\beta) h\bigr)+E_{x} \bigl( \varphi(Y_{h})-\varphi(x) \bigr) e^{-h (
\beta+c ) }/h\nonumber\\
&&{} +c \bigl( e^{-h ( c+\beta) }-1 \bigr) V(x)/\bigl( (
c+\beta) h\bigr)\nonumber\\[-8pt]\\[-8pt]
&&{} +(\beta/h)\int_{0}^{h} \biggl( \int_{0}^{\infty}%
E_{x} \bigl( V(Y_{s}-\alpha)-V(x) \bigr) \,dF(\alpha) \biggr) e^{-(c+\beta
)s}\,ds\nonumber\\
&&{} -l_{0}P(\widetilde{\tau}<h).\nonumber
\end{eqnarray}
But, since $\varphi$ is twice continuously, we get, from It\^{o}'s formula,
%
%
\begin{eqnarray}\label{DeltaPhi}\hspace*{35pt}
\varphi(Y_{h})-\varphi(x) & = & \int_{0}^{h}\varphi^{\prime}(Y_{s}%
)\,dY_{s}+(\sigma^{2}\gamma_{0}^{2}/2)\int_{0}^{h}\varphi^{\prime\prime}%
(Y_{s})Y_{s}^{2}\,ds\nonumber\\
& = & \int_{0}^{h} \bigl( \varphi^{\prime}(Y_{s}) \bigl( (p-l_{0})+r\gamma
_{0}Y_{s} \bigr) +\varphi^{\prime\prime}(Y_{s})Y_{s}^{2}\sigma^{2}\gamma
_{0}^{2}/2 \bigr) \,ds\\
& & +\int_{0}^{h}\varphi^{\prime}(Y_{s})\sigma\gamma_{0}Y_{s}\,dW_{s}.\nonumber
\end{eqnarray}

Note that the last term of (\ref{DeltaPhi}) is a martingale. Letting
$h$ go to
$0^{+}$ in (\ref{DesigDivH}), we obtain from Lemma \ref{LemmaPYt}(b)
and (\ref{DeltaPhi}) that
\begin{eqnarray*}
0 & \geq& l_{0} \bigl( 1-\varphi^{\prime}(x) \bigr) \\
&&{} + \biggl( \sigma^{2}\gamma_{0}^{2}x^{2}\varphi^{\prime\prime
}(x)/2+(p+r\gamma_{0}x)\varphi^{\prime}(x)\\
&&\hspace*{17.3pt}{}-(\beta+c)V(x) + \beta\int
_{0}^{\infty}V(x-\alpha)\,dF(\alpha) \biggr).
\end{eqnarray*}
Since this inequality holds for all $l_{0}\geq0$, we have that $\varphi
^{\prime}(x)\geq1$, and taking $l_{0}=0$ we get $\mathcal{L}_{\gamma_{0}
}(V,\varphi)(x)\leq0$, so
\[
\max\Bigl\{1-\varphi^{\prime}(x),\sup_{\gamma\in[0,1]}\mathcal
{L}%
_{\gamma}(V,\varphi)(x)\Bigr\}\leq0
\]
and we have the result.

The proof that $V$ is a viscosity subsolution at any $x>0$ is similar
to the one of Proposition 3.8 of Azcue and Muler (\citeyear{AM05}), but
in this case we should also consider a martingale that involves the
Brownian motion $W_{t}$.
\end{pf}

From Propositions \ref{supersemiconcave} and \ref{Vviscositysolution} we
get the following corollary.
\begin{corollary}
The optimal value function $V$ is semiconcave in any interval
$[x_{0},x_{1}]\subset(0,\infty)$ and so $V^{\prime \prime}$ exists a.e.
\end{corollary}

\section{Comparison principle for viscosity solutions}\label{sec4}

We prove in this section a \textit{comparison principle} for viscosity
solutions of (\ref{InicDif}), and as a consequence we obtain the uniqueness
with the boundary condition $u(0)$ among all the functions $u$ which
satisfy the
following regularity and growth assumptions:

(A.1) $u\dvtx[0,\infty)\rightarrow\mathbf{R}$ is locally Lipschitz.

(A.2) If $0\leq x<y$, then $u(y)-u(x)\geq y-x$.

(A.3) There exists a constant $k>0$ such that $u(x)\leq x+k$ for
all $x\in[0,\infty)$.
\begin{proposition}
\label{Superarribasub}
If $\underline{u}$ is a subsolution
and $\overline{u}$ is a supersolution of (\ref{InicDif}) in
$ ( 0,\infty) $  with $\underline{u}(0)\leq\overline
{u}(0)$ and they satisfy the conditions \textup{(A.1)}, \textup{(A.2)}
and \textup{(A.3)}, then $\underline{u}\leq\overline{u}$ in
$(0,\infty)$.
\end{proposition}
\begin{pf}
The first part of this proof is similar to the proof of Proposition 4.2 of
Azcue and Muler (\citeyear{AM05}) although in this case we should also
use the tools
provided by Crandall, Ishii and Lions (\citeyear{CIL92}) to prove
comparison principles
for second-order differential equations and adapt them to
integro-differential equations.

Assume that $\underline{u}(x_{0})-\overline{u}(x_{0})>0$ for some point
$x_{0}>0$. It is straightforward to show that the functions $\overline
{u}%
^{s}(x)=s\overline{u}(x)$ with $s>1$ are also a supersolution and satisfy
$\overline{u}^{s}(0)\geq\underline{u}(0)$. If $\varphi$ is a continuously
differentiable function such that the minimum of $\overline
{u}^{s}-\varphi$ is
attained at $x_{0}$ then $1-\varphi^{\prime}(x_{0})\leq1-s<0$. Let us take
$s_{0}>1$ with $\underline{u}(x_{0})-\overline{u}^{s_{0}}(x_{0})>0$ and
define%
%
%
\begin{equation} \label{defM}%
M=\sup_{x\geq0} \bigl( \underline{u}(x)-\overline{u}^{s_{0}}(x) \bigr).
\end{equation}
From assumptions (A.2) and (A.3) we obtain, as in
Proposition 4.2 of Azcue and Muler (\citeyear{AM05}), that%
%
%
\begin{equation} \label{desmax}%
0<\underline{u}(x_{0})-\overline{u}^{s_{0}}(x_{0})\leq M=\max
_{x\in
[ 0,b ] } \bigl( \underline{u}(x)-\overline{u}^{s_{0}}(x) \bigr),
\end{equation}
where $b=k/(s_{0}-1)$. Call $x^{\ast}=\arg\max_{x\in[ 0,b ]
} ( \underline{u}(x)-\overline{u}^{s_{0}}(x) ) $. Since
$\underline{u}(x)$ and $\overline{u}^{s_{0}}(x)$ satisfy assumption
(A.1), there exists a constant $m>0$ such that%
%
%
\begin{equation}\label{acotacion-con-m}%
\frac{\underline{u}(x_{1})-\underline{u}(x_{2})}{x_{1}-x_{2}}\leq
m,\qquad
\frac{\overline{u}^{s_{0}}(x_{1})-\overline
{u}^{s_{0}}(x_{2})}{x_{1}-x_{2}%
}\leq m
\end{equation}
for $0\leq x_{2}\leq x_{1}\leq b$.

Let us consider%
\[
A= \{ ( x,y ) \dvtx0\leq y\leq b, 0\leq x\leq y \}
\]
and for any $\lambda>0$ the functions
%
%
\begin{eqnarray} \label{DefinicionPhi}%
\Phi^{\lambda} ( x,y ) &=& \lambda( x-y ) ^{2}%
/2+2m/\bigl(\lambda^{2} ( y-x ) +\lambda\bigr),
\\
\label{def-sigma-definitiva-por-hoy}%
\Sigma^{\lambda} ( x,y ) &=& \underline{u}(x)-\overline{u}^{s_{0}%
}(y)-\Phi^{\lambda} ( x,y ).
\end{eqnarray}
Calling $M_{\lambda}=\max_{A}\Sigma^{\lambda}$ and $ ( x_{\lambda
},y_{\lambda} ) =\arg\max_{A}\Sigma^{\lambda}$, we obtain that
$M_{\lambda}\geq\Sigma^{\lambda}(x^{\ast},x^{\ast})=M-2m/\lambda$ and
so, from
(\ref{desmax}) we get that $M_{\lambda}>0$ for $\lambda\geq4m/M$ and%
%
%
\begin{equation}\label{liminf-mlambda}%
\liminf_{\lambda\rightarrow\infty}M_{\lambda}\geq M.
\end{equation}
Since $ ( x_{\lambda},y_{\lambda} ) \in A$, we have that%
%
%
\begin{equation} \label{relxalfayalfa}%
y_{\lambda}\geq x_{\lambda}.
\end{equation}

As in Proposition 4.2 of Azcue and Muler (\citeyear{AM05}), we can show
that for any
$\lambda\geq\lambda_{0}=\max\{2m/\delta,4m/M\}$ the point $ ( x_{\lambda
},y_{\lambda} )\notin\partial A$.

Using Theorem 3.2 of Crandall, Ishii and Lions (\citeyear{CIL92}), it
can be proved that for any
$\delta>0$, there exist real numbers $A_{\delta}$ and $B_{\delta}$ such that
%
%
\begin{equation} \label{Adelta}%
( \Phi_{x}^{\lambda}(x_{\lambda},y_{\lambda}),A_{\delta} ) \in
D^{+}\underline{u}(x_{\lambda})
\end{equation}
and
%
%
\begin{equation} \label{Bdelta}%
( -\Phi_{y}^{\lambda}(x_{\lambda},y_{\lambda}),B_{\delta} ) \in
D^{-}\overline{u}^{s}(y_{\lambda})
\end{equation}
with
%
%
\begin{equation} \label{MatrizDelta}\quad
D^{2}\Phi^{\lambda}(x_{\lambda},y_{\lambda})+\delta( D^{2}\Phi^{\lambda
}(x_{\lambda},y_{\lambda}) ) ^{2}-
\pmatrix{
A_{\delta} & 0\cr
0 & -B_{\delta}%
} \geq\pmatrix{
0 & 0\cr
0 & 0},
\end{equation}
where $D^{2}\Phi^{\lambda}$ corresponds to the matrix of second
derivatives of
$\Phi^{\lambda}$, and $D^{+}$ and $D^{-}$ are defined in Definition
\ref{NovisimaDefinicionSubySuper}. The inequality in (\ref
{MatrizDelta}) means
that the matrix on the left-hand side is positive-semidefinite. So, we
obtain from
(\ref{Adelta}) and (\ref{Bdelta}) that%
%
%
\begin{equation} \label{superxalfa}%
\max\Bigl\{1-\Phi_{x}^{\lambda}(x_{\lambda},y_{\lambda}),\sup_{\gamma
\in[0,1]}\mathcal{L}_{\gamma}(\underline{u},\Phi_{x}^{\lambda
}(x_{\lambda},y_{\lambda}),A_{\delta})(x_{\lambda})\Bigr\}\geq0
\end{equation}
and%
%
%
\begin{equation}\label{subyalfa}\quad
\max\Bigl\{1+\Phi_{y}^{\lambda}(x_{\lambda},y_{\lambda}),\sup_{\gamma
\in[0,1]}\mathcal{L}_{\gamma}(\overline{u}^{s_{0}},-\Phi
_{y}^{\lambda
}(x_{\lambda},y_{\lambda}),B_{\delta})(y_{\lambda})\Bigr\}\leq0.
\end{equation}

From (\ref{MatrizDelta}), we obtain that
%
%
\begin{eqnarray}\label{DesigualdarSegundoOrden}%
&&
A_{\delta}x_{\lambda}^{2}-B_{\delta}y_{\lambda}^{2}\nonumber\\%
&&\qquad
\leq\bigl( \bigl( \lambda+(4m\lambda)/ \bigl( \lambda( y_{\lambda
}-x_{\lambda} ) +1 \bigr) ^{3} \bigr)\\
&&\qquad\quad\hspace*{2.3pt}{} +2\delta\bigl( \lambda
+(4m\lambda)/ \bigl( \lambda( y_{\lambda}-x_{\lambda} ) +1 \bigr)
^{3} \bigr) ^{2} \bigr) ( x_{\lambda}-y_{\lambda} ) ^{2}.\nonumber
\end{eqnarray}
We also have from (\ref{DefinicionPhi}) that
%
%
\begin{equation} \label{SumaDerivasPhi}%
\Phi_{x}^{\lambda} ( x_{\lambda},y_{\lambda} ) +\Phi_{y}^{\lambda
} ( x_{\lambda},y_{\lambda} ) =0
\end{equation}
and
%
%
\begin{eqnarray}\label{CasiSumaDerivasPhi}%
&&x_{\lambda}\Phi_{x}^{\lambda} ( x_{\lambda},y_{\lambda} )
+y_{\lambda}\Phi_{y}^{\lambda} ( x_{\lambda},y_{\lambda}
)\nonumber\\[-8pt]\\[-8pt]
&&\qquad
=\lambda(x_{\lambda}-y_{\lambda})^{2}+2m(x_{\lambda}-y_{\lambda})/ \bigl(
\lambda( y_{\lambda}-x_{\lambda} ) +1 \bigr) ^{2}.\nonumber
\end{eqnarray}
But $ ( -\Phi_{y}^{\lambda}(x_{\lambda},y_{\lambda}),B_{\delta} )
\in D^{-}\overline{u}^{s}(y_{\lambda})$, so we obtain that $-\Phi_{y}%
^{\lambda} ( x_{\lambda},y_{\lambda} ) \geq s_{0}>1$, and so we
conclude from (\ref{superxalfa}) and (\ref{SumaDerivasPhi}) that%
%
%
\begin{equation}\label{subyalfa1}%
\sup_{\gamma\in[0,1]}\mathcal{L}_{\gamma}(\underline
{u},\Phi
_{x}^{\lambda}(x_{\lambda},y_{\lambda}),A_{\delta})(x_{\lambda})\geq0.
\end{equation}
Therefore, taking $\gamma_{\lambda}=\arg\max\mathcal{L}_{\gamma
}(\underline
{u},\Phi_{x}^{\lambda}(x_{\lambda},y_{\lambda}),A_{\delta})(x_{\lambda
})$ we
get from (\ref{subyalfa}) and (\ref{subyalfa1}) that%
\[
0\leq\mathcal{L}_{\gamma_{\lambda}}(\underline{u},\Phi_{x}^{\lambda
}(x_{\lambda},y_{\lambda}),A_{\delta})(x_{\lambda})-\mathcal{L}_{\gamma
_{\lambda}}(\overline{u}^{s_{0}},-\Phi_{y}^{\lambda}(x_{\lambda
},y_{\lambda
}),B_{\delta})(y_{\lambda})
\]
and so%
%
%
\begin{eqnarray}\label{desdifsup}%
&&(c+\beta) \bigl( \underline{u}(x_{\lambda})-\overline{u}^{s_{0}}(y_{\lambda
}) \bigr) \nonumber\\
&&\qquad \leq \sigma^{2}\gamma_{\lambda}^{2}(A_{\delta}x_{\lambda}%
^{2}-B_{\delta}y_{\lambda}^{2})/2\nonumber\\
&&\qquad\quad{} +p\bigl(\Phi_{x}^{\lambda}(x_{\lambda},y_{\lambda})+\Phi_{y}^{\lambda
}(x_{\lambda},y_{\lambda})\bigr)\\
&&\qquad\quad{} +r\gamma_{\lambda}\bigl(\Phi_{x}^{\lambda}(x_{\lambda},y_{\lambda
})x_{\lambda
}+\Phi_{y}^{\lambda}(x_{\lambda},y_{\lambda})y_{\lambda}\bigr)\nonumber\\
&&\qquad\quad{} +\beta\biggl( \int_{0}^{x_{\lambda}}\underline{u}(x_{\lambda
}-\alpha)\,dF(\alpha)-\int_{0}^{y_{\lambda}}\overline{u}^{s_{0}%
}(y_{\lambda}-\alpha)\,dF(\alpha) \biggr) .\nonumber
\end{eqnarray}

Using the inequality
\[
\Sigma^{\lambda} ( x_{\lambda},x_{\lambda} ) +\Sigma^{\lambda
} ( y_{\lambda},y_{\lambda} ) \leq2\Sigma^{\lambda} (
x_{\lambda},y_{\lambda} ),
\]
we obtain that%
\[
\lambda( x_{\lambda}-y_{\lambda} ) ^{2}\leq\underline
{u}(x_{\lambda})-\underline{u}(y_{\lambda})+\overline
{u}^{s_{0}}(x_{\lambda
})-\overline{u}^{s_{0}}(y_{\lambda})+4m(y_{\lambda}-x_{\lambda});
\]
then we have from (\ref{acotacion-con-m}) that%
%
%
\begin{equation} \label{desxalfa-yalfac}%
\lambda( x_{\lambda}-y_{\lambda} ) ^{2}\leq6m | x_{\lambda
}-y_{\lambda} | .
\end{equation}

We can find a sequence $\lambda_{n}\hspace*{-0.15pt}\rightarrow\hspace*{-0.15pt}\infty$ such that $ (
x_{\lambda_{n}},y_{\lambda_{n}} ) \hspace*{-0.15pt}\rightarrow\hspace*{-0.15pt}( \overline
{x},\overline{y} ) \in A$. From~(\ref{desxalfa-yalfac}), we get that
$ \vert x_{\lambda_{n}}-y_{\lambda_{n}} \vert\leq6m/\lambda_{n}$ and
this gives $\overline{x}=\overline{y}$ and so $\lim_{n\rightarrow
\infty}\lambda_{n} ( x_{\lambda_{n}}-y_{\lambda_{n}} ) ^{2}=0$.
Taking $\delta=1/\lambda$, we get using that $y_{\lambda_{n}}\geq
x_{\lambda_{n}}$ for all $n$, (\ref{DesigualdarSegundoOrden}),
(\ref{SumaDerivasPhi}), (\ref{CasiSumaDerivasPhi}) and (\ref{desdifsup})
%
%
\begin{equation}\label{Uriel}
(c+\beta) \bigl( \underline{u}(\overline{x})-\overline{u}^{s_{0}}(\overline
{x}) \bigr) \leq\beta\int_{0}^{C} \bigl( \underline
{u}(\overline{x}-\alpha)-\overline{u}^{s_{0}}(\overline{x}-\alpha)
\bigr)\,
dF(\alpha),
\end{equation}
where $C$ can be equal to either $\overline{x}$ or $\overline{x}^{-}$.

From (\ref{liminf-mlambda}) and (\ref{Uriel}) we obtain $M\leq\beta
M/(c+\beta)$. This is a contradiction because $M>0$ and $\beta/(c+\beta)<1$.
\end{pf}

From the previous proposition, we conclude the following corollary.
\begin{corollary}
\label{Unicidad}
For any $u_{0}>0$, there is at most one
viscosity solution of (\ref{InicDif}) in $(0,+\infty)$ satisfying
assumptions \textup{(A.1)}, \textup{(A.2)} and \textup{(A.3)} with the boundary
condition $u(0)=u_{0}$.
\end{corollary}

\section{Characterization of $V$ as the smallest supersolution and a
verification result}\label{sec5}

In Sections \ref{sec2} and \ref{sec3}, we have proved that the optimal value function $V$ is
well defined and that it is a viscosity solution of (\ref{InicDif}). In
Section \ref{sec4}, we have proved that (\ref{InicDif}) has a comparison
principle that
gives us uniqueness of viscosity solutions with a given boundary
condition. As
it can be seen in the next remark there are infinitely many classical
solutions of the HJB equation satisfying (A.1), (A.2) and
(A.3).
\begin{remark}
\label{Nohayunicidad} Note that $u(x)=k+x$ is a viscosity solution
of (\ref{InicDif}) in $[0,\infty)$ for any $k\geq p/c$ because
$u^{\prime
}=1$ and $\mathcal{L}^{\ast}(u)\leq0$.
\end{remark}

Our main goal in this section is to characterize $V$ among all the viscosity
solutions of (\ref{InicDif}). We show that the optimal value function
$V$ is
the smallest of the absolutely continuous supersolutions of the HJB equation.
We use this result to prove a \textit{verification theorem} that states that
if a supersolution of the HJB equation is obtained, either as a value function
of an admissible strategy, or as a limit of value functions of admissible
strategies, then this supersolution should be the optimal value function.

Later in this section, using the Corollary \ref{Unicidad}, we also
characterize $V$ as the viscosity solution of the HJB equation with the
smallest possible boundary condition at zero.

To prove Proposition \ref{MenorSuper} we need the following technical lemma.
\begin{lemma}
\label{A.1}
Let $\overline{u}$ be an absolutely continuous
nonnegative supersolution of (\ref{InicDif}) in $(0,+\infty)$.
Given any pair of real numbers $x_{1}>x_{0}>0$, we can find a
sequence of nonnegative functions $u_{n}\dvtx\mathbf{R}\rightarrow\mathbf{R}$
such that:

\textup{(a)} $u_{n}$ is twice continuously differentiable,

\textup{(b)} $u_{n}$ converges uniformly to $\overline{u}$
in $[0,x_{1}]$,

\textup{(c)} $u_{n}^{\prime}\geq1$ in $[x_{0},x_{1}]$,

\textup{(d)} $\limsup_{n\rightarrow\infty}\mathcal
{L}%
^{\ast}(u_{n})(x)\leq\beta\overline{u}(0) ( F(x)-F(x^{-}) ) $
for $x\in[x_{0},x_{1}]$.
\end{lemma}
\begin{pf}
Let us consider an even and twice-continuously differentiable function
$\phi$
with support included in $(-1,1)$, with integral one, such that $\phi
^{\prime
}\geq0$ in $(-1,0)$ and $\phi^{\prime}\leq0$ in $(0,1)$. Consider $\phi
_{n}(x)=n\phi(n ( x-1/n ) )$ and define $u_{n}$ as the left-sided
convolution $u_{n}(x)= ( \overline{u}\ast\overline{\phi}_{n} )
(x)$. The results (a) and (b) follow using standard techniques [see, for
instance, Wheeden and Zygmund (\citeyear{WZ77})]; (c) follows because
$\overline{u}{}
^{\prime}\geq1$ a.e.

Let us prove (d). By Proposition \ref{supersemiconcave}, $\overline
{u}$ is
semiconcave and so $\overline{u}^{\prime\prime}$ exists a.e., and the possible
jumps of $\overline{u}$ are downward. So, the left-sided convolution $u_{n}$
satisfies $u_{n}^{\prime\prime}(x)\leq( \overline{u}^{\prime\prime}%
\ast\overline{\phi}_{n} ) (x)$. The result (d) follows because
$\mathcal{L}_{\gamma}(\overline{u})(x)\leq0$ a.e. for any $\gamma\in
[0,1]$, and it can be shown that
\[
\limsup_{n\rightarrow\infty} \bigl( \mathcal{L}%
_{\gamma}(u_{n})(x)- \bigl( \mathcal{L}_{\gamma}(\overline{u})\ast
\overline{\phi}_{n} \bigr) (x) \bigr) \leq\beta\overline{u}(0) \bigl(
F(x)-F(x^{-}) \bigr)
\]
for all $x\in[ x_{0},x_{1}]$.
\end{pf}
\begin{proposition}
\label{MenorSuper}
Let $\overline{u}$ be an absolutely
continuous nonnegative supersolution of (\ref{InicDif}) in $(0,+\infty
)$, then $\overline{u}\geq V$ in $[0,+\infty)$.
\end{proposition}
\begin{pf}
Let us define $S$ as the set of discontinuity points of the claim-size
distribution $F$. Since $F$ is increasing $S$ is a countable set. Take $x>0$,
by Lemmas~\ref{admissiblestrategiesacotadas} and \ref{LemaEvitarSaltosF}
(included in the \hyperref[app]{Appendix}), it is enough to prove that
for any pair
$(x_{0},x_{1})$ such that $0<x_{0}\leq x\leq x_{1}$, we have
\[
\sup_{\pi\in\Pi_{x}^{ [ x_{0},x_{1} ] }\cap\Pi_{x}(S)}%
V_{\pi}(x)\leq\overline{u}(x),
\]
where $\Pi_{x}^{ [ x_{0},x_{1} ] } = \{ \pi
\in\Pi_{x}\dvtx x_{0}\leq X_{t}^{\pi}\leq x_{1},t\geq0 \} $ and $\Pi
_{x}(S)$ is the set of all the admissible strategies $\pi\in\Pi_{x}$ such
that the measure of
\[
\{ (\omega,t)\in\Omega\times\mathbf{[}0\mathbf{,\infty)}\dvtx X_{t}^{\pi
}(\omega)\in S \}
\]
is zero.

Take $\pi=(\gamma_{t},L_{t})\in\Pi_{x}^{ [ x_{0},x_{1} ] }\cap
\Pi_{x}(S)$. Consider the functions $u_{n}$ defined in Lemma \ref{A.1}; since
they are twice continuously differentiable, we can write%
%
%
\begin{eqnarray}\label{SPS1(a)}\qquad
&&
u_{n}(X_{\tau^{\pi}})e^{-c\tau^{\pi}}-u_{n}(x) \nonumber\\
&&\qquad = \int_{0}%
^{\tau^{\pi}}u_{n}^{\prime}(X_{s})e^{-cs}\,dX_{s}-c\int_{0}^{\tau
^{\pi
}}u_{n}(X_{s})e^{-cs}\,ds\\
&&\qquad\quad{} +(\sigma^{2}/2)\int_{0}^{\tau^{\pi}}u_{n}^{\prime\prime}
(X_{s})\gamma_{s}^{2}X_{s}^{2}e^{-cs}\,ds\nonumber
\end{eqnarray}
for any $t\geq0$.

Note that, since $L_{t}$ is nondecreasing and left-continuous, it can be
written as
%
%
\begin{equation} \label{Lt}%
L_{t}=\int_{0}^{t}dL_{s}^{c}+\sum_{X_{s^{+}}\neq
X_{s},s<t}(L_{s^{+}%
}-L_{s}),
\end{equation}
where $L_{s}^{c}$ is a continuous and nondecreasing function. Hence, using
expressions (\ref{Xt}) and (\ref{Lt}), we get
%
%
\begin{eqnarray}\label{SPS2(a)}%
&&\int_{0}^{\tau^{\pi}}u_{n}^{\prime}(X_{s})e^{-cs}\,dX_{s}\nonumber\\%
&&\qquad=\int_{0}^{\tau^{\pi}}u_{n}^{\prime}(X_{s})e^{-cs}p\,ds+\int
_{0}^{\tau^{\pi}}u_{n}^{\prime}(X_{s})e^{-cs}rX_{s}\gamma
_{s}\,ds\nonumber\\
&&\qquad\quad{}+\int_{0}^{\tau^{\pi}}u_{n}^{\prime}(X_{s})e^{-cs}\sigma X_{s}%
\gamma_{s}\,dW_{s}\nonumber\\[-8pt]\\[-8pt]
&&\qquad\quad{}-\int_{0}^{\tau^{\pi}}u_{n}^{\prime}(X_{s}%
)e^{-cs}\,dL^{c}(s)\nonumber\\
&&\qquad\quad{}+\sum_{X_{s^{-}}\neq X_{s},s\leq\tau^{\pi}} \bigl(
u_{n}(X_{s})-u_{n}(X_{s^{-}}) \bigr) e^{-cs}\nonumber\\
&&\qquad\quad{} + \sum_{X_{s^{+}}\neq
X_{s},s<\tau^{\pi}} \bigl( u_{n}(X_{s^{+}})-u_{n}(X_{s}) \bigr) e^{-cs}.\nonumber
\end{eqnarray}
We have that $X_{s^{+}}\neq X_{s}$ only at the discontinuities of
$L_{s}$, so
$X_{s^{+}}-X_{s}=- ( L_{s^{+}}-L_{s} ) $ and%
\begin{eqnarray*}
&&\sum_{X_{s^{+}}\neq X_{s},s<\tau^{\pi}} \bigl( u_{n}(X_{s^{+}}%
)-u_{n}(X_{s}) \bigr) e^{-cs} \\
&&\qquad= -\sum_{L_{s^{+}}\neq L_{s}%
,s<\tau^{\pi}} \biggl( \int_{0}^{L_{s^{+}}-L_{s}}u_{n}^{\prime}%
(X_{s}-\alpha)\,d\alpha\biggr) e^{-cs}.
\end{eqnarray*}
From Lemma \ref{A.1}(c), $u_{n}^{\prime}\geq1$, so we obtain
%
%
\begin{eqnarray}\label{SPSvaloractual(a)}%
&&-\int_{0}^{\tau^{\pi}}u_{n}^{\prime}(X_{s})e^{-cs}\,dL_{s}^{c}%
\nonumber\\
&&\quad{}+\sum_{X_{s^{+}}\neq X_{s},s<\tau^{\pi}} \bigl( u_{n}(X_{s^{+}}%
)-u_{n}(X_{s}) \bigr) e^{-cs}\nonumber\\[-8pt]\\[-8pt]
&&\qquad\leq- \biggl( \int_{0}^{t\wedge\tau^{\pi}}%
e^{-cs}\,dL_{s}^{c}+\sum_{L_{s^{+}}\neq L_{s},s<\tau^{\pi}} \biggl(
\int_{0}^{L_{s^{+}}-L_{s}}d\alpha\biggr) e^{-cs} \biggr)
\nonumber\\
&&\qquad=-\int_{0}^{\tau^{\pi}}e^{-cs}\,dL_{s}.\nonumber
\end{eqnarray}

Since $X_{s}\neq X_{s^{-}}$ only at the arrival of a claim, the process
%
\begin{eqnarray}\label{SPSmartingala(a)}%
M_{t}^{(1)} &=&
{ \sum_{X_{s^{-}}\neq X_{s},s\leq t}}
\bigl( u_{n}(X_{s})-u_{n}(X_{s^{-}}) \bigr) e^{-cs}\nonumber\\[-8pt]\\[-8pt]
&&{} -\beta\int_{0}^{t}e^{-cs}\int_{0}^{\infty} \bigl(
u_{n}(X_{s^{-}}-\alpha)-u_{n}(X_{s^{-}}) \bigr) \,dF(\alpha)\,ds\nonumber
\end{eqnarray}
is a martingale with zero-expectation.

From (\ref{SPS1(a)}), (\ref{SPS2(a)}), (\ref{SPSvaloractual(a)}) and
(\ref{SPSmartingala(a)}), we obtain
%
%
\begin{eqnarray}\label{limite0}%
&& u_{n}(X_{\tau^{\pi}})e^{-c\tau^{\pi}}-u_{n}(x)\nonumber\\[-8pt]\\[-8pt]
&&\qquad{}\leq\int
_{0}^{\tau
^{\pi}}\mathcal{L}_{\gamma_{s}}(u_{n})(X_{s^{-}})e^{-cs}\,ds-\int
_{0}%
^{\tau^{\pi}}e^{-cs}\,dL_{s}+M_{\tau^{\pi}}^{(1)}+M_{\tau^{\pi}}^{(2)},\nonumber
\end{eqnarray}
where
\[
M_{t}^{(2)}=\int_{0}^{t}u_{n}^{\prime}(X_{s})e^{-cs}\sigma
X_{s}\gamma_{s}\,dW_{s}%
\]
is a martingale with zero-expectation.

We have $E_{x} (
{ \int_{0}^{\tau^{\pi}}}
e^{-cs}\,dL_{s} ) =V_{\pi}(x)$, $E_{x} ( u_{n}(X_{\tau^{\pi}%
})e^{-c\tau^{\pi}} ) \geq0$ and from Lem\-ma~\ref{A.1}(d), since $\pi
\in\Pi_{x}(S)$, we have that
\[
\lim_{n\rightarrow\infty}E_{x} \biggl( \int_{0}^{t\wedge
\tau^{\pi}}\mathcal{L}_{\gamma_{s}}(u_{n})(X_{s^{-}})e^{-cs}\,ds \biggr) \leq0
\]
for all $t$. Then, from Lemma \ref{A.1}(b), we obtain $\overline{u}%
(x)=\lim_{n\rightarrow\infty}u_{n}(x)\geq V_{\pi}(x)$.
\end{pf}

In order to state the \textit{verification theorem}, we need to extend the
concept of strategies by the following definition.
\begin{definition}
(a) Fix $x\geq0$, let us define the map $\mathcal{V}%
_{x}\dvtx\Pi_{x}\rightarrow[0,\infty)$ as $\mathcal{V}_{x}%
(\pi)=V_{\pi}(x)$. We give to $\Pi_{x}$ the initial
topology of $\mathcal{V}_{x}$ and define $\widetilde{\Pi}_{x}
$ as the completion of $\Pi_{x}$ under this topology [see,
for instance, Kelley (\citeyear{K55})]. We say that the elements of
$\widetilde{\Pi}%
_{x}$ are limit strategies.

(b) Given $\widetilde{\pi}\in\widetilde{\Pi}_{x}$, there exists
a sequence $\pi^{k}\in\Pi_{x}$ such that $\lim_{k\rightarrow
\infty}\pi^{k}=\widetilde{\pi}$, we define $V_{\widetilde{\pi}}(x)=\lim
_{k\rightarrow\infty}V_{\pi^{k}}(x)$.
\end{definition}

From Proposition \ref{MenorSuper}, we get the following \textit{verification
theorem}.
\begin{theorem}
\label{VerificationTheorem}
Let $\pi$ be a limit strategy
such that the corresponding value function $V_{\widetilde{\pi}}$
is an absolutely continuous supersolution of (\ref{InicDif}) in $ (
0,\infty) $, then $V_{\widetilde{\pi}}=V$.
\end{theorem}

We conclude from Remark \ref{Nohayunicidad} and Proposition \ref{MenorSuper}
that the optimal value function $V$ satisfies
%
%
\begin{equation} \label{CrecimientoLineal}%
V(x)\leq x+p/c\qquad\mbox{for }x\geq0,
\end{equation}
and so it satisfies (A.3). By Propositions \ref{1} and
\ref{2}, the
optimal value function $V$ also satisfies (A.1) and (A.2).
Therefore, from Corollary \ref{Unicidad} and Proposition \ref
{MenorSuper} we
get the following corollary.
\begin{corollary}
The function $V$ can be also characterized as the unique
viscosity solution of (\ref{InicDif}) satisfying assumptions \textup{(A.1)},
\textup{(A.2)} and \textup{(A.3)} with the boundary condition,
\[
V(0)=\inf\{u(0)\dvtx u\mbox{ viscosity supersolution of (\ref{InicDif})
satisfying
\textup{(A.3)}}\}.
\]
\end{corollary}

\section{Solutions of the second-order differential equation}\label{sec6}

In the previous sections we have characterized the optimal value
function $V$
without assuming any regularity conditions on the claim-size distribution
function $F$. To find the optimal value function $V$ and the value
function of
barrier strategies, we need some technical results about the solutions
of%
%
%
\begin{equation}\label{EcuDif(w)}%
\mathcal{L}^{\ast}(W)=0
\end{equation}
on open sets. In order to have classical solutions of this equation, we assume,
from this section on, that the claim-size distribution function $F$ has a
bounded density. If we do not assume this, we would have to deal with
viscosity solutions of (\ref{EcuDif(w)}) and this adds some technical problems.

Equation (\ref{EcuDif(w)}) is similar to the HJB equation that arises in
the problem of maximizing the survival probability of an insurance company
whose uncontrolled reserve follows the classical Cram\'{e}r--Lundberg process
and where the management has the possibility of investing in the financial
market. Azcue and Muler (\citeyear{AM09}) considered this problem and
showed that the
optimal survival probability function $\delta$ is a classical solution of
$\mathcal{L}^{\ast}(\delta)=0$ in $(0,\infty)$, but with parameter $c$ equal
to zero.

The existence and uniqueness of classical solutions of (\ref
{EcuDif(w)}) is
not straightforward since the ellipticity of $\mathcal{L}^{\ast}$ degenerates
at $0$ and could degenerate at any positive point. However, we prove in this
section that the optimal $\gamma$ in (\ref{EcuDif(w)}) is not zero for
positive points. On the other hand, the degeneracy of the ellipticity
of the
operator at zero gives the uniqueness of twice continuously differentiable
solutions of (\ref{EcuDif(w)}) in $ ( 0,\infty) $ with only one
boundary condition at zero.

In the next proposition we construct, via a fixed-point argument, the unique
twice continuously differentiable solution of (\ref{EcuDif(w)}) in $ (
0,\infty) $ with the boundary condition $W(0)=1$.
\begin{proposition}
\label{WesC2} \textup{(a)}
There exists a unique increasing classical solution
$W$ of (\ref{EcuDif(w)}) in $(0,\infty)$ with the boundary
condition, $W(0)=1$ and $\mathcal{L}^{\ast}(W)=\mathcal{L}%
_{\widetilde{\gamma}(W)}(W)=0$, where
%
%
\begin{equation}\label{DefinicionGammaEstrella}%
\widetilde{\gamma}(W)(x)= \cases{
-rW^{\prime}(x)/(\sigma^{2}xW^{\prime\prime}(x)), \cr
\hspace*{32.29pt}\mbox{if
$0<-rW^{\prime
}(x)/(\sigma^{2}xW^{\prime\prime}(x))\leq1$},\cr
1, \qquad\mbox{otherwise}.}
\end{equation}

\textup{(b)} The function $W$ can be written as $W(x)=1+\int
_{0}^{x}w(s)\,ds$, where $w$ is the unique
nonnegative fixed point of the operator,
%
%
\begin{equation} \label{FixPointU}%
T(w)(x)=\inf_{\Gamma\in\mathcal{G}}\frac{2\int_{0}^{x}A_{\Gamma} (
s ) M(W)(s)\,ds}{\sigma^{2}x^{2}\Gamma(x)^{2}A_{\Gamma} ( x )}.
\end{equation}
Here
%
%
\begin{eqnarray}\label{DefinicionM(w)}%
\mathcal{G} &=& \{ \Gamma\dvtx[0,\infty)\rightarrow( 0,1 ]\nonumber\\
&&\hspace*{5.7pt}\mbox{piecewise continuous with }\inf( \Gamma) >0 \},
\\
M(W)(x) &=&
(c+\beta)W(x)-\beta\int_{0}^{x}W(x-\alpha)\,dF(\alpha)\nonumber
\end{eqnarray}
and
\[
A_{\Gamma} ( x ) =e^{\int_{1}^{x}2(p+r\Gamma(s)s)/(\sigma^{2}%
\Gamma(s)^{2}s^{2})\,ds}/(\Gamma(x)^{2}x^{2}).
\]
\end{proposition}
\begin{pf}
We give here a sketch of the proof and refer to Sections 3 and 4 in
Azcue and
Muler (\citeyear{AM09}) for details since the proof is similar.

It can be proved that if $U$ is any classical increasing solution of
(\ref{EcuDif(w)}) with $U(0)=1$, then $u=U^{\prime}$ is a fixed-point of
(\ref{FixPointU}) and also that there is a unique continuous nonnegative
fixed-point $w$ of (\ref{FixPointU}). It can be proved that $w$ is locally
Lipschitz, and so $W(x)=1+\int_{0}^{x}w(s)\,ds$ is semiconcave
in any
compact set included in $(0,\infty)$. The next step consists of proving that
$W$ is twice continuously differentiable and so it is a classical
solution of
(\ref{EcuDif(w)}). To do that, we construct twice continuously differentiable
increasing solutions of the second-order integro-differential equations,
$\mathcal{L}_{1}(W_{1})=0$ and $\sup_{\gamma\in\mathbf{R}}%
\mathcal{L}_{\gamma}(W_{2})=0$, and show then that $W$ coincides
locally with
one or the other and also that $W$ is obtained by gluing smoothly
solutions of
these equations. Hipp and Plum (\citeyear{HP00}) and Schmidli (\citeyear
{S02}) studied and found
classical solutions of the equation $\sup_{\gamma\in\mathbf{R}%
}\mathcal{L}_{\gamma}(W)=0$ with $c=0$ for the problem of minimizing
the ruin
probability of an insurance company without borrowing constraints.

Finally, since $\mathcal{L}_{\gamma}(W)(x)$ is a quadratic function on
$\gamma$ and $W$ is increasing, the maximum is attained at $\gamma=1$
or at
the vertex $\gamma=-rW^{\prime}(x)/(\sigma^{2}xW^{\prime\prime}(x))$.
It can
be shown that the vertex cannot be zero.
\end{pf}
\begin{remark}
Given $\Gamma\in\mathcal{G}$, consider the related problem of finding the
survival probability $S(x)$ of an insurance company with initial
surplus $x$,
whose uncontrolled reserve follows the classical Cram\'{e}r--Lundberg process
and where the management invests a proportion $\Gamma$ of the current surplus
in the financial market. The function $S$ can be called the scale
function of
the surplus process as in Revuz and Yor (\citeyear{RY99}), and it is a
solution of
$\mathcal{L}_{\Gamma}(S)=0$ but with parameter $c$ equal to zero [see Azcue
and Muler (\citeyear{AM09})]. So, as in Proposition \ref{WesC2}, we have
\[
S^{\prime}(x)=\frac{2\int_{0}^{x}A_{\Gamma} ( s ) \beta(
S(s)-\int_{0}^{s}S(s-\alpha)\,dF(\alpha) ) \,ds}{\sigma^{2}x^{2}%
\Gamma(x)^{2}A_{\Gamma} ( x ) }.
\]
\end{remark}
\begin{proposition}
\label{WesC2(segunda)}
\textup{(a)} The function $\widetilde{\gamma}(W)$
defined in (\ref{DefinicionGammaEstrella}) can be written as
\[
\widetilde{\gamma}(W)(x)=\min\bigl\{1,2 \bigl( M(W)(x)-pW^{\prime}(x) \bigr)
/(rxW^{\prime}(x))\bigr\}.
\]

\textup{(b)} There exists $\varepsilon>0$ such that
$\widetilde{\gamma}(W)(x)=1$ for $x\in[0,\varepsilon)$.

\textup{(c)} $W^{\prime}(0^{+})= ( c+\beta) /p$ and
$W^{\prime\prime}(0^{+})= ( c+\beta-r ) ( c+\beta)
/p^{2}-F^{\prime}(0)\beta/p$.
\end{proposition}
\begin{pf}
We obtain (a) by   replacing the value of $W^{\prime\prime}(x)$ obtained
from the
equation $\mathcal{L}_{\widetilde{\gamma}(W)}(W)=0$ in the definition
(\ref{DefinicionGammaEstrella}). To prove (b) and (c) consider $W_{1}$ the
unique increasing twice continuously differentiable solution of
$\mathcal{L}%
_{1}(W_{1})=0$. It can be proved using L'H\^{o}pital's rule that
\[
W_{1}^{\prime}(0^{+})= ( c+\beta) /p\quad\mbox{and}\quad
W_{1}^{\prime\prime}(0^{+})= ( c+\beta-r ) (
c+\beta) /p^{2}-F^{\prime}(0)\beta/p,
\]
and so%
\[
{\lim_{x\rightarrow0^{+}}} \vert rW_{1}^{\prime}(x)/(\sigma^{2}%
xW_{1}^{\prime\prime}(x)) \vert=+\infty.
\]
We conclude that there exists $\varepsilon>0$ such that %
$\widetilde{\gamma}(W_{1})=1$ in $[0,\varepsilon)$.
Therefore, using that $\mathcal{L}^{\ast}(W_{1})=\mathcal{L}_{\widetilde
{\gamma}(W_{1})}(W_{1})$, we obtain that $W_{1}$ satisfies $\mathcal
{L}^{\ast
}(W_{1})=0$ in $[0,\varepsilon)$ and so $W(x)=W_{1}(x)$ for small
values of $x$.
\end{pf}

In an analogous way, given a positive $x_{0}$ and an increasing
continuous function $W_{0}$ defined in $[0,x_{0}]$ such that
$W_{0}$ is differentiable at $x_{0}$, we can construct the unique twice
continuously differentiable solution of%
%
%
\begin{equation} \label{EcuDif(w,w0)}
\mathcal{L}^{\ast}(U,W_{0})(x)=0\qquad\mbox{for }x>x_{0}
\end{equation}
with boundary conditions $U(x_{0})=W_{0}(x_{0})$ and $U^{\prime}(x_{0}%
)=W_{0}^{\prime}(x_{0})$ where%
%
%
\begin{eqnarray}
\label{Lestrellacontinuacion}
\mathcal{L}^{\ast}(U,W_{0})&=&\sup_{\gamma\in[0,1]}\mathcal
{L}%
_{\gamma}(U,W_{0}),
\\
\label{LgammaContinuacion}\qquad%
\mathcal{L}_{\gamma}(U,W_{0})(x) & = & \sigma^{2}\gamma^{2}x^{2}%
U^{\prime\prime}(x)/2+ ( p+r\gamma x ) U^{\prime}(x)-M(U,W_{0})(x)
\end{eqnarray}
and
%
%
\begin{eqnarray}\label{DefinicionMContinuacion}%
M(U,W_{0})(x)&=&(c+\beta)U(x)-\beta\int_{0}^{x-x_{0}}U(x-\alpha
)\,dF(\alpha)\nonumber\\[-8pt]\\[-8pt]
&&{}-\beta\int_{x-x_{0}}^{x}W_{0}(x-\alpha)\,dF(\alpha).\nonumber
\end{eqnarray}

The next proposition is analogous to Propositions \ref{WesC2} and
\ref{WesC2(segunda)}(a); the proof follows by using a fixed-point argument
similar to the one used in Proposition \ref{WesC2}.
\begin{proposition}
\label{NuevaProposicion1}
Assume that $W_{0}$ is a
continuous, positive and increasing function in $[0,x_{0}]$
and that $W_{0}$ is differentiable at $x_{0}$.

\textup{(a)} There exists a unique twice continuously differentiable solution
$U$ of (\ref{EcuDif(w,w0)}) in $ ( x_{0},\infty) $ with
$U(x_{0})=W_{0}(x_{0})$ and $U^{\prime}(x_{0})=W_{0}^{\prime}(x_{0})$.

\textup{(b)} If we define
\[
\widetilde{\gamma}(U,W_{0})(x)=\min\bigl\{1,2 \bigl( M(U,W_{0})(x)-pU^{\prime
}(x) \bigr) /(rxU^{\prime}(x))\bigr\}
\]
we have that
\[
\mathcal{L}^{\ast}(U,W_{0})=\mathcal{L}_{\widetilde{\gamma}(U,W_{0})}%
(U,W_{0})=0.
\]
\end{proposition}

\section{Barrier strategies}\label{sec7}

A dividend payment policy is called \textit{barrier with level} $y$
when all
excess surplus above $y$ is paid out immediately as dividends, but
there is no
dividends payment when surplus is less than $y$. In this section we
would like
to obtain the \textit{optimal barrier strategy}, that is, the admissible
strategy that maximizes the cumulative expected discounted dividends
among all
the strategies whose dividend policies are barrier. We would also like to
prove that the optimal barrier strategy is \textit{stationary}, in the sense
that the decision on how much dividend to pay and how to invest at any time
depends only on the current surplus. Note that a stationary strategy
$\pi$
determines an admissible strategy $\pi_{x}\in\Pi_{x}$ for each initial surplus
$x$.

In the classical Cram\'{e}r--Lundberg model without the possibility of
investment, there exists an optimal barrier strategy. Let $y^{\ast}$ be the
optimal level. It has been proved [for instance, in Azcue and Muler
(\citeyear{AM05})]
that the optimal policy for current surplus $y^{\ast}$ is to pay all the
incoming premium as dividends in order to maintain the surplus at level
$y^{\ast}$ until the arrival of the next claim.

In the model with investment, it is possible to define similar barrier
strategies for any level $y$ (if the current surplus is $y$, pay all the
incoming premium as dividends and keep all the surplus in bonds), but these
barrier strategies are never optimal. In fact there is not a stationary
barrier strategy which is optimal, since it is not possible to
determine the
dividends payment policy when the current surplus coincides with the threshold.
We construct in this section a candidate of optimal barrier strategy as an
explicit limit of stationary admissible barrier strategies and find its value
function. In the next sections we will prove that this strategy is
indeed the
optimal barrier strategy, also we will show that the optimal strategy in
(\ref{V}) could be nonbarrier, but this optimal strategy and the optimal
barrier strategy coincide for small surpluses.

First in this section we use the function $W$, constructed in Section
\ref{sec6}, to
obtain the value function of a limit barrier strategy with a given
level $y$
and the best investment policy. Later we find the optimal level of these
strategies. In all the cases, the optimal investment policy is
stationary in
the sense that the decision on how to invest depends only on the
current surplus.
\begin{definition}
\label{Definiciondetauestrellita}
Given a predictable process
$\gamma_{t}\in[0,1]$ and points\break \mbox{$0<z<y$},
we define recursively for initial surplus $x\geq0$, the
admissible strategy $\pi_{x}^{(\gamma_{t},z,y)}\in\Pi_{x}$ as:

\begin{enumerate}
\item If $x>y$, pay immediately the surplus $x-y$
as dividends and follow the strategy $\pi_{y}^{(\gamma_{t},z,y)}\in\Pi_{y}$.

\item If $x\leq y$, follow the admissible strategy
$(\gamma_{t},0)$ up to the exit time $\tau^{\ast}=\min\{\tau
_{y},\tau^{\pi}\}$ where
\[
\tau_{y}=\min\bigl\{t\dvtx X_{t}^{\pi_{x}^{(\gamma_{t},z,y)}}=y\bigr\}
\]
and $\tau^{\pi}$ is the ruin time. When $\tau^{\ast}%
=\tau_{y}$, pay immediately $y-z$ as dividends and follow
the strategy $\pi_{z}^{(\gamma_{t},z,y)}$ with initial surplus
$z$.
\end{enumerate}
\end{definition}

Let us call $\overline{\Pi}{}^{z,y}_{x}$ the set of all these
strategies, and
let us consider for all $x\in[0,y]$ the function
%
%
\begin{equation}\label{DefWyz}%
W_{z,y}(x)=\sup_{\pi\in\overline{\Pi}_{x}^{z,y}}V_{\pi}(x).
\end{equation}
We define $W_{z,y}(x)=0$ for $x<0$. We first state some basic
properties of
the function $W_{z,y}$. The proof of the next proposition is similar to the
proof of Propositions~\ref{1} and \ref{2}.
\eject
\begin{proposition}
\label{Teoremawyz} We have that:

\textup{(a)} The value function $W_{z,y}$ is well defined.

\textup{(b)} If $y\geq x_{2}>x_{1}$, then
\[
W_{z,y}(x_{2})-W_{z,y}(x_{1})\leq\bigl( e^{(c+\beta)(x_{2}-x_{1}%
)/p}-1 \bigr) W_{z,y}(x_{1}).
\]

\textup{(c)} $W_{z,y}(y)=W_{z,y}(z)+(y-z)$.

\textup{(d)} $W_{z,y}$ is increasing in $[0,y]$.

\textup{(e)} $W_{z,y}$ is absolutely continuous in
$[0,y]$.
\end{proposition}

Let us state now a dynamic programming principle for these value functions.
\begin{proposition}
\label{DPPw}
Given $x\in[0,y]$ and any stopping time
$\tau$, we have that
\[
W_{z,y}(x)=\sup_{ ( \gamma_{t} )\ \mathrm{admissible}}%
E_{x} \bigl( e^{-c ( \tau\wedge\tau^{\ast} ) }W_{z,y}\bigl(X_{\tau
\wedge\tau^{\ast}}^{(\gamma_{t},0)}\bigr) \bigr),
\]
where $\tau^{\ast}$ is the stopping time defined in
Definition \ref{Definiciondetauestrellita}.
\end{proposition}

In the next proposition, we show that all the functions $W_{z,y}$ are
multiples of the function $W$ obtained in Proposition \ref{WesC2}; this allows
us to describe the optimal investment policy for (\ref{DefWyz}).
\begin{proposition}
\label{CaracterizaciondeWz,y}
\textup{(a)} We have that
\[
W_{z,y}(x)= \cases{
\dfrac{W(x)}{(W(y)-W(z))/(y-z)}, &\quad if $0\leq x<y$,\vspace*{2pt}\cr
\dfrac{W(y)}{(W(y)-W(z))/(y-z)}+(x-y), &\quad if $x\geq y$,}
\]
where $W$ is the function obtained in Proposition
\ref{WesC2}.

\textup{(b)} $W_{z,y}(x)$ is the value function of the admissible
stationary strategy $\pi_{x}\in\overline{\Pi}_{x}^{z,y}$, the optimal
investment policy depends only on the current surplus $X_{t^{-}}^{\pi}%
$ and it is given by $\overline{\gamma}_{t}=\widetilde{\gamma
}(W)(X_{t^{-}}^{\pi})$ where the function $\widetilde{\gamma}%
(W)$ is defined in Proposition \ref{WesC2}.
\end{proposition}
\begin{pf}
We extend the definition of $W$ as $W(x)=0$ for $x<0$. Let us take any
admissible strategy $\pi=(\gamma_{t},L_{t})\in\overline{\Pi}_{x}^{z,y}$ and
consider the stopping times $\tau_{y}$ and $\tau^{\ast}$ defined in
Definition \ref{Definiciondetauestrellita}. Up to time $\tau^{\ast
}$, the
dividend payment policy $L_{t}$ is zero, so the strategy $\pi$ only depends
on the investment policy $\gamma=(\gamma_{t})$. To simplify notation, we
denote $X_{t}^{\gamma}$ the corresponding controlled risk process
starting at
$x$. This process satisfies up to $\tau^{\ast}$ the following stochastic
differential equation:
%
%
\begin{equation}\label{XtsinL}%
dX_{s}^{\gamma}= ( p+rX_{s}^{\gamma}\gamma_{s} ) \,ds+\sigma
X_{s}^{\gamma}\gamma_{s}\,dW_{s}-d\Biggl(\sum_{i=1}^{N_{s}}U_{i}\Biggr).
\end{equation}
Since the function $W(x)$ is twice continuously differentiable, using the
expressions (\ref{XtsinL}) and the It\^{o}'s formula for semimartingales
[see Protter (\citeyear{P92})], it can be shown with arguments similar
to the
proof of
Proposition \ref{MenorSuper} that%
%
%
\begin{equation} \label{igualdaddeIto}\quad
W(X_{\tau^{\ast}}^{\gamma})e^{-c\tau^{\ast}}-W(x)=\int_{0}%
^{\tau^{\ast}}\mathcal{L}_{\gamma_{s}}(W)(X_{s^{-}}^{\gamma})e^{-cs}%
\,ds+M_{\tau^{\ast}}^{(1)}+M_{\tau^{\ast}}^{(2)},
\end{equation}
where
%
%
\begin{eqnarray}\label{SPSmartingala(barrier)}%
M_{t}^{(1)} &=&
{ \sum_{X_{s^{-}}\neq X,s\leq t}}
\bigl( W(X_{s}^{\gamma})-W(X_{s^{-}}^{\gamma}) \bigr)
e^{-cs}\nonumber\\[-8pt]\\[-8pt]
&&{} -\beta\int_{0}^{t}e^{-cs}\int_{0}^{\infty} \bigl(
W(X_{s^{-}}^{\gamma}-\alpha)-W(X_{s^{-}}^{\gamma}) \bigr)
\,dF(\alpha)\,ds\nonumber
\end{eqnarray}
and
%
%
\begin{equation} \label{SPSotramartingala(barrier)}%
M_{t}^{(2)}=\int_{0}^{t}W^{\prime}(X_{s}^{\gamma})e^{-cs}\sigma
X_{s}^{\gamma}\gamma_{s}\,dW_{s}
\end{equation}
are martingales with zero-expectation.

Note that we have%
\[
E_{x}(W(X_{\tau^{\ast}}^{\gamma})e^{-c\tau^{\ast}})=E_{x}\bigl(W(X_{\tau
^{\ast}%
}^{\gamma})e^{-c\tau^{\ast}}\chi_{ \{ \tau^{\ast}=\tau_{y} \}
}\bigr)=E_{x}\bigl(W(y)e^{-c\tau^{\ast}}\chi_{ \{ \tau^{\ast}=\tau_{y} \}
}\bigr).
\]
From (\ref{SPSmartingala(barrier)}), (\ref{SPSotramartingala(barrier)}) and
(\ref{igualdaddeIto}), by Proposition \ref{WesC2}, we get that
\[
\sup_{\gamma\ \mathrm{admissible}}E_{x}(W(X_{\tau^{\ast}}^{\gamma
})e^{-c\tau^{\ast}})=E_{x}(W(X_{\tau^{\ast}}^{\overline{\gamma}}%
)e^{-c\tau^{\ast}})=W(x)
\]
and so $\sup_{\gamma\ \mathrm{admissible}}E_{x}(e^{-c\tau^{\ast}}%
\chi_{ \{ \tau^{\ast}=\tau_{y} \} })=W(x)/W(y)$. The supremum is
reached at the process $\overline{\gamma}=(\overline{\gamma}_{t})$. On the
other hand, from Proposition \ref{DPPw}, we obtain that
\[
\sup_{\gamma\ \mathrm{admissible}}E \bigl( e^{-c\tau^{\ast}}\chi_{ \{
\tau^{\ast}=\tau_{y} \} } \bigr) =W_{z,y}(x)/W_{z,y}(y),
\]
and the result follows from $W_{z,y}(y)=W_{z,y}(z)+(y-z)$.
\end{pf}

Note that the optimal investment policy of all the strategies defined above
does not depend on the value of $z$. The corresponding controlled risk process
with initial surplus $x\leq y$ never exceeds the threshold $y$. In the next
definition we define the limit dividend barrier strategies $\widetilde
{\pi
}_{x}^{y}$ for any $x\in[ 0,y ) $.
\begin{definition}
\label{ValueFunctionPiY}
Given a sequence $z_{n}\nearrow y$
and any current surplus $x\in[ 0,y ) $, take
$\pi_{x}^{ ( \overline{\gamma}_{t},z_{n},y ) }\in\overline{\Pi
}_{x}^{z,y}$. We define $\widetilde{\pi}_{x}^{y}=\lim
_{n\rightarrow
\infty}\pi_{x}^{ ( \overline{\gamma}_{t},z_{n},y ) } $.
\end{definition}

In the next proposition we obtain the expression for the limit value function;
the proof follows immediately from Proposition \ref{CaracterizaciondeWz,y}.
\begin{proposition}
\label{PropositionValueFunctionPiy}
We have that
\[
V_{\widetilde{\pi}_{x}^{y}}(x)=\lim_{n\rightarrow\infty}W_{z_{n}%
,y}(x)= \cases{
W(x)/W^{\prime}(y), &\quad if $0\leq x<y$,\cr
W(y)/W^{\prime}(y)+(x-y), &\quad if $x\geq y$.}
\]
\end{proposition}

Note that the function $V_{\widetilde{\pi}_{x}^{y}}$ is twice continuously
differentiable in $ ( 0,y ) \cup( y,\infty) $ and
differentiable at $y$. We show now that $W^{\prime}$ reaches
the minimum.
\begin{proposition}
\label{MinimoWsombrero}
Consider the function $W$
defined in
Proposition \ref{WesC2}, then
\[
w_{1}=\inf W^{\prime}=W^{\prime}(x)>0
\]
for some $x\geq0$. Call $x_{\ast}=\min\{x\geq
0\dvtx W^{\prime
}(x)=w_{1}\}$.
\end{proposition}
\begin{pf}
Define for $u\geq0$, the function $G(u)=\inf_{x\in[
0,u]}W^{\prime}(x)$. Since $W^{\prime}$ is a continuous positive function,
then $G$ is continuous, nonincreasing and positive. We want to prove that
there exists $u_{0}$ such that $G(u)$ is constant for $u\geq u_{0}$. Suppose
that this is not the case, then there exists $u_{2}>u_{1}>p/(c-r)$ such that
$G(u_{2})<G(u_{1})<G(p/(c-r))$. Consider
\[
x_{1}=\min\{x\dvtx W^{\prime}(x)=G(u_{1})\},\qquad x_{2}=\min\{x\dvtx W^{\prime
}(x)=G(u_{2}%
)\}.
\]
Note that $x_{2}>u_{1}\geq x_{1}>p/(c-r)$. Let us consider the value functions
of the limit barrier strategies,%
\[
\mathcal{U}_{x_{i}}(x)= \cases{
W(x)/W^{\prime}(x_{i}), &\quad if $x<x_{i}$,\cr
W(x_{i})/W^{\prime}(x_{i})+(x-x_{i}), &\quad if $x\geq x_{i}$,}
\]
for $i=1,2$.

We prove now that $\mathcal{U}_{x_{i}}$ is a supersolution of (\ref{InicDif})
in $x>0$. Since $W$ is a solution of (\ref{EcuDif(w)}), $W^{\prime
}(x_{i}) \leq W^{\prime}(x)${} for $x\in(0,x_{i}]$ and $\mathcal
{U}_{x_{i}}^{\prime
}=1$ in $(x_{i},\infty)$ we only need to show that $\mathcal
{U}_{x_{i}}$ is a
supersolution of (\ref{EcuDif(w)}) in $[x_{i},\infty)$. Let us show
first that
$\mathcal{U}_{x_{i}}$ is a supersolution at $x>x_{i}$, take any $\gamma
\in[0,1]$, since $\mathcal{U}_{x_{i}}$ is increasing and $\mathcal
{U}%
_{x_{i}}\geq x_{i}$, we have that $\mathcal{L}_{\gamma}(\mathcal
{U}_{x_{i}%
})<0$. Let us show now that $\mathcal{U}_{x_{i}}$ is a supersolution at
$x_{i}$. We have that $\mathcal{U}_{x_{i}}^{\prime}(x_{i})=1$, take $q$ such
that%
\begin{eqnarray*}
q/2&\leq&\lim\inf_{h\rightarrow0}\frac{(\mathcal{U}_{x_{i}}%
(x_{i}+h)-\mathcal{U}_{x_{i}}(x_{i}))/h-1}{h}\\
&\leq&\lim
_{h\rightarrow
0^{+}}\frac{(\mathcal{U}_{x_{i}}(x_{i}+h)-\mathcal
{U}_{x_{i}}(x_{i}))/h-1}%
{h}=0.
\end{eqnarray*}
Since $\mathcal{U}_{x_{i}}$ is a supersolution for $x>x_{i}$ and
$\sup_{\gamma\in[0,1]}\mathcal{L}_{\gamma}(\mathcal
{U}_{x_{i}%
},1,q)(x)$ is continuous for $x\geq x_{i}$ we have that $\sup
_{\gamma
\in[0,1]}\mathcal{L}_{\gamma}(\mathcal{U}_{x_{i}},1,q)(x)\leq0$.

Since $\mathcal{U}_{x_{i}}$ is the value function of a limit strategy
we have
that $\mathcal{U}_{x_{i}}\leq V$, and since $\mathcal{U}_{x_{i}}$ is a
supersolution of (\ref{InicDif}), we have that $\mathcal{U}_{x_{i}}\geq V$.
Then $\mathcal{U}_{x_{i}}=V$ for $i=1,2$, and this is a
contradiction since
$\mathcal{U}_{x_{1}}\neq\mathcal{U}_{x_{2}}$.
\end{pf}

In the next proposition we see that the value $x_{\ast}$ defined in
Proposition \ref{MinimoWsombrero} is the optimal threshold of the dividend
barrier strategies given in Definition \ref{ValueFunctionPiY}. We also
give a
test to see whether the value function of the limit barrier strategy
$\widetilde{\pi}_{x}^{_{x_{\ast}}}$ is the optimal value function $V$
at $x$.
The proof follows directly from Proposition \ref{MinimoWsombrero} and Theorem
\ref{VerificationTheorem}.
\begin{proposition}
\label{optimalBarrier}
Define $V_{1}(x)$ as the value
function of the limit barrier strategy obtained in Proposition
\ref{PropositionValueFunctionPiy} with barrier $x_{\ast}:=\arg\min
W^{\prime}
$. Then:

\textup{(a)} $V_{1}(x)=\max_{y\geq0}V_{\widetilde{\pi
}_{x}^{y}}(x) $
for all $x\geq0$ and the function $V_{1}$ is
twice continuously differentiable.

\textup{(b)} If $V_{1}$ is a viscosity supersolution of
(\ref{InicDif}), then $V_{1}$ coincides with the optimal value
function $V$.
\end{proposition}

In Remark \ref{optimalBarrierStrategy} of the next section, we will see that
the limit stationary barrier strategy $\widetilde{\pi}^{_{x_{\ast}}}$
defined as $\widetilde{\pi}_{x}^{_{x_{\ast}}}\in\Pi_{x}$ for any initial
surplus $x\geq0$ is the optimal barrier strategy. Note that the investment
policy corresponding to this strategy is stationary and it is given by
\[
\gamma^{\ast}(u)=\min\bigl\{1,2 \bigl( M(W)(u)-pW^{\prime}(u) \bigr)
/(ruW^{\prime}(u))\bigr\}
\]
for any current surplus $u\in[0,x_{\ast}]$. Also note that, by
Proposition \ref{WesC2(segunda)}(b), $\gamma^{\ast}=1$ for small surpluses.
This means that the whole surplus should be invested in stocks. In the
unconstrained case where it is allowed to borrow money to buy risky
assets, it
can be seen that optimal investment policy tends to infinite as the surplus
goes to zero, that is, for small surpluses the company should always
borrow money
to buy stocks.

\section{Band structure of the optimal dividend strategy}\label{sec8}

We will show in Section \ref{sec9} that the optimal value function $V$ is not always
the value function of a limit barrier strategy. Nevertheless, we prove
in this
section that the optimal dividend payment policy has a band structure.
As in
the case of the optimal barrier strategy, $V$ is not the value function
of a
stationary admissible strategy, but it can be written explicitly as a
limit of
value functions of admissible stationary strategies.

We have shown in Section \ref{sec3} that $V$ is a viscosity solution of equation
(\ref{InicDif}). In this section we see that $V$ can be
obtained by
gluing, in a smooth way, classical solutions of $\mathcal{L}^{\ast
}(V)=0$ on
an open set $\mathcal{C}_{0}$ with solutions of $V^{\prime}=1$ on a set
$\mathcal{B}_{0}$. The set $\mathcal{B}_{0}$ is a disjoint union of left-open,
right-closed intervals. These sets will be defined in Proposition
\ref{Conjuntossub0}.

When the current surplus $x$ is in the set $\mathcal{B}_{0}$, the optimal
dividend payment policy should be to pay out immediately a positive sum of
dividends, and when the current surplus $x$ is in the set $\mathcal{C}_{0}$,
the optimal strategy should be to pay no dividends and to follow the
investment policy $\gamma(x)=\break\arg\max_{\gamma\in[
0,1]}\mathcal{L}_{\gamma}(V)(x)$ which depends only on the current surplus
$x$. In the simplest case, when the optimal value function $V$ is the solution
of $\mathcal{L}^{\ast}(V)=0$ in $\mathcal{C}_{0}=(0,y^{\ast})$ and
$V^{\prime
}=1$ in $\mathcal{B}_{0}=(y^{\ast},\infty)$, the optimal dividend payment
policy is barrier.

We see that $V$ is continuously differentiable; it is twice continuously
differentiable in $\mathcal{B}_{0}$ and $\mathcal{C}_{0}$, but at some points
outside $\mathcal{B}_{0}\cup\mathcal{C}_{0}$, the second derivative
could not
exist. So we still need the notion of viscosity solutions to
characterize $V$
as a solution of the associated HJB equation.

We also prove in this section that, for small surpluses, the optimal strategy
coincides with the optimal barrier obtained in Section \ref{sec7}, and for large
surpluses, the optimal strategy is to pay out as dividends the surplus
exceeding some level.

In the next proposition, we give conditions under which the optimal value
function $V$ is the supremum of the value functions corresponding to
admissible strategies with surplus not exceeding $\widehat{x}$.
\begin{proposition}
\label{LocalVerificationTheorem}
Assume there exists $\widehat{x}>0$
with $V^{\prime}(\widehat{x})=1$; then
\[
V(x)=\sup_{\pi\in\Pi_{x}^{\widehat{x}}}V_{\pi}(x)
\qquad\mbox{for all } x\leq\widehat{x}.
\]
\end{proposition}
\begin{pf}
Given any $\varepsilon>0$, let us consider the twice continuously
differentiable solution $g$ of the equation $\mathcal{L}^{\ast}(g)=0$
for the special case $\beta=0$. From Proposition~\ref{MinimoWsombrero},
we get that $\inf_{x\geq0}g^{\prime}(x)=g^{\prime}(x_{\ast})>0$ for
some $x_{\ast}\geq0$. So $\lim_{x\rightarrow\infty}g(x)=\infty$ and we
can find a number $D$ such that
$g(D)\geq2g(\widehat{x})V(\widehat{x})/\varepsilon$. Consider
$x_{n}=\widehat{x}-D/n$, and define $h_{n}=(V(x_{n})-V(\widehat
{x}))/(x_{n}-\widehat{x})-1$. Since $V^{\prime}(\widehat{x})=1$, we
have that $h_{n}$ goes to $0$ as $n$ goes to infinity, and so we can
find an integer $n_{0}$ large enough such that
$h_{n_{0}}<\varepsilon/(8D)$.

We can find points $0=y_{0}<y_{1}<\cdots<y_{M}=\widehat{x}$ such that
$V(y_{j+1})-V(y_{j})\leq\varepsilon/(16n_{0})$ and admissible
strategies
$\pi_{y_{j}}\in\Pi_{y_{j}}$ such that $V(y_{j})-V_{\pi_{y_{j}}}(y_{j}%
)\leq\varepsilon/(16n_{0})$. Consider, for any $x\in[0,\widehat{x}]$,
the point $y(x)=\max\{y_{j}\dvtx y_{j}\leq x\}$ and the strategy $\pi
_{x}\in
\Pi
_{x}$ which pays out immediately $x-y(x)$ as dividends and then follows the
strategy $\pi_{y(x)}\in\Pi_{y(x)}$. We obtain that $V(x)-V_{\pi_{x}}%
(x)\leq\varepsilon/(8n_{0})$ for any $x\in[0,\widehat{x}]$.

For any $x\in[0,\widehat{x}]$, we define recursively strategies $\pi
_{x}^{k}\in\Pi_{x}$ as follows. For $k=0$, take $\pi_{0}=\pi_{x}$. For
$k>0$ and for the initial surplus $x\leq x_{n_{0}}$, follow the
strategy $\pi_{x}$ while $X_{t}^{\pi}<\widehat{x}$, when the surplus
$X_{t}^{\pi}$ reaches $\widehat{x}$, pay out immediately the difference
$\widehat {x}-x_{n_{0}}$ as dividend and then follow the strategy
$\pi_{x_{n_{0}}}^{k-1} \in\Pi _{x_{n_{0}}}$. For $k>0$ and for the
initial surplus $x\in
(x_{n_{0}},\widehat{x}%
]$, pay out immediately the difference $x-x_{n_{0}}$ as dividend and then
follow the strategy $\pi_{x_{n_{0}}}^{k-1} \in\Pi_{x_{n_{0}}}$.

With arguments similar  to Lemma A.5 in Azcue and Muler (\citeyear
{AM05}) it can be seen that, for any $x\in[0,\widehat{x}]$ and $k\geq0$
the strategy $\pi _{x}^{k}\in\Pi_{x}$ is admissible and
%
%
\begin{equation} \label{PrimerDesigualdad}%
V(x)-V_{\pi_{x}^{n_{0}}}(x)<\varepsilon/2\qquad\mbox{for all }x\in[
0,\widehat{x}].
\end{equation}

Let us prove now that, for any $x\in[0,\widehat{x}]$, there
exists an
admissible strategy $\widetilde{\pi}\in\Pi_{x}^{\widehat{x}}$ such that
%
%
\begin{equation} \label{desig3}%
V_{\pi_{x}^{n_{0}}}(x)-V_{\widetilde{\pi}}(x)<\varepsilon/2\qquad\mbox{for all
}x\in[0,\widehat{x}].
\end{equation}
Let us define $\widehat{\tau}=\inf\{ t>0\dvtx X_{t}^{\pi_{x}^{n_{0}%
}}>\widehat{x} \} $. Consider the process $Y_{t}^{\pi_{x}^{n_{0}}}$
defined in Lemma \ref{LemmaPYt}, as the process corresponding to $X_{t}%
^{\pi_{x}^{n_{0}}}$ without claims and without paying dividends, but starting
at $Y_{0}^{\pi_{x}^{n_{0}}}=x$. Since the process $X_{t}^{\pi
_{x}^{n_{0}}}%
$ should pass at least $n_{0}$ times through the interval $[x_{n_{0}%
},\widehat{x}]$ before surpassing $\widehat{x}$, we obtain that
%
%
\begin{equation}\label{DistintosTaus}%
\widehat{\tau}\geq\tau_{n_{0}}^{Y}:=\inf\{ t>0\dvtx Y_{t}^{\pi
_{x}^{n_{0}}}>x_{n_{0}}+n_{0}(\widehat{x}-x_{n_{0}}) \} .
\end{equation}
To prove this, consider $X_{t}^{\widetilde{\pi}_{x}^{n_{0}}}$ the
corresponding process without the dividends payment $\widehat
{x}-x_{n_{0}}$ in
each step, then%
\begin{eqnarray*}
&&
\inf\biggl\{ t>0\dvtx X_{t}^{\pi_{x}^{n_{0}}}>\widehat{x}=x_{n_{0}%
}\biggl(1+\frac{\widehat{x}-x_{n_{0}}}{x_{n_{0}}}\biggr) \biggr\} \\
&&\qquad
=\inf\biggl\{ t>0\dvtx
X_{t}^{\widetilde{\pi}_{x}^{n_{0}}}>x_{n_{0}} \biggl( \frac{\widehat{x}%
}{x_{n_{0}}} \biggr) ^{n_{0}} \biggr\},
\end{eqnarray*}
and since $x_{n_{0}} ( \widehat{x}/x_{n_{0}} ) ^{n_{0}}\geq
x_{n_{0}}+n_{0}(\widehat{x}-x_{n_{0}})$ and $Y_{t}^{\pi_{x}^{n_{0}}}\geq
X_{t}^{\widetilde{\pi}_{x}^{n_{0}}}$, we obtain that $\widehat{\tau}\geq
\tau_{n_{0}}^{Y}$.

Since $\mathcal{L}^{\ast}(g)=0$ and $Y_{\tau_{n_{0}}^{Y}}^{\pi}=x_{n_{0}
}+n_{0}(\widehat{x}-x_{n_{0}})$ we have, using It\^{o}'s formula, that
\[
g\bigl(x_{n_{0}}+n_{0}(\widehat{x}-x_{n_{0}})\bigr)E(e^{-c\tau_{n_{0}}^{Y}})\leq g(x).
\]
So, we have from the fact that $g$ is increasing and (\ref{DistintosTaus})
that
%
%
\begin{eqnarray} \label{AcotacionEsperanza}%
E(e^{-c\widehat{\tau}}) &\leq& E(e^{-c\tau_{n_{0}}^{Y}})\leq\frac{g(x)}%
{g(x_{n_{0}}+n_{0}(\widehat{x}-x_{n_{0}}))}
\nonumber\\[-8pt]\\[-8pt]
&\leq&\frac{g(\widehat
{x})}{g(D)}%
\leq\frac{\varepsilon}{2V(\widehat{x})}.\nonumber
\end{eqnarray}

Again, with arguments similar  to Lemma A.5 in Azcue and Muler (\citeyear{AM05}),
we obtain%
\[
V_{\pi_{n_{0}}}(x)-V_{\widetilde{\pi}}(x)\leq E(e^{-c\widehat{\tau}%
})\bigl(V(\widehat{x})-\widehat{x}\bigr).
\]
So using (\ref{AcotacionEsperanza}), we conclude (\ref{desig3}). From
(\ref{PrimerDesigualdad}) and (\ref{desig3}) we get the
result.
\end{pf}

We have to introduce some auxiliary sets to define precisely the sets
$\mathcal{B}_{0}$ and $\mathcal{C}_{0}$ mentioned above.
\begin{definition}
\label{Conjuntos}
Let us define the continuous function
%
%
\begin{equation} \label{Lambda}%
\Lambda(x)=(p+rx)-M(V)(x),
\end{equation}
where the operator $M$ is defined in (\ref{DefinicionM(w)}),
and the sets:

\begin{itemize}
\item$\mathcal{A}= \{ x\in[0,\infty)\mbox{ such that
}V^{\prime}(x^{+})=1\mbox{ and }\Lambda(x)=0 \}$,

\item$\mathcal{B}= \{ x\in(0,\infty)\mbox{ such that }
V^{\prime}(x)=1\mbox{ and }\Lambda(x)<0 \}$,

\item$\mathcal{C}=[0,\infty)- ( \mathcal{A}\cup\mathcal{B} )
$.
\end{itemize}
\end{definition}
\begin{lemma}
\label{LambdaNegativo}
The following situations are not
possible:

\begin{enumerate}
\item$V^{\prime}(x^{+})=1$ and $\Lambda(x)>0$.

\item$1=V^{\prime}(x^{+})<V^{\prime}(x^{-})$ and $\Lambda(x)=0$.
\end{enumerate}

So, we conclude that
\begin{eqnarray*}
\mathcal{A} & = & \{ x\in[0,\infty)\mbox{ such that
}V^{\prime}(x)=1\mbox{ and }\Lambda(x)=0 \},\\
\mathcal{B} & = & \{ x\in(0,\infty)\mbox{ such that }V^{\prime
}(x)=1\mbox{ and }\Lambda(x)<0 \},\\
\mathcal{C} & = & \{x\in(0,\infty)\mbox{ such that }V^{\prime}%
(x^{+})>1\}\\
& &{} \cup\{x\in(0,\infty)\mbox{ such that }V^{\prime}(x^{-}%
)>V^{\prime}(x^{+})=1 \mbox{ and }\Lambda(x)<0\}.
\end{eqnarray*}
\end{lemma}
\begin{pf}
Let us prove first that given $x\geq0$, if $V^{\prime}(x^{+})=1$ then
$\Lambda(x)\leq0$. Assume that $\Lambda(x)>0$, then we can find $\delta>0$
such that $\Lambda(y)>0$ for all $y\in[ x,x+\delta)$. Let us
define $D$
as the set of points in $(x,x+\delta)$ where $V^{\prime}$ and $V^{\prime
\prime}$ exist, since $V$ is semiconcave the set $D$ has full measure. The
function $V$ is a supersolution of (\ref{InicDif}), then for any $y\in
D$ we
have%
\[
0\geq\mathcal{L}^{\ast}(V)(y)\geq\sigma^{2}y^{2}V^{\prime\prime}%
(y)/2+\Lambda(y)
\]
and so $V^{\prime\prime}(y)\leq-2 \Lambda(y)/(\sigma^{2}y^{2})<0$.
Then, since
$V$ is semiconcave, we have that for any $y\in D$%
\[
V^{\prime}(y)-1=V^{\prime}(y)-V^{\prime}(x^{+})\leq\int_{x}^{y}V^{\prime
\prime}(s)\,ds<0
\]
and this is a contradiction because $V^{\prime}(y)\geq1$.

Let us prove now that if $x\in\mathcal{A}$ and $x>0$, then $V$ is
differentiable at $x$ and $V^{\prime}(x)=1$. If we have that
$1=V^{\prime
}(x^{+})<V^{\prime}(x^{-})$, take any $d\in(V^{\prime
}(x^{+}),V^{\prime
}(x^{-}))$, then%
\[
\limsup_{h\rightarrow0}\frac{(V(x+h)-V(x))/h-d}{h}=-\infty
\]
and so, for any $q$, we have that%
\[
\max\Bigl\{1-d,\max_{\gamma\in[0,1]} \bigl( \sigma^{2}x^{2}\gamma
^{2}q/2+(p+rx\gamma)d-M(V)(x) \bigr) \Bigr\}\geq0
\]
and then, since $d>1$, so%
\[
\max_{\gamma\in[0,1]} \bigl( \sigma^{2}x^{2}\gamma^{2}q/2+(p+rx\gamma
)d-M(V)(x) \bigr) \geq0.
\]
Since this holds for any $q$, taking a sequence $q_{n}\rightarrow-\infty
$, we
obtain that $pd-M(V)(x)\geq0$ for any $d\in(1,V^{\prime}(x^{-}))$. This
implies that $p-M(V)(x)\geq0$ and so $\Lambda(x)>0$, which is a contradiction.
\end{pf}
\begin{definition}
\label{Conjuntossub0}
We define the sets $\mathcal{A}_{0}%
$, $\mathcal{B}_{0}$ and $\mathcal{C}_{0}$ as:

\begin{itemize}
\item$\mathcal{B}_{0}=\mathcal{B}\cup\{a\in\mathcal{A}\dvtx(a-\vartheta
,a)\subset\mathcal{A}\cup\mathcal{B}$ for some
$\vartheta>0\}$,

\item$\mathcal{C}_{0}=\mathcal{C}\cup\{a\in\mathcal{A}\dvtx(a-\vartheta
,a)\cup(a,a+\vartheta)\subset\mathcal{C}$ for some $\vartheta
>0\}$,

\item$\mathcal{A}_{0}=[0,\infty)- ( \mathcal{C}_{0}\cup\mathcal{B}%
_{0} ) $.
\end{itemize}
\end{definition}
\begin{proposition}
\label{PropiedadesConjuntossub0}
The sets introduced in Definition
\ref{Conjuntossub0} satisfy the following properties:

\textup{(a)} $\mathcal{B}_{0}$ is a disjoint union of
intervals that
are left-open and right-closed.

\textup{(b)} If $(x_{0},\widehat{x}]\subset\mathcal{B}_{0}$ and
$x_{0}\notin\mathcal{B}_{0}$, then $x_{0}\in\mathcal{A}_{0}%
$.

\textup{(c)} There exists $x^{\ast}\geq0$ such that $(x^{\ast
},\infty)\subset\mathcal{B}_{0}$.

\textup{(d)} $\mathcal{C}_{0}$ is an open set in $[0,\infty
)$, that is, if $0\in\mathcal{C}_{0}$, there exists
$\delta>0$ such that $[0,\delta)\subset\mathcal{C}_{0}$
and
if a positive $x\in\mathcal{C}_{0}$ there exists $\delta
>0$ such that $(x-\delta,x+\delta)\subset\mathcal
{C}_{0}$.

\textup{(e)} Both $\mathcal{A}_{0}$ and $\mathcal{B}_{0}%
$ are nonempty.
\end{proposition}
\begin{pf}
The proof follows immediately from Definition \ref{Conjuntossub0} and Lemmas
\ref{Continuacioncomosupersolution} and \ref{PropiedadesConjuntos} included
in the \hyperref[app]{Appendix}.
\end{pf}

From the previous proposition we can conclude that the upper boundary
of any
connected component of $\mathcal{C}_{0}$ belongs to $\mathcal{A}_{0}$
and also
that the the lower boundary of any connected component of $\mathcal{B}_{0}$
belongs to $\mathcal{A}_{0}$.

The next proposition describes the optimal value function $V$ for small
initial surpluses.
\begin{proposition}
\label{W1yY0}
Consider the function $W$ defined in
Proposition \ref{WesC2} and the values $w_{1}$ and $x_{\ast}$
defined in Proposition \ref{MinimoWsombrero}, then the optimal value
function $V(x)$ coincides with $W(x)/w_{1}$ for all
$x\in[0,x_{\ast}]$. In particular, $V$ is twice
continuously differentiable in $[0,x_{\ast}]$.
\end{proposition}
\begin{pf}
By Lemma \ref{PropiedadesConjuntos}(b) included in the \hyperref
[app]{Appendix},
$\mathcal{A}$ is left closed, so there exists $m=\min\mathcal{A}$. Note that,
by Proposition \ref{MinimoWsombrero}, $w_{1}$ and $x_{\ast}$ are well defined.
Consider $V_{1}$ the value function of the limit strategy $\widetilde
{\pi}%
_{x}^{x_{\ast}}$ obtained in Proposition \ref{PropositionValueFunctionPiy}.
From (\ref{V}), we have that $V_{1}(x)\leq V(x)$.

If $m>x_{\ast}$, we have from Proposition \ref{MenorSuper} that
$V(x)\leq
W(x)/W^{\prime}(x_{\ast})$ in $[0,\infty)$ because $W(x)/W^{\prime
}(x_{\ast})$
is a supersolution of (\ref{InicDif}). So $V(x)=\break W(x)/W^{\prime}(x_{\ast
})=V_{1}(x)$ in $[0,x_{\ast}]$. Then $V^{\prime}(x_{\ast})=1$ and this implies
that $x_{\ast}\in\mathcal{A}\cup\mathcal{B}$; this is a contradiction
since in
both cases there would exist a point in $\mathcal{A}$ smaller that $m$. In
particular, if $x_{\ast}=0$, then $m=0$.

If $0<m<x_{\ast}$, since $V_{1}^{\prime}\geq1$ and $\mathcal{L}^{\ast}%
(V_{1})=0$ in $(0,m)$, we have that $V_{1}$ is a supersolution of
(\ref{InicDif}) in $(0,m)$ and since $V^{\prime}(m)=1$, by Proposition
\ref{LocalVerificationTheorem}, $W(x)/W^{\prime}(x_{\ast})=V(x)$ in $[0,m]$,
but then $1=V^{\prime}(m)=W^{\prime}(m)/W^{\prime}(x_{\ast})$ and this
is a
again a contradiction because by definition of $x_{\ast}$, $W^{\prime
}(m)/W^{\prime}(x_{\ast})>1$.

Finally, in the case that $m=0$, since $0\in\mathcal{A}$ we have from
(\ref{Lambda}) that $V(0)=(c+\beta)/p$, but from Proposition
\ref{WesC2(segunda)}(c) we have that $W^{\prime}(0)=(c+\beta)/p$, and
so we
get $V(0)=W(0)/W^{\prime}(0)$. This implies that $x_{\ast}=0$ because if
$x_{\ast}$ were positive, we would obtain%
\[
V(0)=W(0)/W^{\prime}(0)<W(0)/W^{\prime}(x_{\ast})\leq V(0).
\]

Therefore, $m=x_{\ast}$ and $V=V_{1}$ in $[0,x_{\ast}]$.
\end{pf}

The previous proposition allows us to obtain $V$ for small surpluses using
only the function $W$. In the case that $x_{\ast}=0$, we only obtain
from this
proposition the value at zero, $V(0)=(c+\beta)/p$. Hence, using Corollary
\ref{Unicidad}, we can conclude that $V$ is the unique viscosity
solution of
(\ref{InicDif}) with the boundary condition $V(0)=W(0)/w_{1}$.
\begin{remark}
\label{optimalBarrierStrategy} The limit stationary strategy $\widetilde
{\pi}^{_{x_{\ast}}}$ defined in Section \ref{sec7} is the optimal barrier strategy.
In effect, the optimal barrier strategy is the one with maximum value function
al $0$ and, by Proposition \ref{W1yY0}, the value function of this limit
stationary strategy is $V_{1}(0)=W(0)/w_{1}=V(0)$.
\end{remark}

Let us show now that $V$ is a classical solution of $\mathcal{L}^{\ast
}(V)=0 $
in $\mathcal{C}_{0}$.
\begin{proposition}
\label{VywenC0}
\textup{(a)} Let $(x_{1},x_{2})$ with $x_{1}>0$ be a
connected component of~ $C_{0}$. Consider $U$ the unique
classical solution of
%
%
\begin{equation} \label{SupdeLgamadeVyw}%
\mathcal{L}^{\ast}(U,V)(x)=0
\end{equation}
in $(x_{1},\infty)$ with $U(x_{1})=V(x_{1})$
and $U^{\prime}(x_{1})=V^{\prime}(x_{1})=1$. Then $V=U$ in
$[x_{1},x_{2}]$.

\textup{(b)} The optimal value function $V$ is a classical solution
of $\mathcal{L}^{\ast}(V)=0$ in the open set $\mathcal{C}_{0}$.
\end{proposition}
\begin{pf}
Using Lemma \ref{VywenC} included in the \hyperref[app]{Appendix}, it
only remains to
prove that $V$ is twice continuously differentiable at the points
$a\in\mathcal{A}$ such that, there exists $\delta>0$ with $(a-\delta
,a)\cup(a,a+\delta)\subset\mathcal{C}$. The number
\[
\gamma^{\ast}(a)=\min\bigl\{1,2 \bigl( M(V)(a)-pV^{\prime}(a) \bigr)
/(raV^{\prime}(a))\bigr\}
\]
is positive because $M(V)(a)-pV^{\prime}(a)>\Lambda(a)=0$ and $V^{\prime
}(a)=1$. Take any sequence $u_{n}\rightarrow a$ with $u_{n}\in\mathcal
{C};$ we
have from Propositions \ref{WesC2} and \ref{NuevaProposicion1} that
\[
V^{\prime\prime}(u_{n})=2 \bigl( (p+ru_{n}\gamma_{n})V^{\prime}(u_{n}%
)-M(V)(u_{n}) \bigr) /(\sigma^{2}u_{n}^{2}\gamma_{n}^{2}),
\]
where%
\[
\gamma_{n}=\min\bigl\{1,2 \bigl( M(V)(u_{n})-pV^{\prime}(u_{n}) \bigr)
/(ru_{n}V^{\prime}(u_{n}))\bigr\}.%
\]
Since $V$ is semiconcave we get that%
\[
\lim_{n\rightarrow\infty}V^{\prime\prime}(u_{n})=2 \bigl(
\bigl(p+ra\gamma^{\ast}(a)\bigr)V^{\prime}(a)-M(V)(a) \bigr) /(\sigma^{2}a^{2}%
\gamma^{\ast}(a)^{2}),
\]
so $V$ is twice continuously differentiable at $a$.
\end{pf}
\begin{remark}
\label{Viscontinuouslydifferentiable.}The optimal value function $V$ is
continuously differentiable at $(0,\infty)$ because it is continuously
differentiable both in $\mathcal{C}_{0}$ and in the interior of
$\mathcal{B}%
_{0}$. At any other point $x$ we have that $V^{\prime}$ is
continuously differentiable since
\[
\lim_{y\in\mathcal{B}_{0},y\rightarrow x}V^{\prime}(y)=\lim
_{y\in\mathcal{C}_{0},y\rightarrow x}V^{\prime}(y)=1=V^{\prime}(x).
\]
\end{remark}

We prove now that $V$ can be written as a limit of value functions of
admissible stationary strategies. All of these admissible strategies
coincide on
$\mathcal{B}_{0}$ and $\mathcal{C}_{0}$. If the current surplus is in
$\mathcal{B}_{0}$, the optimal strategy is to pay out as dividends the amount
exceeding the lower boundary of the connected component of $\mathcal{B}_{0}$.
If the current surplus is $x\in\mathcal{C}_{0}$, the optimal
strategy is
to pay no dividends and to invest $\gamma(x)=\arg\max_{\gamma
\in[0,1]}\mathcal{L}_{\gamma}(V)(x)$. Finally, if the current
surplus is in $\mathcal{A}_{0}$, we need to consider a limit of admissible
strategies similar to the one we used to obtain barrier strategies in
Section \ref{sec7}.

We define admissible stationary strategies $\pi$ based upon the sets
$\mathcal{A}_{0}$, $\mathcal{B}_{0}$ and $\mathcal{C}_{0}$ introduced in
Definition \ref{Conjuntossub0}. Since these strategies are
stationary, for
any $x\geq0$ we can denote $\pi(x)\in\Pi_{x}$ the corresponding
strategy with
initial surplus $x$.
\begin{definition}
\label{DefiniciondePidependiendodeAprima}
Given a finite subset
$\mathcal{A}^{\prime}\subset\mathcal{A}_{0}$ and a number
$u>0$ satisfying the following conditions:

\begin{enumerate}
\item if $\min\mathcal{A}_{0}=0$ then $0\in\mathcal{A}
^{\prime}$,

\item$c_{a}=a/e^{u}\in\mathcal{C}_{0}$ for all positive
$a\in\mathcal{A}^{\prime}$,
\end{enumerate}

we define recursively the admissible stationary strategy $\pi
$ in the following way:

\begin{itemize}
\item If the current surplus $x\in\mathcal{C}_{0}$,
pay no dividends and take
\[
\gamma^{\ast}(x)=\min\bigl\{1,2 \bigl( M(V)(x)-pV^{\prime}(x) \bigr)
/(rxV^{\prime}(x))\bigr\}
\]
up to the exit time $\tau$ of $C_{0}$.
Then follow the strategy $\pi(x_{1})\in\Pi_{x_{1}}$ where
$x_{1}=X_{\tau}^{\pi(x)}\in\mathcal{A}_{0}\cup\mathcal{B}_{0}$.

\item If the current surplus $x\in\mathcal{B}_{0}$,
by Proposition \ref{PropiedadesConjuntossub0}(a) and (b), there exists
$a\in\mathcal{A}_{0}$ such that $(a,x]\subset\mathcal{B}_{0}$.
In this case pay out immediately $x-a$ as dividends, and
follow the strategy $\pi(a)\in\Pi_{a}$ described
below.

\item If the current surplus $x\in\mathcal{A}_{0}\setminus
\mathcal{A}^{\prime}$, pay out immediately $x-a$ as
dividends where $a$ is the maximum element of $\mathcal
{A}^{\prime}$ smaller than $x$, and then follow the strategy
$\pi(a)\in\Pi_{a}$.

\item If the current surplus is $a\in\mathcal{A}^{\prime}$,
pay out immediately $a-c_{a}$ as dividends and then follow
the strategy $\pi(c_{a})\in\Pi_{c_{a}}$.

\item In the case that the current surplus is $0\in\mathcal
{A}%
^{\prime}$, pay out all the incoming premium as dividends up
to the
ruin time.
\end{itemize}
\end{definition}

In the case that $\mathcal{A}_{0}$ is finite, $V$ can be written as the limit
(with $u$ going to zero) of the value functions of the admissible strategies
defined above taking $\mathcal{A}^{\prime}=\mathcal{A}_{0}$; but in the case
that $\mathcal{A}_{0}$ is infinite, we have to consider finite subsets
$\mathcal{A}^{\prime}\subset\mathcal{A}_{0}$. This result is proved in the
next theorem.
\begin{theorem}
\label{Vcomolimitedeestrategias}
Given $\varepsilon
>0$, we can find a finite set $\mathcal{A}^{\prime}\subset
\mathcal{A}_{0}$ and a number $u>0$ such that the
admissible stationary strategy introduced in Definition
\ref{DefiniciondePidependiendodeAprima} satisfies $V(x)-V_{\pi
(x)}(x)<\varepsilon$ for all $x\geq0$. In
the case
that $\mathcal{A}_{0}$ is finite, we can take $\mathcal
{A}^{\prime
}=\mathcal{A}_{0} $.
\end{theorem}
\begin{pf}
We assume that $\min\mathcal{A}_{0}>0$, in the case $\min\mathcal{A}_{0}=0$
the proof is similar. Let us consider $\widehat{x}=\max\mathcal{A}_{0}$ and
the twice continuously differentiable solution $g$ of the equation
$\mathcal{L}^{\ast}(g)=0$ for the special case $\beta=0$. From Proposition
\ref{MinimoWsombrero}, we get that $\inf_{x\geq0}g^{\prime
}(x)=g^{\prime}(x_{\ast})>0$ for some $x_{\ast}\geq0$. Since $\lim
_{x\rightarrow\infty}g(x)=\infty$, we can find a number $M$ such that
$g(1)/g(e^{M})\leq\varepsilon/(4V(\widehat{x}))$.

We can find $\delta>0$ such that, if $h\leq\delta$ then
%
%
\begin{equation} \label{DefinicionDelta}%
0\leq\bigl(V(a+h)-V(a)\bigr)/h-1\leq\varepsilon/(4\widehat{x}).
\end{equation}
In effect, $V^{\prime}$ is absolutely continuous in $[0,\widehat{x}]$,
$V^{\prime}(a)=1$ for all $a\in\mathcal{A}_{0}\cup\mathcal{B}_{0}$, and from
Proposition \ref{PropiedadesConjuntossub0}(b) and (c) we have that $ [
\widehat{x},\infty) \subseteq\mathcal{A}_{0}\cup\mathcal{B}_{0}$.

Given $\delta$, take the finite set $\mathcal{A}_{\delta}$ and the number
$\varsigma>0$ given by Lemma \ref{ConstrucciondeAdelta} included in the
\hyperref[app]{Appendix}, and take $u>0$ such that%
%
%
\begin{equation} \label{CeroCondicionu}%
u\leq\delta/(2\widehat{x}),\qquad a-\varsigma<a/e^{u}
\end{equation}
and
%
%
\begin{equation}\label{TercierCondicionu}%
0\leq\frac{V(a)-V(a/e^{u})}{a-a/e^{u}}-1\leq\varepsilon/\bigl(8(M+2)\widehat{x}\bigr)
\end{equation}
for all $a\in\mathcal{A}_{\delta}$. Take $N=\#\mathcal{A}_{\delta}$,
%
%
\begin{equation} \label{Definicionk}%
k_{0}= [ M/u+N ] +1
\end{equation}
and admissible strategies $\overline{\pi}(a)\in\Pi_{a}$ with $a\in
\mathcal{A}_{\delta}$ such that%
%
%
\begin{equation} \label{DefinicionPibarra}%
V(a)-V_{\overline{\pi}(a)}(a)\leq\varepsilon/\bigl(4(2k_{0}+3)\bigr)\qquad\mbox{for
all }%
a\in\mathcal{A}_{\delta}.
\end{equation}
Let us define $c_{a}=a/e^{u}$ for all $a\in\mathcal{A}_{\delta}$, then, by
(\ref{CeroCondicionu}), $c_{a}\in\mathcal{C}_{0}$. Take the admissible
stationary strategy $\pi$ associated with $u>0$ and the finite set
$\mathcal{A}_{\delta}$ given by Definition~\ref{DefiniciondePidependiendodeAprima}.

We define recursively a family of admissible strategies $\overline{\pi}%
_{k}(x)\in\Pi_{x}$ for all $x\geq0$ and $k\geq0$, in the following way:

\begin{itemize}
\item Take $\overline{\pi}_{0}(a)$ as the admissible strategy $\overline
{\pi
}(a)$ defined in (\ref{DefinicionPibarra}) for all $a\in\mathcal
{A}_{\delta}$.

\item If the surplus $x\in\mathcal{C}_{0}$, pay no dividends and take
\[
\gamma^{\ast}(x)=\min\bigl\{1,2 \bigl( M(V)(x)-pV^{\prime}(x) \bigr)
/(rxV^{\prime}(x))\bigr\}
\]
up to the exit time $\tau_{0}$ of $\mathcal{C}_{0}$. Then follow the strategy
$\overline{\pi}_{k}(x_{1})\in\Pi_{x_{1}}$ starting at $x_{1}$ where
$x_{1}=X_{\tau_{0}}^{\overline{\pi}_{k}(x)}\in\mathcal{A}_{0}\cup
\mathcal{B}_{0}$.

\item If the surplus $x\in\mathcal{B}_{0}$, by Proposition
\ref{PropiedadesConjuntossub0}(a) and (b), there exists $a\in
\mathcal{A}_{0}$ such that $(a,x]\subset\mathcal{B}_{0}$. In this case, pay
out immediately $x-a$ as dividends and follow the strategy $\overline
{\pi}%
_{k}(a)\in\Pi_{a}$ described below.

\item If the surplus $x\in\mathcal{A}_{0}\setminus\mathcal
{A}_{\delta} $,
pay out immediately $x-a$ as dividends where $a$ is the maximum element of
$\mathcal{A}_{\delta}$ smaller than $x$, and then follow the strategy
$\overline{\pi}_{k}(a)\in\Pi_{a}$.

\item If the surplus is $a\in\mathcal{A}_{\delta}$ with $a>0$,
pay out
immediately $a-c_{a}$ as dividends and then follow the strategy
$\overline
{\pi}_{k-1}(c_{a})\in\Pi_{c_{a}}$.
\end{itemize}

To simplify notation we write $V_{\overline{\pi}_{k}}(x)$ instead of
$V_{\overline{\pi}_{k}(x)}(x)$. Let us prove first that%
%
%
\begin{equation} \label{Parte1}%
\max_{x\geq0} \bigl( V(x)-V_{\overline{\pi}_{k}}(x) \bigr)
\leq3\varepsilon/4.
\end{equation}
Given any initial surplus $x\geq0$, note that all the processes $X_{t}%
^{\overline{\pi}_{k}}$ with $k\geq0$ coincide for $t\leq\tau\wedge
\widehat{\tau}$ where $\tau$ is the time of arriving to $\mathcal
{A}_{\delta}$
and $\widehat{\tau}$ the ruin time. So, using the dynamic programing
principle, we have that
%
%
\begin{eqnarray}\label{DiferenciasdeV-1}
&&\vert V_{\overline{\pi}_{k_{0}}}(x)-V_{\overline{\pi}_{0}}(x) \vert
\nonumber\\
&&\qquad = \bigl\vert E_{x}\bigl(e^{-c ( \tau\wedge\widehat{\tau} ) } \bigl(
V_{\overline{\pi}_{k_{0}}}(X_{\tau\wedge\widehat{\tau}}^{\overline{\pi}%
_{k_{0}}})-V_{\overline{\pi}_{0}}(X_{\tau\wedge\widehat{\tau
}}^{\overline{\pi
}_{k_{0}}}) \bigr) \bigr) \bigr\vert\nonumber\\[-8pt]\\[-8pt]
&&\qquad \leq E_{x}\bigl( \bigl\vert e^{-c ( \tau\wedge\widehat{\tau} )
} \bigl( V_{\overline{\pi}_{k_{0}}}(X_{\tau\wedge\widehat{\tau}}%
^{\overline{\pi}_{k_{0}}})-V_{\overline{\pi}_{0}}(X_{\tau\wedge\widehat
{\tau}%
}^{\overline{\pi}_{k_{0}}}) \bigr) \chi_{\{\tau<\widehat{\tau}\}} \bigr\vert
\bigr)\nonumber\\
&&\qquad \leq {\max_{a\in\mathcal{A}_{\delta}}} \vert V_{\overline{\pi
}_{k_{0}}}(a)-V_{\overline{\pi}_{0}}(a) \vert.\nonumber
\end{eqnarray}
Consider $a\in\mathcal{A}_{\delta}$, the processes $X_{t}^{\overline{\pi
}_{k}%
}$ starting at $a$ and $\widehat{\tau}$ the ruin time, we define as usual
$X_{t}^{\overline{\pi}_{k}}=X_{\widehat{\tau}}^{\overline{\pi}_{k}}$ for
$t\geq\widehat{\tau}$. Let $\tau_{k}$ be the $k$th time that $X_{t}%
^{\overline{\pi}_{k_{0}+1}}$ reaches $\mathcal{A}_{\delta}$ and let%
\[
K= \{ k\geq0 \mbox{ such that }\tau_{k+1}<\widehat{\tau}\mbox{ and
}X_{\tau_{k}}^{\overline{\pi}_{k_{0}}}=X_{\tau_{k+1}}^{\overline{\pi
}_{k_{0}}%
} \}.
\]
Since the processes $X_{t}^{\overline{\pi}_{k}}$ and $X_{t}^{\overline
{\pi
}_{k-1}}$ coincide until $\tau_{k-1}\wedge\widehat{\tau}$, we have using
(\ref{DefinicionPibarra}) and (\ref{TercierCondicionu}) that%
%
%
\begin{eqnarray}\label{DiferenciasdeV-2}%
&&\vert V_{\overline{\pi}_{k_{0}}}(a)-V_{\overline{\pi}_{0}}(a) \vert
\nonumber\\
&&\qquad \leq \sum_{k=0}^{k_{0}-1} \vert V_{\overline{\pi}_{k+1}%
}(a)-V_{\overline{\pi}_{k}}(a) \vert\\
&&\qquad =  E_{a} \Biggl( \sum_{k=0}^{k_{0}-1} \bigl( e^{-c\tau_{k}} \vert
V_{\overline{\pi}_{1}}(X_{\tau_{k}}^{\overline{\pi
}_{k_{0}}})-V_{\overline
{\pi}_{0}}(X_{\tau_{k}}^{\overline{\pi}_{k_{0}}}) \vert\bigr)
\chi_{ \{ \tau_{k}<\widehat{\tau} \} } \Biggr) .\nonumber
\end{eqnarray}
We denote $a_{0}=X_{\tau_{k}}^{\overline{\pi}_{k_{0}}}\in\mathcal
{A}_{\delta}%
$. We define $\widetilde{\tau}_{k}$ as the first time that $X_{t}%
^{\overline{\pi}_{k_{0}}}$ leaves $\mathcal{C}_{0}$ after~$\tau_{k}$,
and we
denote $a_{2}=X_{\widetilde{\tau}_{k}}^{\overline{\pi}_{k_{0}}}$. We
obtain,
using It\^{o}'s formula, Proposition~\ref{VywenC0}(b) and the definition
of $\overline{\pi}_{0}$,%
%
%
\begin{eqnarray}\label{DiferenciasdeV-2-a}\quad
&&
\vert V_{\overline{\pi}_{1}}(a_{0})-V_{\overline{\pi}_{0}}(a_{0})\vert
\nonumber\\
&&\qquad
= \vert V_{\overline{\pi}_{0}}(c_{a_{0}})+(a_{0}-c_{a_{0}%
})-V_{\overline{\pi}_{0}}(a_{0}) \vert\nonumber\\
&&\qquad = \bigl\vert E \bigl( \bigl( V_{\overline{\pi}_{0}}(a_{2}%
)-V(a_{2}) \bigr) e^{-c ( \widetilde{\tau}_{k}-\tau_{k} )
} \vert\mathcal{F}_{\tau_{k}} \bigr)\nonumber\\[-8pt]\\[-8pt]
&&\qquad\quad\hspace*{1.5pt}{} +a_{0}-c_{a_{0}}-V(a_{0}%
)+V(c_{a_{0}})+V(a_{0})-V_{\overline{\pi}_{0}}(a_{0}) \bigr\vert\nonumber\\
&&\qquad \leq E \bigl( \bigl( V(a_{2})-V_{\overline{\pi}_{0}}(a_{2}) \bigr)
e^{-c ( \widetilde{\tau}_{k}-\tau_{k} ) } \vert\mathcal{F}%
_{\tau_{k}} \bigr) + \bigl( V(a_{0})-V_{\overline{\pi}_{0}}(a_{0}) \bigr) \nonumber\\
&&\qquad\quad{} +(a_{0}-c_{a_{0}}) \biggl( \frac{V(a_{0})-V(c_{a_{0}})}{a_{0}-c_{a_{0}}%
}-1 \biggr) .\nonumber
\end{eqnarray}
From (\ref{DefinicionPibarra}), (\ref{CeroCondicionu}),
(\ref{TercierCondicionu}) and using that $e^{-u}\geq1-u$, we obtain that
%
%
\begin{eqnarray}\label{DiferenciasdeV-3}%
&& \bigl( V(a_{0})-V_{\overline{\pi}_{0}}(a_{0}) \bigr) +(a_{0}-c_{a_{0}%
}) \biggl( \frac{V(a_{0})-V(c_{a_{0}})}{a_{0}-c_{a_{0}}}-1 \biggr) \nonumber\\%
&&\qquad
\leq\frac{\varepsilon}{4(2k_{0}+3)}+\frac{\varepsilon}{8(M+2)\widehat
{x}%
}(a_{0}-c_{a_{0}})\\
&&\qquad\leq\frac{\varepsilon}{4(2k_{0}+3)}+\frac{\varepsilon}{8(M+2)}u.\nonumber
\end{eqnarray}
If $k\notin K$ and $a_{2}\geq0$ denote $a_{1}=X_{\tau_{k+1}}^{\overline
{\pi
}_{k_{0}}}\in\mathcal{A}_{\delta}$. We obtain that $a_{0}>a_{1}$, and by
(\ref{DefinicionPibarra}),%
%
%
\begin{eqnarray}\label{DiferenciasdeV-4}\hspace*{28pt}
&&
E \bigl( \bigl( V(a_{2})-V_{\overline{\pi}_{0}}(a_{2}) \bigr)
e^{-c ( \widetilde{\tau}_{k}-\tau_{k} ) } \vert\mathcal{F}%
_{\tau_{k}} \bigr) \nonumber\\%
&&\qquad
= E \bigl( \bigl( V(a_{2})-V_{\overline{\pi}_{0}}(a_{2}) \bigr)
e^{-c ( \widetilde{\tau}_{k}-\tau_{k} ) }\chi_{\{\widetilde{\tau
}_{k}<\widehat{\tau}\}} \vert\mathcal{F}_{\tau_{k}} \bigr) \nonumber\\
&&\qquad
\leq E\bigl( \bigl( V(a_{2})-V_{\overline{\pi}_{0}}(a_{2}) \bigr) \chi
_{\{\widetilde{\tau}_{k}<\widehat{\tau}\}}|\mathcal{F}_{\tau
_{k}}\bigr)\nonumber
\\
&&\qquad = E\bigl( \bigl( V(a_{0})-V_{\overline{\pi}_{0}}(a_{0}) \bigr) \chi_{ \{
k\in K \} }|\mathcal{F}_{\tau_{k}}\bigr)\nonumber\\[-8pt]\\[-8pt]
&&\qquad\quad{} +E\bigl( \bigl( V(a_{2})-V_{\overline{\pi}_{0}}(a_{2}) \bigr) \chi_{ \{
k\notin K \} }\chi_{\{\widetilde{\tau}_{k}<\widehat{\tau}\}}%
|\mathcal{F}_{\tau_{k}}\bigr)\nonumber\\
&&\qquad
\leq\frac{\varepsilon}{4(2k_{0}+3)}+E\bigl( \bigl( V(a_{2})-V(a_{1}%
)-(a_{2}-a_{1})\nonumber\\
&&\qquad\quad\hspace*{111.7pt}{}+V(a_{1})-V_{\overline{\pi}_{0}}(a_{1}) \bigr) \chi_{ \{
k\notin K \} }\chi_{\{\widetilde{\tau}_{k}<\widehat{\tau}\}}%
|\mathcal{F}_{\tau_{k}}\bigr)\nonumber\\
&&\qquad
\leq\frac{2\varepsilon}{4(2k_{0}+3)}+E\bigl( \bigl( V(a_{2})-V(a_{1}%
)-(a_{2}-a_{1}) \bigr) \chi_{ \{ k\notin K \} }\chi_{\{\widetilde
{\tau}_{k}<\widehat{\tau}\}}|\mathcal{F}_{\tau_{k}}\bigr).\nonumber
\end{eqnarray}
Note that $(a_{1},a_{2})\cap\mathcal{A}_{\delta}=\phi$, and so there is no
connected component $(r_{1},r_{2})$ of $\mathcal{C}_{0}$ included in
$[a_{1},a_{2}]$ with length greater than $\delta$. In effect, if such
component exists, then $r_{2}\in\mathcal{A}_{0}\setminus\mathcal{A}%
_{\delta}$, and this contradicts Lemma \ref{ConstrucciondeAdelta}(b)
included in the \hyperref[app]{Appendix}. Then we can find
$a_{1}=x_{1}\leq x_{2}\leq
\cdots\leq x_{n}=a_{2}$ such that $x_{i}\in\mathcal{A}_{0}\cup\mathcal{B}_{0}$
and $x_{i+1}-x_{i}<\delta$. So we get, by (\ref{DefinicionDelta}),
%
%
\begin{eqnarray}\label{DiferenciasdeV-7}\qquad
&&
E \bigl( \bigl( V(a_{2})-V(a_{1})-(a_{2}-a_{1}) \bigr) \chi_{ \{
k\notin K \} }|\mathcal{F}_{\tau_{k}} \bigr) \nonumber\\
&&\qquad
= E \Biggl( \Biggl( \sum_{i=1}^{n}V(x_{i+1})-V(x_{i})-(x_{i+1}%
-x_{i}) \Biggr) \chi_{ \{ k\notin K \} }\chi_{\{\widetilde{\tau
}_{k}<\widehat{\tau}\}}|\mathcal{F}_{\tau_{k}} \Biggr) \nonumber\\[-8pt]\\[-8pt]
&&\qquad
\leq\frac{\varepsilon}{4\widehat{x}}E \Biggl( \sum_{i=1}^{n}%
(x_{i+1}-x_{i})\chi_{ \{ k\notin K \} }\chi_{\{\widetilde{\tau}%
_{k}<\widehat{\tau}\}}|\mathcal{F}_{\tau_{k}} \Biggr) \nonumber\\
&&\qquad
\leq\frac{\varepsilon}{4\widehat{x}}E \bigl( (X_{\tau_{k}}^{\overline{\pi
}_{k_{0}}}-X_{\tau_{k+1}}^{\overline{\pi}_{k_{0}}})\chi_{\{\widetilde
{\tau}_{k}<\widehat{\tau}\}}|\mathcal{F}_{\tau_{k}} \bigr) .\nonumber
\end{eqnarray}
From (\ref{DiferenciasdeV-2-a})--(\ref{DiferenciasdeV-7}) and from
(\ref{DefinicionPibarra}), (\ref{CeroCondicionu}) and (\ref
{TercierCondicionu}%
) we obtain that%
%
%
\begin{eqnarray} \label{DiferenciasdeV-8}\qquad
&&
\vert V_{\overline{\pi}_{1}}(X_{\tau_{k}}^{\overline{\pi}_{k_{0}}%
})-V_{\overline{\pi}_{0}}(X_{\tau_{k}}^{\overline{\pi}_{k_{0}}}) \vert
\nonumber\\[-8pt]\\[-8pt]
&&\qquad\leq\frac{3\varepsilon}{4(2k_{0}+3)}+\frac{\varepsilon}{8(M+2)}%
u+\frac{\varepsilon}{4\widehat{x}}E \bigl( ( X_{\tau_{k}}%
^{\overline{\pi}_{k_{0}}}-X_{\tau_{k+1}}^{\overline{\pi}_{k_{0}}} )
\chi_{\{\widetilde{\tau}_{k}<\widehat{\tau}\}}|\mathcal{F}_{\tau_{k}}
\bigr).\nonumber
\end{eqnarray}
And so from (\ref{DiferenciasdeV-2}) and (\ref{DiferenciasdeV-8}), we
have using (\ref{Definicionk}) and Lemma \ref{ConstrucciondeAdelta}(c)
included in the \hyperref[app]{Appendix}, that%
\begin{eqnarray*}
\vert V_{\overline{\pi}_{k_{0}}}(a)-V_{\overline{\pi}_{0}}(a) \vert
& \leq& E_{a} \Biggl( \sum_{k=0}^{k_{0}-1} \vert V_{\overline{\pi
}_{1}}(X_{\tau_{k}}^{\overline{\pi}_{k_{0}}})-V_{\overline{\pi}_{0}}%
(X_{\tau_{k}}^{\overline{\pi}_{k_{0}}}) \vert\Biggr) \\
& \leq& k_{0} \biggl( \frac{3\varepsilon}{4(2k_{0}+3)}+\frac{\varepsilon
}{8(M+2)}u \biggr)\\
&&{} +\frac{\varepsilon}{4\widehat{x}}E_{a} \bigl( (
X_{\tau_{k}}^{\overline{\pi}_{k_{0}}}-X_{\tau_{k+1}}^{\overline{\pi
}_{k_{0}}%
} ) \chi_{\{\widetilde{\tau}_{k}<\widehat{\tau}\}} \bigr) .
\end{eqnarray*}
So we have proved (\ref{Parte1}).

Let us prove now that%
%
%
\begin{equation} \label{Parte2}%
{\max_{x\geq0}} \vert V_{\pi}(x)-V_{\overline{\pi}_{k_{0}}%
}(x) \vert\leq\varepsilon/4.
\end{equation}
Given any initial surplus $x\geq0$, consider the process
$X_{t}^{\overline
{\pi}_{k_{0}}}$ with initial value $x$. Since the processes $X_{t}%
^{\overline{\pi}_{k_{0}}}$ and $X_{t}^{\pi}$ coincide up to $\tau_{k_{0}
}\wedge\widehat{\tau}$,
%
%
\begin{eqnarray}\label{Primercuenta}%
&&
{\max_{x\geq0}} \vert V_{\pi}(x)-V_{\overline{\pi}_{k_{0}}%
}(x) \vert\nonumber\\
&&\qquad \leq E_{x}\bigl(e^{-c(\tau_{k_{0}}\wedge\widehat{\tau}%
)} \vert V_{\pi}(X_{\tau_{k_{0}}\wedge\widehat{\tau}}^{\pi
})-V_{\overline
{\pi}_{0}}(X_{\tau_{k_{0}}\wedge\widehat{\tau}}^{\pi})
\vert\bigr)\nonumber\\[-8pt]\\[-8pt]
&&\qquad = E_{x}\bigl(e^{-c(\tau_{k_{0}}\wedge\widehat{\tau})} \vert V_{\pi}%
(X_{\tau_{k_{0}}\wedge\widehat{\tau}}^{\pi})-V_{\overline{\pi}_{0}}%
(X_{\tau_{k_{0}}\wedge\widehat{\tau}}^{\pi}) \vert\chi_{\{\tau_{k_{0}%
}<\widehat{\tau}\}}\bigr)\nonumber\\
&&\qquad = E_{x}(e^{-c\tau_{k_{0}}})V(\widehat{x}).\nonumber
\end{eqnarray}
Consider the process $Y_{t}^{\pi}$ defined in Lemma \ref{LemmaPYt}, as the
process corresponding to $X_{t}^{\pi}$ without claims and without paying
dividends, but starting at $Y_{0}^{\pi}=1$. When the process $X_{t}^{\pi}$
arrives the $k_{0}$th time to $\mathcal{A}_{\delta}$, it should have already
passed $k_{0}-N$ times through intervals of the form $ ( c_{a},a )
$ with $a\in\mathcal{A}_{\delta}$. So%
\[
\tau_{k_{0}}\geq T_{k_{0}-N}:=\min\bigl\{t\dvtx Y_{t}^{\pi}\geq
e^{(k_{0}-N)u}\bigr\}.
\]
Let $\gamma_{t}$ be the investment policy corresponding to the strategy
$\pi$.
We have using It\^{o}'s formula that%
\begin{eqnarray*}
&&
g\bigl(e^{(k_{0}-N)u}\bigr)E(e^{-cT_{k_{0}-N}})-g(1)\\%
&&\qquad
= E\biggl(\int_{0}^{T_{k_{0}-N}}e^{-cs} \bigl( \sigma^{2}\gamma_{s}^{2} (
Y_{s}^{\pi} ) ^{2}g^{\prime\prime}(Y_{s}^{\pi})/2\\
&&\qquad\quad\hspace*{71.1pt}{} + ( p+r\gamma
_{s}Y_{s}^{\pi} ) g^{\prime}(Y_{s}^{\pi})-cg(Y_{s}^{\pi}) \bigr) \,ds\biggr)
\leq0,
\end{eqnarray*}
so we get%
\[
E(e^{-c\tau_{k_{0}}})\leq E(e^{-cT_{k_{0}-N}})\leq
g(1)/g\bigl(e^{(k_{0}-N)u}\bigr)\leq
g(1)/g(e^{M})\leq\varepsilon/(4V(\widehat{x}))
\]
and from (\ref{Primercuenta}) we obtain (\ref{Parte2}).

We get the result combining (\ref{Parte1}) and (\ref{Parte2}).
\end{pf}
\begin{remark}
From Propositions \ref{PropiedadesConjuntossub0} and
\ref{Vcomolimitedeestrategias}, we conclude that the optimal
strategy for
large surpluses is to pay out as dividends the amount exceeding $a^{\ast
}%
=\max\mathcal{A}_{0}$.
\end{remark}
\begin{remark}
Propositions \ref{W1yY0} and \ref{VywenC0}, Remark
\ref{Viscontinuouslydifferentiable.}, and the fact that the
derivative of
the optimal value function should be one in $\mathcal{A}_{0}\cup
\mathcal{B}_{0}$, suggest a method to construct the optimal value
function in
the case that the optimal dividend payment policy has the structure of a
finite band: we could construct the value function of the best one-band
strategy, the best two-band strategy, etc. as candidates of the optimal value
function. If any of these candidates is a viscosity solution of (\ref
{InicDif}%
), it should be $V$. We use this method to find the optimal value
function in
the examples of the next section.
\end{remark}

\section{Numerical examples and final remarks}\label{sec9}

In this section we present numerical approximations of the optimal value
function $V$. In order to do this, we obtain as a first step an approximation
of the function $W$ using the fixed-point operator defined in Proposition
\ref{WesC2}; it is not possible to use an standard approximation scheme
because of the lack of both the ellipticity of the equation (\ref{EcuDif(w)})
and the boundary condition at zero.

We construct two examples of optimal value functions. In one example the
optimal dividend payment policy is barrier and in the other it is not.
\begin{example}\label{Example91}
We consider the exponential distribution $F(x)=1-e^{-x}$ and parameters $p=4$,
$\beta=1$, $c=0.5$, $r=0.3$, $\sigma=2$. We first obtain numerically, using
Proposition \ref{WesC2}, the function $W$ and we get that the derivative
reaches the minimum at $y=4.846$. Then, we prove that the value
function $V_{1}$ of the optimal barrier strategy is a solution of
(\ref{InicDif}) and so, by Proposition \ref{optimalBarrier}, $V=V_{1}
$ and $V
$ is twice continuously differentiable.

\begin{figure}

\includegraphics{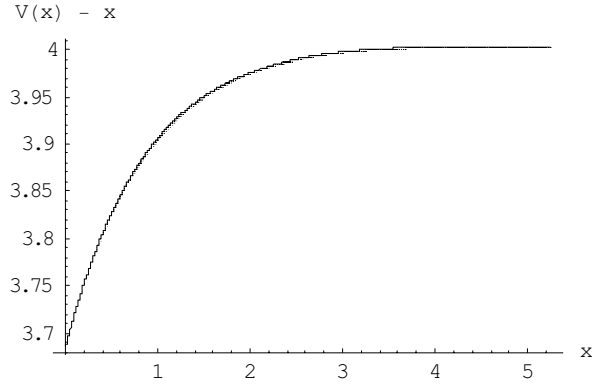}

\caption{$V(x)-x$ for an exponential distribution.}
\label{figure1}
\end{figure}

\begin{figure}[b]

\includegraphics{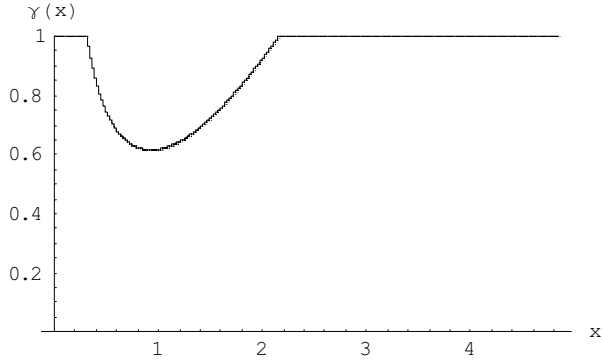}

\caption{$\gamma$ for an exponential distribution.}
\label{figure2}
\end{figure}

We show in Figure \ref{figure1} the function $V(x)-x$ and in Figure \ref{figure2} the optimal
investment policy $\gamma^{\ast}(x)$ for $x\in[ 0,y ] $. Note
that, according to Proposition \ref{WesC2(segunda)}(b), $\gamma^{\ast
}=1$ for
small surpluses.
\end{example}
\begin{example}\label{Example92}
We consider the following claim distribution:%
\[
F(x)= \cases{
0, &\quad if $x\in[0,7/10]$,\cr
(10/3)(x-7/10), &\quad if $x\in(7/10,1]$,\cr
1, &\quad if $x>1$,}
\]
and parameters $p=1.6$, $\beta=1$, $c=0.3$, $r=0.2$,
$\sigma=1$.

We prove that the derivative of the function $W$ of Proposition \ref{WesC2}
reaches the minimum at zero, so the value function of the optimal barrier
strategy is $V_{1}(x)=x+(c+\beta)/p$, but in this case $V_{1}$ is not a
supersolution of (\ref{InicDif}).

We now look for the best two-band strategy. First we obtain
numerically, using
Proposition \ref{NuevaProposicion1}, the function
\[
W_{y}(x)= \cases{
x+(c+\beta)/p, &\quad if $x\leq y$,\cr
U_{1}(x), &\quad if $y>x$,}
\]
for each $y>0$, where $U_{1}$ is the unique solution of $\mathcal
{L}^{\ast
}(U_{1},W_{y})=0$ in $(y,\infty)$ with boundary conditions $U_{1}(y)=W_{y}(y)$
and $U_{1}^{\prime}(y)=1$. Take
\[
y_{1}=\min\{y\dvtx\mbox{there exists }z>y\mbox{ with }V_{y}^{\prime
}(z)=1\}
\]
and $z_{1}$ with $V_{y_{1}}^{\prime}(z_{1})=1$. We get $y_{1}=0.291$,
$z_{1}=2.926$ and we can prove that
\[
V_{y_{1}}(x)= \cases{
W_{y_{1}}(x), &\quad if $x\leq z_{1}$,\cr
W_{y_{1}}(z_{1})+(x-z_{1}), &\quad if $y>z_{1}$,}
\]
is a viscosity solution of (\ref{InicDif}). Hence $V=V_{y_{1}}$ because
$V_{y_{1}}$ is the value function of a limit strategy corresponding to the
sets $\mathcal{A}_{0}=\{0,z_{1}\}$, $\mathcal{B}_{0}=(0,y_{1}]\cup
(z_{1},\infty)$ and $\mathcal{C}_{0}=(y_{1},z_{1})$.

We show in Figure \ref{figure3} the function $V(x)-x$, in Figure \ref{figure4} the derivative
of $V$
and in Figure \ref{figure5} the optimal investment policy $\gamma^{\ast}(x)$ for
$x\in(y_{1},z_{1})$. It can be seen in Figure \ref{figure4} that $V$ is not twice
continuously differentiable at $y_{1}$.

\begin{figure}[b]

\includegraphics{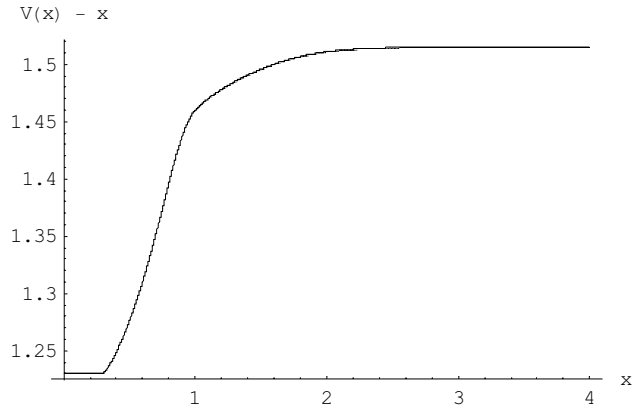}

\caption{$V(x)-x$ for a non-monotone density distribution.}
\label{figure3}
\end{figure}

\begin{figure}

\includegraphics{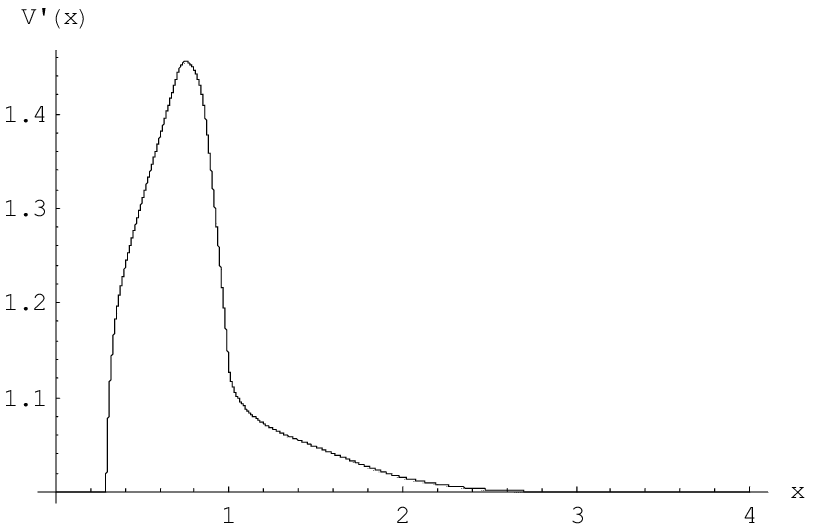}

\caption{$V'$ for a non-monotone density distribution.}
\label{figure4}
\end{figure}

\begin{figure}[b]

\includegraphics{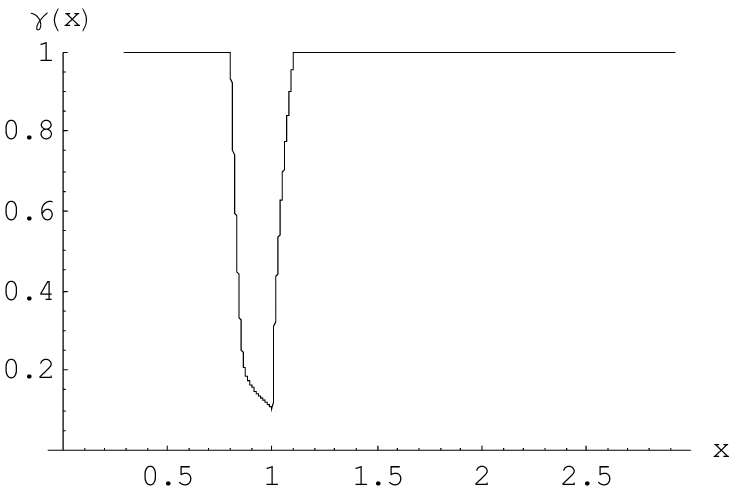}

\caption{$\gamma$ for a non-monotone density distribution.}
\label{figure5}
\end{figure}

Let us finally note that in the setting of diffusion approximation
[see, for
instance, H\o jgaard and Taksar (\citeyear{HT04})], the optimal value function
$V$ is
always twice continuously differentiable, concave and comes from an optimal
barrier strategy. We see in the last example that this is not always
the case
in the Cram\'{e}r--Lundberg setting.%
\end{example}

\begin{appendix}\label{app}
\section*{Appendix: Technical lemmas}

\subsection*{Lemmas for Proposition \protect\ref{MenorSuper}}

In the following two lemmas we show that, in order to define $V$ as a supremum
of value function of admissible strategies, we can discard the strategies
where the surplus stays a positive time at the points of a given countable
set, and also that $V$ can be written as a limit of value functions of
strategies whose surpluses are confined in compact subsets of $(0,\infty)$.
\setcounter{lemma}{0}
\begin{lemma}
\label{admissiblestrategiesacotadas}
\textup{(a)} Given $x\geq0$ and $x_{1}>x$, let us
define $\Pi_{x}^{x_{1}}$ as the set of $\pi\in$
$\Pi
_{x}$ such that $X_{t}^{\pi}\leq x_{1}$ for all
$t\geq
0$ and $\mathcal{V}^{x_{1}}(x)=\sup\{V_{\pi}(x)$
with
$\pi\in\Pi_{x}^{x_{1}}\}$, then
\[
\lim_{x_{1}\rightarrow\infty}\mathcal{V}^{x_{1}}(x)=V(x).
\]

\textup{(b)} Given $x_{0}\geq x\geq0$, let us define $\Pi
_{x}^{ [ x_{0},\infty) }$ as the set of $\pi\in
\Pi_{x}$ such that $X_{t}^{\pi}\geq x_{0}$
for all $t\geq0$ and $\mathcal{V}_{x_{0}}(x)=\sup\{V_{\pi}%
(x)$ with $\pi\in\Pi_{x}^{ [ x_{0},\infty) }\}$,
then
\[
\lim_{x_{0}\searrow0}\mathcal{V}_{x_{0}}(x)=V(x).
\]
\end{lemma}
\begin{pf}
(a) Given $\varepsilon>0$ consider $\pi\in\Pi_{x}$ such that
$V(x)<V_{\pi}(x)+\varepsilon$ and consider for any $x_{1}>0$ the admissible
strategy $\pi_{x_{1}}\in\Pi_{x}^{x_{1}}$ which coincides with the strategy
$\pi$ while the surplus is less than $x_{1}$, and pay out $x_{1} $ as dividends
at the moment $\tau_{x_{1}}$when the surplus reaches $x_{1}$. Since
\[
\lim_{x_{1}\rightarrow\infty}E_{x} \biggl( \int_{0}%
^{\tau_{x_{1}}\wedge\tau^{\pi}}e^{-cs}\,dL_{s}^{\pi} \biggr) =V_{\pi}(x),
\]
there exists $x_{1}$ large enough such that $V(x)-V_{\pi_{x_{1}}%
}(x)<2\varepsilon$.

(b) Take $x_{0}\in(0,x)$. We can find an admissible strategy $\pi\in
\Pi_{x-x_{0}}$ such that $V(x-x_{0})<V_{\pi}(x-x_{0})+x_{0}$. Define the
admissible strategy $\pi_{0}\in\Pi_{x}$ which invests $x_{0}$ in
bonds and
then follows the strategy $\pi$ corresponding to initial surplus
$x-x_{0}$ up
to the time $\tau_{x_{0}}=\inf\{ t\dvtx X_{t}^{\pi_{0}}%
<x_{0} \} $. Then we have that $\mathcal{V}_{x_{0}}(x)\geq V_{\pi_{0}%
}(x)=V_{\pi}(x-x_{0})$ and the result follows from the continuity of
$V$ at
$x$.
\end{pf}
\begin{lemma}
\label{LemaEvitarSaltosF}
Given $x\geq0$ and
a countable set $S\subset[0,\infty)$, let $\Pi_{x}%
(S)$ be the set of all the admissible strategies $\pi\in\Pi
_{x}$ such that the set
\[
\{ (\omega,t)\in\Omega\times\mathbf{[0,\infty)}\dvtx X_{t}^{\pi
}(\omega)\in S \},
\]
has zero measure. Then $V(x)=\sup_{\pi\in\Pi_{x}(S)}%
V_{\pi}(x)$.
\end{lemma}
\begin{pf}
Given $\varepsilon>0$, take $\pi=(\gamma_{t},L_{t})\in\Pi_{x}$ such that
$V(x)-V_{\pi}(x)<\varepsilon/2$. Given any $a\in(0,\varepsilon/2)$, consider
the stopping times $\tau_{a}=\inf\{t\dvtx L_{t}\geq a\}$ and $\tau_{0}%
=\inf\{t\dvtx L_{t}\geq0\}$ and the admissible strategy $\pi_{a}=(\gamma
_{t}%
^{a},L_{t}^{a})$ such that the dividend policy consists in paying no dividends
up to time $\tau_{a}$ and following the dividend policy $L_{t}-a$
afterward, and such that the amount of the surplus invested in stocks
coincides with the amount of the surplus invested in stocks in the original
strategy. We have that $X_{t}^{\pi_{a}}$ coincides with $X_{t}^{\pi}$ for
$t\in[0,\tau_{0}\wedge\tau^{\pi}]$, that $X_{t}^{\pi
_{a}}-X_{t}^{\pi}%
\in(0,a)$ if $t\in(\tau_{0},\tau_{a}\wedge\tau^{\pi})$ and that $X_{t}%
^{\pi_{a}}-X_{t}^{\pi}=a$ if $t\in[\tau_{a},\tau^{\pi}]$. We
obtain that
$\tau^{\pi_{a}}\geq\tau^{\pi}$ and that $V_{\pi}(x)-V_{\pi_{a}}(x)\leq
a<\varepsilon/2$, and so $V(x)-V_{\pi_{a}}(x)\leq\varepsilon$ for all
$a\in(0,\varepsilon/2)$. Note that, fixing $x_{i}\in S$ we have that%
\[
1\geq P\biggl(\bigcup_{a\in(0,\varepsilon/2)} \{ (\omega,t)\dvtx X_{t}^{\pi}%
=x_{i}-a \} \biggr)\geq P\biggl(\bigcup_{a\in(0,\varepsilon/2)} \{
(\omega,t)\dvtx X_{t}^{\pi_{a}}=x_{i},\tau_{a}\leq t \} \biggr),
\]
and the last union is disjoint. Then the set of $a\in(0,\varepsilon/2)$ such
that
\[
P\bigl( \{ (\omega,t)\dvtx X_{t}^{\pi_{a}}=x_{i},\tau_{a}\leq
t \} \bigr)>0
\]
is countable. So, since $S$ is countable, there exists $a_{0}\in
(0,\varepsilon/2)$ such that
\[
P\bigl( \{ (\omega,t)\dvtx X_{t}^{\pi_{a}}\in S,\tau_{a_{0}}\leq t \}
\bigr)=0.
\]
If $t<\tau_{a_{0}}$, then $L_{t}^{a_{0}}=0$ and $X_{t}^{\pi_{a_{0}}}%
=X_{t}^{(\gamma_{t},0)}$ which does not depend on $a_{0}$. Define $\tau^{0}=0$
and call $\tau^{i}$ the time of the $i$th claim, we obtain that%
\[
\{ (\omega,t)\dvtx X_{t}^{\pi_{a_{0}}}\in S, t\leq\tau_{a} \} =%
{\bigcup_{i=0}^{\infty}}
\{ (\omega,t)\dvtx X_{t}^{\pi_{a_{0}}}\in S,t\in[\tau
^{i}\wedge\tau_{a_{0}},\tau^{i+1}\wedge\tau_{a_{0}}) \},
\]
but if $t\in[\tau^{i}\wedge\tau_{a_{0}},\tau^{i+1}\wedge\tau_{a_{0}})$,
we have that $X_{t}^{(\gamma_{t},0)}$ is a linear diffusion [see, for instance,
Borodin and Salminen (\citeyear{BS02})], and so
\[
P \bigl( \{ (\omega,t)\dvtx X_{t}^{\pi_{a_{0}}}\in S, t\in[
\tau^{i}\wedge\tau_{a_{0}},\tau^{i+1}\wedge\tau_{a_{0}}) \} \bigr)
=0.
\]
We conclude that $P ( (\omega,t)\dvtx X_{t}^{\pi_{a_{0}}}\in S )
=0$.
\end{pf}

\subsection*{Lemmas for Proposition \protect\ref{PropiedadesConjuntossub0}}

We need the following result in order to prove Lemma \ref{PropiedadUy}.
\begin{lemma}
\label{MenorSuperLocal}
Assume that $V^{\prime}(\widehat{x}%
)=1$ for some $\widehat{x}>0$ and $\overline{u}%
$ is an absolutely continuous supersolution of (\ref{InicDif}) in
$(0,\widehat{x})$, then $\overline{u}\geq V$ in
$[0,\widehat{x}]$.
\end{lemma}
\begin{pf}
The argument coincides with the one used to prove Proposition \ref{MenorSuper},
but taking admissible strategies $\pi$ such that the corresponding controlled
risk process $X_{t}$ satisfies $X_{t}\leq\widehat{x}$.
\end{pf}

The following lemma gives conditions under which the optimal value function
$V$ is linear in some interval.
\begin{lemma}
\label{PropiedadUy}
Given any $y>0$, we define
%
%
\begin{equation}\label{Uy}%
\mathcal{U}_{y}(x)= \cases{
V(x), &\quad if $x\leq y$,\cr
V(y)-y+x, &\quad if $x>y$.}
\end{equation}

\textup{(a)} If $\mathcal{U}_{y}$ is supersolution of (\ref
{InicDif}%
) in $(y,\infty)$, then $\mathcal{U}_{y}=V$ in
$[0,\infty)$.

\textup{(b)} Assume that $V^{\prime}(\widehat{x})=1$ for some
$\widehat{x}>0$ and there exists $y<\widehat{x}$
such that $\mathcal{U}_{y}$ is supersolution of (\ref{InicDif}) in
$(y,\widehat{x}]$ then $\mathcal{U}_{y}=V$ in
$[0,\widehat{x}]$.
\end{lemma}
\begin{pf}
(a) Let us prove first that $\mathcal{U}_{y}$ is a supersolution of
(\ref{InicDif}). We only need to check it at $y$. In the case that
$\mathcal{U}_{y}^{\prime}(y^{-})=V^{\prime}(y^{-})>1=\mathcal
{U}_{y}^{\prime
}(y^{+})$, there is no test for viscosity supersolution at $y$ and in
the case
that $\mathcal{U}_{y}^{\prime}(y)=1$. Take $q$ such that%
\begin{eqnarray*}
q/2&\leq&\lim\inf_{h\rightarrow0}\frac{ ( \mathcal{U}_{y}%
(y+h)-\mathcal{U}_{y}(y) ) /h-1}{h}\\
&\leq&\lim_{h\rightarrow0^{+}%
}\frac{ ( \mathcal{U}_{y}(y+h)-\mathcal{U}_{y}(y) ) /h-1}{h}=0.
\end{eqnarray*}
Since $\mathcal{U}_{y}$ is a supersolution for $x>y$ and $\sup
_{\gamma\in[0,1]}\mathcal{L}_{\gamma}(\mathcal{U}_{y},1,q)(x)$
is right continuous for $x\geq y$ we have that $\sup_{\gamma
\in[0,1]}\mathcal{L}_{\gamma}(\mathcal{U}_{y},1,q)(y)\leq0$.

From Proposition \ref{MenorSuper} we get that $\mathcal{U}_{y}\geq V$.
Let us
prove now that $\mathcal{U}_{y}(x)\leq V(x)$ for all $x>y$. Given any
$\varepsilon>0$, take an admissible strategy $\pi\in\Pi_{y}$ such that
$V_{\pi}(y)\geq V(y)-\varepsilon$. For any initial surplus $x\geq
y$, we
define a new strategy $\pi_{x}\in\Pi_{x}$ as follows: pay out
immediately the
excedent $x-y$ as dividend, and then use the strategy $\pi$. Since $\pi
$ is
admissible, $\pi_{x}$ is also admissible. We get that, for all $x>y$ and
$\varepsilon>0$,
\[
\mathcal{U}_{y}(x)-\varepsilon=x-y+V(y)-\varepsilon\leq x-y+V_{\pi}%
(y)=V_{\pi_{x}}(x)\leq V(x),
\]
and so we get the result.

The proof of (b) is analogous to the proof of (a) using Lemma
\ref{MenorSuperLocal}.
\end{pf}
\begin{lemma}
\label{Continuacioncomosupersolution}
If $\Lambda(x_{0}%
)<0$, there exists $h_{0}>0$, such that the function
$\mathcal{U}_{x_{0}-h_{0}}$ defined in (\ref{Uy}) is a supersolution
of (\ref{InicDif}) in $(x_{0}-h_{0},x_{0}+h_{0})$.
\end{lemma}
\begin{pf}
Since $V$ is locally Lipschitz, for a small $h>0$ and $x\in(
x_{0}-h_{0},x_{0}+h_{0} )$, there exists $K>1$ such that $V(x)-V(x_{0}%
-h)\leq K ( x-x_{0}+h ) $. By definition $\mathcal{U}_{x_{0}%
-h}(x)=V(x_{0}-h)+x-x_{0}+h$, and so $V(x)-\mathcal{U}_{x_{0}-h}%
(x)\leq(K-1) ( x-x_{0}+h ) $. Then we obtain that $ \vert
\Lambda(x)-\mathcal{L}^{\ast}(\mathcal{U}_{x_{0}-h})(x) \vert
\leq(c+2\beta)(K-1) ( x-x_{0}+h ) $. By assumption, $\Lambda
(x_{0})<0$, since $\Lambda$ is continuous for $h$ small enough and
$x\in( x_{0}-h_{0},x_{0}+h_{0} ) $ we have that $\Lambda(x)<0$.
Therefore, there exists $h_{0}$ small enough such that $\mathcal
{L}^{\ast
}(\mathcal{U}_{x_{0}-h_{0}})(x)<0$ for $x\in( x_{0}-h_{0},x_{0}%
+h_{0} ) $, and so we have the result.
\end{pf}
\begin{lemma}
\label{PropiedadesConjuntos}The sets introduced in Definition
\ref{Conjuntos} satisfy the following properties:

\textup{(a)} $\mathcal{B}$ is a left-open set, that is if
$x\in\mathcal{B}$ there exists $\delta>0$ such that
$(x-\delta,x]\subset\mathcal{B}$.

\textup{(b)} $\mathcal{A}$ is a left closed set, that is if
$x_{n}\in\mathcal{A}$ and $x_{n}\searrow x$ then
$x\in\mathcal{A}$.

\textup{(c)} If $(x_{0},\widehat{x}]\subset\mathcal{B}$ and
$x_{0}\notin\mathcal{B}$ then $x_{0}\in\mathcal{A}$.

\textup{(d)} There is a $x^{\ast}$ such that $(x^{\ast}%
,\infty)\subset\mathcal{B}$.

\textup{(e)} $\mathcal{C}$ is an open set in $[0,\infty
)$,
that is if $0\in\mathcal{C}$, there exists $\delta>0$
such that $(0,\delta)\subset\mathcal{C}$ and if a positive
$x\in\mathcal{C}$ there exists $\delta>0$ such that
$(x-\delta,x+\delta)\subset\mathcal{C}$.

\textup{(f)} Both $\mathcal{A}$ and $\mathcal{B}$ are
nonempty.
\end{lemma}
\begin{pf}
(a) Assume that $x_{0}\in\mathcal{B}$. By Lemma
\ref{Continuacioncomosupersolution}, we can find $h_{0}>0$, such that the
function $U_{x_{0}-h_{0}}$ defined in (\ref{Uy}) is a supersolution of
(\ref{InicDif}) in $(x_{0}-h_{0},x_{0}]$, and then, by Lemma \ref
{PropiedadUy}(b), since $V^{\prime}(x_{0})=1$, we have $\mathcal
{U}_{x_{0}-h_{0}}=V$ at
$[0,x_{0})$ and so $(x_{0}-h_{0},x_{0}]\subset\mathcal{B}$.

(b) It follows from the right continuity of the function
$\Lambda(x)$ and $V^{\prime}(x^{+})$.

(c) Since $\Lambda(x)$ is continuous and $V^{\prime}(x^{+})$ is
right continuous, we have that $V^{\prime}(x_{0}^{+})=1$ and $\Lambda
(x_{0})\leq0$. But $x_{0}\notin\mathcal{B}$, so either $\Lambda
(x_{0})=0$ and
$V^{\prime}(x_{0}^{+})=1$ or $V^{\prime}(x_{0}^{-})>V^{\prime}(x_{0}^{+})=1$
and $\Lambda(x_{0})<0$. In the first case $x_{0}\in\mathcal{A}$, let us see
that the second case is not possible. Since $\Lambda(x_{0})<0$, by Lemma
\ref{Continuacioncomosupersolution}, we can find $h_{0}>0$, such that the
function $U_{x_{0}-h_{0}}$ defined in (\ref{Uy}) is a supersolution of
(\ref{InicDif}) in $(x_{0}-h_{0},x_{0}+h_{0})$ and $x_{0}+h_{0}\in
\mathcal{B}$. Since $V^{\prime}(x_{0}+h_{0})=1$, we have from Lemma
\ref{PropiedadUy}(b) that $\mathcal{U}_{x_{0}-h_{0}}=V$ at $[0,x_{0}+h_{0})$
and so $x_{0}\in(x_{0}-h_{0},x_{0}+h_{0}]\subset\mathcal{B}$; this is a
contradiction.

(d) For each $y>0$ let us consider the functions $\mathcal{U}_{y}$
defined in (\ref{Uy}). We will show that, if $y\geq p/(c-r)$, then
$\mathcal{U}_{y}$ is a viscosity supersolution of (\ref{InicDif}) for all
$x\in(y,\infty)$, and the result follows from Lemma \ref{PropiedadUy}(a).
Since $\mathcal{U}_{y}^{\prime}=1$ in $(y,\infty)$ we only need to show that
$\mathcal{L}^{\ast}(\mathcal{U}_{y})\leq0$ in $(y,\infty)$. Take any
$\gamma\in[0,1]$, since $\mathcal{U}_{y}$ is increasing, we have that
$\mathcal{L}_{\gamma}(\mathcal{U}_{y})(x)\leq p+(r-c)y$. Hence, the result
follows with $x^{\ast}=p/(c-r)$.

(e) Take $x\in\mathcal{C}$, if there is no $\delta>0$ such that
$[x,x+\delta)\subset\mathcal{C}$, then we can find a sequence $x_{n}%
\in\mathcal{A}\cup\mathcal{B}$ such than $x_{n}\searrow x$. If there is a
subsequence $x_{n_{k}}\in\mathcal{A}$, then by (b) we get that $x\in
\mathcal{A}$, and if a subsequence $x_{n_{k}}\in\mathcal{B}$, by (c) we can
find a sequence $y_{k}\in\mathcal{A}$ with $x<y_{k}<x_{n_{k}}$; then
again by
(b) we get that $x\in\mathcal{A}$. Take a positive $x\in\mathcal{C}$.
If there
is no $\delta>0$ such that $[x-\delta,x)\subset\mathcal{C}$, then we
can find
a sequence $x_{n}\in\mathcal{A}\cup\mathcal{B}$ such than
$x_{n}\nearrow x$.
Then, $V^{\prime}(x^{-})=1$ and then $V^{\prime}(x)=1$. Then, since
$x\in\mathcal{C}$, $\Lambda(x)>0$ but since $\Lambda$ is continuous
\mbox{$\Lambda(x)=\lim_{n\rightarrow\infty}\Lambda(x_{n})\leq0$}, and
this is
a contradiction.

(f) It follows from (c) and (d).
\end{pf}

\subsection*{Lemmas for Proposition \protect\ref{VywenC0}}

\begin{lemma}
\label{L=0enC} The optimal value function $V$ is a
viscosity solution of $\mathcal{L}^{\ast}(V)=0$ on the open set
$\mathcal{C}$.
\end{lemma}
\begin{pf}
It follows from (\ref{InicDif}) that $V$ is a viscosity supersolution of
$\mathcal{L}^{\ast}(V)=0$. Let us prove that it is a viscosity
subsolution of
$\mathcal{L}^{\ast}(V)=0$ in $\mathcal{C}$. First consider $x\in\mathcal{C}$
with $1\leq V^{\prime}(x^{+})<V^{\prime}(x^{-})$. Take any $d\in
(V^{\prime
}(x^{+}),V^{\prime}(x^{-}))$; we have that%
\[
\limsup_{h\rightarrow0}\frac{ ( V(x+h)-V(x) ) /h-d}%
{h}=-\infty
\]
and then, for any $q$,%
\[
\max\Bigl\{1-d,\max_{\gamma\in[0,1]} \bigl( \sigma^{2}x^{2}\gamma
^{2}q/2+(p+rx\gamma)d-M(V)(x) \bigr) \Bigr\}\geq0,
\]
so, since $d>1$, we have that%
%
%
\begin{equation} \label{expresion-sub-d-q}%
\max_{\gamma\in[0,1]} \bigl( \sigma^{2}x^{2}\gamma^{2}q/2+(p+rx\gamma
)d-M(V)(x) \bigr) \geq0.
\end{equation}
Since this holds for any $q$, taking a sequence $q_{n}\rightarrow-\infty$,
\[
pd-M(V)(x)\geq0 \qquad\mbox{for any
}d\in(V^{\prime}(x^{+}),V^{\prime}(x^{-}))
\]
that implies $pV^{\prime}(x^{+})-M(V)(x)\geq0$, and so
(\ref{expresion-sub-d-q}) holds for any $d\in[ V^{\prime}(x^{+}%
)$,\break $V^{\prime}(x^{-})]$ and any $q$. So $V$ is a viscosity subsolution of
$\mathcal{L}^{\ast}(V)=0$ at $x$.

Next consider $x\in\mathcal{C}$ such that $V$ is differentiable with
$1<V^{\prime}(x)$. We have $d=V^{\prime}(x)>1$, and then%
\[
\max\Bigl\{1-d,\max_{\gamma\in[0,1]} \bigl( \sigma^{2}x^{2}\gamma
^{2}q/2+(p+rx\gamma)d-M(V)(x) \bigr) \Bigr\}\geq0
\]
implies that%
\[
\max_{\gamma\in[0,1]} \bigl( \sigma^{2}x^{2}\gamma^{2}q/2+(p+rx\gamma
)d-M(V)(x) \bigr) \geq0
\]
and so $V$ is a viscosity subsolution of $\mathcal{L}^{\ast}(V)=0$ at $x$.

Finally, the case in which $1=V^{\prime}(x)$ and $\Lambda(x)>0$ cannot happen
by Lem\-ma~\ref{LambdaNegativo}.
\end{pf}
\begin{lemma}
\label{VywenC}
\textup{(a)} Given $x_{1}>0$, there exists a unique absolutely
continuous, increasing viscosity solution of
%
%
\begin{equation} \label{nuevasup}%
\mathcal{L}^{\ast}(U,V)=0
\end{equation}
in $ ( x_{1},\infty) $ that is differentiable
at $x_{1}$, with boundary conditions $U(x_{1})=V(x_{1}%
)$ and $U^{\prime}(x_{1})=V^{\prime}(x_{1})=1$.

\textup{(b)} Let $(x_{1},x_{2})$ with
$x_{1}>0$ be
a connected component of $\mathcal{C}$, the function $U$
defined in \textup{(a)} coincides with $V$ in $[x_{1},x_{2}]$.

\textup{(c)} The optimal value function $V$ is a classical solution
of $\mathcal{L}^{\ast}(V)=0$ on the open set $\mathcal{C}$.
\end{lemma}
\begin{pf}
(a) The existence of $U$ follows from Proposition \ref{NuevaProposicion1}.
Let us prove the uniqueness. Given an interval $ ( x_{1},y ) $,
with arguments similar to the ones used in the proof of Proposition
\ref{Superarribasub}, it can be proved that, if a supersolution of
(\ref{SupdeLgamadeVyw}) is greater than a subsolution of
(\ref{SupdeLgamadeVyw}) in the boundaries of the interval, it is also
greater in the interior. From this result we conclude that, if
$\overline{u}$
and $\underline{u}$ are supersolution and subsolution of
(\ref{SupdeLgamadeVyw}) with $\overline{u}(x_{1})=\underline{u}
(x_{1})$, then%
\[
\max_{x\in[ x_{1},y]}\{\underline{u}(x)-\overline{u}(x)\}
\leq
\max\{0,\underline{u}(y)-\overline{u}(y)\}.
\]

Let us take $\overline{w}$ and $\underline{w}$ supersolution and subsolution
of (\ref{SupdeLgamadeVyw}), respectively, with $\overline{w}%
(x_{1})=\underline{w}(x_{1})=V(x_{1})$ and $\overline{w}^{\prime}%
(x_{1})=\underline{w}^{\prime}(x_{1})=V^{\prime}(x_{1})$, and define
$\underline{w}_{\varepsilon}(x)=\underline{w}(x)+\varepsilon(e^{
({c+\beta
})/{p}(x-x_{1})}-1)$. Since $\mathcal{L}_{\gamma}(e^{({c+\beta
})/{p}(x-x_{1}%
)}-1)\geq0$, we obtain that $\underline{w}_{\varepsilon}$ is also a
subsolution with $\underline{w}_{\varepsilon}(x_{1})=V(x_{1})$ and
$\underline{w}_{\varepsilon}^{\prime}(x_{1})=V^{\prime
}(x_{1})+\varepsilon
(c+\beta)/p>V^{\prime}(x_{1})$. Then, since $\underline{w}_{\varepsilon
}(x)-\overline{u}(x)$ is positive for $x\in(x_{1},x_{1}+\delta)$ for some
positive~$\delta$, we have
\[
\max_{x\in[ x_{1},y]}\{\underline{w}_{\varepsilon}%
(x)-\overline{u}(x)\}\leq\max\{0,\underline{w}_{\varepsilon}(y)-\overline
{u}(y)\}=\underline{w}_{\varepsilon}(y)-\overline{u}(y),
\]
so we obtain that $\max_{x\in[ x_{1},y]}\{\underline
{w}_{\varepsilon}(x)-\overline{u}(x)\}=\underline{w}_{\varepsilon}%
(y)-\overline{u}(y)$, and so $\underline{w}_{\varepsilon}(x)-\overline{w}(x)$
is increasing and positive for all $x>x_{1}$ and $\varepsilon>0$. Then
$\underline{w}(x)\geq\overline{w}(x)$ for all $x>x_{1}$.

[(b) and (c)] We showed in Lemma \ref{L=0enC} that $V$ is a viscosity solution
of $\mathcal{L}^{\ast}(V)=0$ on the open set $\mathcal{C}$, so let us
show now
that $V$ is twice continuously differentiable. In the case $x_{1}=0$, the
result follows from Proposition \ref{W1yY0}, and in the case $x_{1}>0$,
we have
that $V^{\prime}(x_{1})=1$, and the result follows from Proposition
\ref{NuevaProposicion1}(a) and (b).
\end{pf}

\subsection*{Lemma for Theorem \protect\ref{Vcomolimitedeestrategias}}

\begin{lemma}
\label{ConstrucciondeAdelta}Given $\delta>0$, we can find a finite set
$\mathcal{A}_{\delta}\subset\mathcal{A}_{0}$ and a number $\varsigma>0$
satisfying:

\textup{(a)} $(a-\varsigma,a)\subset\mathcal{C}_{0}$ for all
$a\in\mathcal{A}_{\delta}$.

\textup{(b)} $ \{ a\in\mathcal{A}_{0}\dvtx a-\max( \mathcal{A}_{0}%
\cap[0,a) ) \geq\delta\} \subset\mathcal{A}_{\delta}$.

\textup{(c)} \#$\mathcal{A}_{\delta}\leq2\widehat{x}/\delta$.
\end{lemma}
\begin{pf}
Consider $\widehat{\mathcal{A}}= \{ a\in\mathcal{A}_{0}\dvtx\mbox{ there
exists }c_{a}<a\mbox{ with }(c_{a},a)\subset\mathcal{C}_{0} \} $ and
\[
\mathcal{D}=\{a\in\mathcal{A}\dvtx(a-\vartheta,a)\subset\mathcal{A}\cup
\mathcal{B}\mbox{ for some }\vartheta>0\}\subset\mathcal{B}_{0}.
\]

Let us prove first that if $(x_{0},x_{1})\cap\mathcal{A}_{0}\neq\phi$, then
$(x_{0},x_{1})\cap\widehat{\mathcal{A}}\neq\phi$. In the case that
$(x_{0},x_{1})\cap\mathcal{C}_{0}=\phi$, then $(x_{0},x_{1})\subset
\mathcal{B}_{0}$ and this is is a contradiction. In the case that
$(x_{0},x_{1})\cap\mathcal{C}_{0}\neq\phi$, since $\mathcal{C}_{0}$ is open,
there exists $c\in(r_{1},r_{2})\subset\mathcal{C}_{0}$ with
$r_{1},r_{2}\notin\mathcal{C}_{0}$; if $r_{1}\leq x_{0}<x_{1}\leq
r_{2}$, we have a
contradiction because $(x_{0},x_{1})\subset\mathcal{C}_{0}$; if $r_{2}<x_{1}$
and $r_{2}\in\mathcal{A}_{0}$, we have that $r_{2}\in\widehat
{\mathcal{A}}%
$; and if $x_{0}<r_{1}<x_{1}\leq r_{2}$, the interval $(x_{0},r_{1})$ cannot
be included in $\mathcal{B}_{0}$ because we would have that $(x_{0}%
,x_{1})\subset\mathcal{C}_{0}\cup\mathcal{B}_{0}$,\vspace*{1pt} so there exists
$c\in\mathcal{C}_{0}\cap(x_{0},r_{1})$, take $a=\sup( \mathcal{C}%
_{0}\cap(x_{0},r_{1}) ) $ then $a\in(x_{0},x_{1})\cap\widehat
{\mathcal{A}}$.

Let us prove now that $ ( \mathcal{A}_{0}\cup\mathcal{D} )
\subset(
{ \bigcup_{a\in\widehat{\mathcal{A}}}}
(c_{a},a+\delta) ) \cup(
{ \bigcup_{d\in\mathcal{D}}}
(d-\delta,d+\delta) ) $. In effect, given $a_{0}\in\mathcal{A}%
_{0}\setminus\widehat{\mathcal{A}}$, we have that $(a_{0}-\delta
,a_{0})$ is
not included in $\mathcal{C}_{0}$. Then $(a_{0}-\delta,a_{0})\cap
\mathcal{A}_{0}\neq\phi$, because if $(a_{0}-\delta,a_{0})\subset
\mathcal{B}_{0}$ then $a_{0}\in\mathcal{B}_{0}$ and if $c\in\mathcal
{C}_{0}\cap(a_{0}-\delta,a_{0})\neq\phi$, the right boundary
of the
connected component of $\mathcal{C}_{0}$ containing $c$ belongs to
$\mathcal{A}_{0}$. Hence, $(a_{0}-\delta,a_{0})\cap\mathcal
{A}_{0}\neq\phi$,
and then $(a_{0}-\delta,a_{0})\cap\widehat{\mathcal{A}}\neq\phi$. Take
$\overline{a}\in(a_{0}-\delta,a_{0})\cap\widehat{\mathcal{A}}$, and we
have that
$a_{0}\in(c_{\overline{a}},\overline{a}+\delta)$.

Since $\mathcal{A}_{0}\cup\mathcal{D}$ is a compact set, we can find finite
sets $\mathcal{A}_{\delta}^{\prime}\subset\mathcal{A}_{0}$ and
$\mathcal{B}%
_{\delta}\subset\mathcal{D}$ such that $ ( \mathcal{A}_{0}\cup
\mathcal{D} ) \subset(
{ \bigcup_{a\in\mathcal{A}_{\delta}^{\prime}}}
(c_{a},a+\delta) ) \cup(
{ \bigcup_{d\in\mathcal{B}_{\delta}}}
(d-\delta,d+\delta) ) $. Finally consider the set $\mathcal{A}_{\delta}$
obtained from $\mathcal{A}_{\delta}^{\prime}$ removing some points in
such a
way that the distance between two consecutive points is larger than
$\delta/2$
and adding the set $ \{ a\in\mathcal{A}_{0}\dvtx a-\max( \mathcal{A}%
_{0}\cap[0,a) ) \geq\delta\} $. Take $\varsigma
=\min_{a\in\mathcal{A}_{\delta}} ( a-c_{a} ) $.
\end{pf}
\end{appendix}

%

%
\printaddresses

\end{document}